\documentclass[10pt]{article}

\usepackage{axodraw}
\usepackage{amssymb}
\begin{document}

%  #] packages:
%  #[ Redefinitions :
\arraycolsep 2pt
\hfuzz 1.2pt

\renewcommand{\topfraction}{1.0}
\renewcommand{\bottomfraction}{1.0}
\renewcommand{\textfraction}{0.0}

\pagestyle{empty} 
\newcommand{\lpar}{\left(}
\newcommand{\rpar}{\right)} 
\newcommand{\parent}[1]{\lpar#1\rpar}

\newcommand{\lrbr}{\left[}
\newcommand{\rrbr}{\right]}
\newcommand{\rbrak}[1]{\lrbr#1\rrbr}

\newcommand{\bq}{\begin{equation}}
\newcommand{\eq}{\end{equation}}
\newcommand{\bqa}{\begin{eqnarray}}
\newcommand{\eqa}{\end{eqnarray}}
\newcommand{\ba}[1]{\begin{array}{#1}}
\newcommand{\ea}{\end{array}}
\newcommand{\nl}{\nonumber\\}

\newcommand{\ord}[1]{{\cal O}\lpar#1\rpar}
\newcommand{\thet}[1]{\theta\lpar#1\rpar}
\newcommand{\delt}[1]{\delta\lpar#1\rpar}

\newcommand{\eqn}[1]{Eq.(\ref{#1})}
\newcommand{\eqns}[2]{Eqs.(\ref{#1}--\ref{#2})}
\newcommand{\fig}[1]{Fig.~\ref{#1}}
\newcommand{\figs}[2]{Figs.~\ref{#1}--\ref{#2}}

\newcommand{\umu}{^{\mu}}
\newcommand{\lmu}{_{\mu}}
\newcommand{\calm}{{\cal M}}
\newcommand{\al}{\alpha}
\newcommand{\be}{\beta}
\newcommand{\ub}{\bar{u}}
\newcommand{\vb}{\bar{v}}
\newcommand{\gp}{(1+\gamma^5)}
\newcommand{\ga}{\gamma}
\newcommand{\ep}{\epsilon}
\newcommand{\sla}[1]{/\!\!\!#1}
\newcommand{\suml}{\sum\limits}
\renewcommand{\to}{\rightarrow}

\newcommand{\ssA}{{\scriptscriptstyle \gamma}}
\newcommand{\ssW}{{\scriptscriptstyle W}}
\newcommand{\ssZ}{{\scriptscriptstyle Z}}
\newcommand{\ssX}{{\scriptscriptstyle X}}
\newcommand{\ssB}{{\scriptscriptstyle B}}
\newcommand{\ssV}{{\scriptscriptstyle V}}

\newcommand{\Umuz}{\hat{\mu}_\ssZ}
\newcommand{\Umuw}{\hat{\mu}_\ssW}
\newcommand{\Umuv}{\hat{\mu}_\ssV}

\newcommand{\muz}{\mu_\ssZ}
\newcommand{\muw}{\mu_\ssW}
\newcommand{\muv}{\mu_\ssV}

\newcommand{\Ue}{\hat{e}}
\newcommand{\Ual}{\hat{\alpha}}

\newcommand{\gw}{g_w}
\newcommand{\gws}{g_w^2}
\newcommand{\Ugw}{\hat{g}_w}
\newcommand{\Ugws}{\hat{g}_w^2}

\newcommand{\Uv}{\hat{v}}
\newcommand{\Ua}{\hat{a}}

\newcommand{\UV}{\hat{V}}

\renewcommand{\iff}{\;\;\Longleftrightarrow\;\;}
\newcommand{\implies}{\;\;\Longrightarrow\;\;}

\newcommand{\stw}{s_w}
\newcommand{\ctw}{c_w}
\newcommand{\stws}{s_w^2}
\newcommand{\stwf}{s_w^4}
\newcommand{\ctws}{c_w^2}

\newcommand{\Ustw}{\hat{s}_w}
\newcommand{\Uctw}{\hat{c}_w}
\newcommand{\Ustws}{\hat{s}_w^2}
\newcommand{\Ustwf}{\hat{s}_w^4}
\newcommand{\Uctws}{\hat{c}_w^2}

\newcommand{\Sgg}{\Sigma_\ssA}
\newcommand{\Szg}{\Sigma_\ssX}
\newcommand{\Szz}{\Sigma_\ssZ}
\newcommand{\Sww}{\Sigma_\ssW}

\newcommand{\USgg}{\hat{\Sigma}_\ssA}
\newcommand{\USzg}{\hat{\Sigma}_\ssX}
\newcommand{\USzz}{\hat{\Sigma}_\ssZ}  
\newcommand{\USww}{\hat{\Sigma}_\ssW}
\newcommand{\USvv}{\hat{\Sigma}_\ssV}

\newcommand{\USggL}{\hat{\Sigma}_{_{\ga,L}}}
\newcommand{\USzgL}{\hat{\Sigma}_{_{X,L}}}
\newcommand{\USzzL}{\hat{\Sigma}_{_{Z,L}}} 
\newcommand{\USwwL}{\hat{\Sigma}_{_{W,L}}}

\newcommand{\Pgg}{\Pi_\ssA}
\newcommand{\Pzg}{\Pi_\ssX}
\newcommand{\Pzz}{\Pi_\ssZ}
\newcommand{\Pww}{\Pi_\ssW}

\newcommand{\UPgg}{\hat{\Pi}_\ssA}
\newcommand{\UPzg}{\hat{\Pi}_\ssX}
\newcommand{\UPzz}{\hat{\Pi}_\ssZ}
\newcommand{\UPww}{\hat{\Pi}_\ssW}

\newcommand{\RPzg}{P_\ssX}
\newcommand{\RPzz}{P_\ssZ}
\newcommand{\RPgg}{P_\ssA}
\newcommand{\RPww}{P_\ssW}

\newcommand{\URPzg}{\hat{P}_\ssX}
\newcommand{\URPzz}{\hat{P}_\ssZ}
\newcommand{\URPgg}{\hat{P}_\ssA}
\newcommand{\URPww}{\hat{P}_\ssW}
\newcommand{\URPvv}{\hat{P}_\ssV}

\newcommand{\GG}{G}

\newcommand{\PW}{$W$}
\newcommand{\PZ}{$Z$}
\newcommand{\gsovermu}{\kappa}

\newcommand{\TeV}{\;\mathrm{TeV}}
\newcommand{\GeV}{\;\mathrm{GeV}}
\newcommand{\MeV}{\;\mathrm{MeV}}

\makeatletter
\ifx\undefined\operator@font
  \let\operator@font=\rm
\fi
\def\Re{\mathop{\operator@font Re}\nolimits}
\def\Im{\mathop{\operator@font Im}\nolimits}

\newcommand{\mut}{m_t^2}
\newcommand{\mw}{m_{_W}}
\newcommand{\gmw}{\Gamma_{_W}}
\newcommand{\mz}{m_{_Z}}
\newcommand{\mzs}{m_{_Z}^2}
\newcommand{\mws}{m_{_W}^2}
\newcommand{\gmz}{\Gamma_{_Z}}
\newcommand{\SA}{S_{\ga}}
\newcommand{\SAl}{S_{\ga,l}}
\newcommand{\SAt}{S_{\ga,t}}
\newcommand{\SW}{S_{_W}}
\newcommand{\TZ}{T_{_Z}}
\newcommand{\TW}{T_{_W}}

\newcommand\pb{\;[\mathrm{pb}]}

\newcommand{\processccten}{$e^-e^+\to \mu^-\bar{\nu}_\mu u\bar{d}$}
\newcommand{\processcceleven}{$e^-e^+\to s\bar{c} u\bar{d}$}
\newcommand{\processcctwenty}{$e^-e^+\to e^-\bar{\nu}_eu\bar{d}$}

\hfuzz 3pt

%  #] Redefinitions : 
%  #[ Some figure definitions :

%  #[ DiagramFermionToBosonFull :

\newcommand{\DiagramFermionToBosonFull}[3][70]{
  \vcenter{\hbox{
  \SetScale{0.8}
  \begin{picture}(#1,50)(15,15)
    \ArrowLine(25,25)(50,50)
    \ArrowLine(50,50)(25,75)
    \Photon(50,50)(90,50){2}{8}   \Text(80,40)[lc]{#2}
    \GCirc(50,50){10}{0}          \Text(60,48)[cb]{#3}
    \Vertex(90,50){2}
  \end{picture}}}
  }

%  #] DiagramFermionToBosonFull : 
%  #[ DiagramFermionToBosonFullWithMomenta :

\newcommand{\DiagramFermionToBosonFullWithMomenta}[8][70]{
  \vcenter{\hbox{
  \SetScale{0.8}
  \begin{picture}(#1,50)(15,15)
    \put(27,22){$\nearrow$}      
    \put(27,54){$\searrow$}    
    \put(59,29){$\to$}    
    \ArrowLine(25,25)(50,50)      \Text(34,20)[lc]{#6} \Text(11,20)[lc]{#3}
    \ArrowLine(50,50)(25,75)      \Text(34,60)[lc]{#7} \Text(11,60)[lc]{#4}
    \Photon(50,50)(90,50){2}{8}   \Text(80,40)[lc]{#2} \Text(55,33)[ct]{#8}
    \GCirc(50,50){10}{0}          \Text(60,48)[cb]{#5} 
    \Vertex(90,50){2}
  \end{picture}}}
  }

%  #] DiagramFermionToBosonFullWithMomenta : 
%  #[ DiagramFermionToBosonFullWithLabels :

\newcommand{\DiagramFermionToBosonFullWithLabels}[6][70]{
  \vcenter{\hbox{
  \SetScale{0.8}
  \begin{picture}(#1,50)(15,15)
    \ArrowLine(25,25)(50,50)      \Text(30,22)[lc]{#4} 
    \ArrowLine(50,50)(25,75)      \Text(30,60)[lc]{#5} 
    \Photon(50,50)(90,50){2}{8}   \Text(80,40)[lc]{#2} \Text(60,35)[ct]{#6}
    \GCirc(50,50){10}{0}          \Text(60,48)[cb]{#3}
    \Vertex(90,50){2}
  \end{picture}}}
  }

%  #] DiagramFermionToBosonFullWithLabels : 
%  #[ DiagramFermionToBosonPropagator :

\newcommand{\DiagramFermionToBosonPropagator}[4][85]{
  \vcenter{\hbox{
  \SetScale{0.8}
  \begin{picture}(#1,50)(15,15)
    \ArrowLine(25,25)(50,50)
    \ArrowLine(50,50)(25,75)
    \Photon(50,50)(105,50){2}{8}   \Text(90,40)[lc]{#2}
    \GCirc(50,50){10}{0.5}         \Text(80,48)[cb]{#3}
    \GCirc(82,50){8}{1}            \Text(55,48)[cb]{#4}
    \Vertex(105,50){2}
  \end{picture}}}
  }

%  #] DiagramFermionToBosonPropagator : 
%  #[ DiagramFermionToBosonEffective :

\newcommand{\DiagramFermionToBosonEffective}[3][70]{
  \vcenter{\hbox{
  \SetScale{0.8}
  \begin{picture}(#1,50)(15,15)
    %\Vertex(15,15){1}
    %\Vertex(15,77.5){1}
    %\Vertex(115,15){1}
    %\Vertex(115,77.5){1}
    \ArrowLine(25,25)(50,50)
    \ArrowLine(50,50)(25,75)
    \Photon(50,50)(90,50){2}{8}   \Text(80,40)[lc]{#2}
    \BBoxc(50,50)(5,5)            \Text(55,48)[cb]{#3}
    \Vertex(90,50){2}
  \end{picture}}}
  }

%  #] DiagramFermionToBosonEffective : 
%  #[ DiagramMa :

\newcommand{\DiagramMa}{
  \SetScale{0.8}
  \begin{picture}(115,35)(0,37)
    \ArrowLine(25,25)(50,50)
    \ArrowLine(50,50)(25,75)      \Text(55,48)[cb]{$\gamma$}
    \Photon(50,50)(90,50){2}{8}   \Text(6,21)[lc]{$e^-$}
    \BBoxc(50,50)(5,5)            \Text(6,63)[lc]{$e^+$}
    \ArrowLine(115,25)(90,50)     \Text(98,21)[lc]{$\bar{f}$}
    \ArrowLine(90,50)(115,75)     \Text(98,62)[lc]{$f$}
    \BBoxc(90,50)(5,5) 
  \end{picture}
  }

%  #] DiagramMa : 
%  #[ DiagramMb :

\newcommand{\DiagramMb}{
  \SetScale{0.8}
  \begin{picture}(115,35)(0,37)
    \ArrowLine(25,25)(50,50)
    \ArrowLine(50,50)(25,75)      \Text(55,48)[cb]{$Z$}
    \Photon(50,50)(90,50){2}{8}   \Text(6,21)[lc]{$e^-$}
    \BBoxc(50,50)(5,5)            \Text(6,63)[lc]{$e^+$}
    \ArrowLine(115,25)(90,50)     \Text(98,21)[lc]{$\bar{f}$}
    \ArrowLine(90,50)(115,75)     \Text(98,62)[lc]{$f$}
    \BBoxc(90,50)(5,5) 
  \end{picture}
  }

%  #] DiagramMb : 
%  #[ DiagramVWWa :

\newcommand{\DiagramVWWa}{
  %\SetScale{1.0}
  \begin{picture}(140,130)(-15,-15)
    \ArrowLine(30,50)(73,75)        \Text(50,75)[lc]{$f$}
    \ArrowLine(73,75)(73,25)        \Text(50,25)[lc]{$f$}
    \ArrowLine(73,25)(30,50)        \Text(77,50)[lc]{$f'$}
    \Vertex(30,50){1.5}
    \Vertex(73,75){1.5}
    \Vertex(73,25){1.5}
    \Photon(  0, 50)(30,50){2}{4}   \Text(-16, 50)[lc]{$B_\mu$}
    \Photon(100,100)(73,75){2}{4}   \Text(104,104)[lc]{$W^-_\lambda$}
    \Photon(100,  0)(73,25){2}{4}   \Text(104, -4)[lc]{$W^+_\kappa$}
    \LongArrow(10,57)(20,57)        \Text( 13, 65)[lc]{$q$}
    \LongArrow(89,98)(82,91)        \Text( 79,103)[lc]{$p_-$}
    \LongArrow(89, 2)(82, 9)        \Text( 79, -3)[lc]{$p_+$}
  \end{picture}
  }

%  #] DiagramVWWa : 
%  #[ DiagramVWWb :

\newcommand{\DiagramVWWb}{
  %\SetScale{1.0}%
  \begin{picture}(140,130)(-15,-15)
    \ArrowLine(73,75)(30,50)        \Text(50,75)[lc]{$f$}
    \ArrowLine(73,25)(73,75)        \Text(50,25)[lc]{$f$}
    \ArrowLine(30,50)(73,25)        \Text(77,50)[lc]{$f'$}
    \Vertex(30,50){1.5}
    \Vertex(73,75){1.5}
    \Vertex(73,25){1.5}
    \Photon(  0, 50)(30,50){2}{4}   \Text(-16, 50)[lc]{$B_\mu$}
    \Photon(100,100)(73,75){2}{4}   \Text(104,104)[lc]{$W^-_\lambda$}
    \Photon(100,  0)(73,25){2}{4}   \Text(104, -4)[lc]{$W^+_\kappa$}
    \LongArrow(10,57)(20,57)        \Text( 13, 65)[lc]{$q$}
    \LongArrow(89,98)(82,91)        \Text( 79,103)[lc]{$p_-$}
    \LongArrow(89, 2)(82, 9)        \Text( 79, -3)[lc]{$p_+$}
  \end{picture}
  }

%  #] DiagramVWWb : 
%  #[ DiagramMaa :

\newcommand{\DiagramMaa}{
  \vcenter{\hbox{
  %\SetScale{0.8}%
  \begin{picture}(150,100)(0,0)
    \ArrowLine(25,25)(50,50)        \Text(10,25)[lc]{$e^-$}
    \ArrowLine(50,50)(25,75)        \Text(10,75)[lc]{$e^+$}
    \GCirc(50,50){8}{0}
    \Photon(50,50)(100,50){1}{9}    \Text(75,55)[bc]{$\gamma,Z$}
    \Photon(100,50)(125,25){1}{7}   \Text(104,30)[lc]{$W$}
    \Photon(100,50)(125,75){1}{7}   \Text(104,70)[lc]{$W$}
    \GCirc(100,50){10}{0.5}
    \ArrowLine(150,10)(125,25)      \Text(165,10)[rc]{$\bar{f}_2'$}
    \ArrowLine(125,25)(150,40)      \Text(165,40)[rc]{$f_2$}
    \GCirc(125,25){8}{0}
    \ArrowLine(150,60)(125,75)      \Text(165,60)[rc]{$\bar{f}_1'$}
    \ArrowLine(125,75)(150,90)      \Text(165,90)[rc]{$f_1$}
    \GCirc(125,75){8}{0}
  \end{picture}}}
  }

%  #] DiagramMaa : 
%  #[ DiagramMab :

\newcommand{\DiagramMab}{
  \vcenter{\hbox{
  %\SetScale{0.8}%
  \begin{picture}(150,100)(0,0)
    \ArrowLine(25,25)(50,50)        \Text(10,25)[lc]{$e^-$}
    \ArrowLine(50,50)(25,75)        \Text(10,75)[lc]{$e^+$}
    \Photon(50,50)(100,50){1}{9}    \Text(75,55)[bc]{$\gamma,Z$}
    \Photon(100,50)(125,25){1}{7}   \Text(104,30)[lc]{$W$}
    \Photon(100,50)(125,75){1}{7}   \Text(104,70)[lc]{$W$}
    \ArrowLine(150,10)(125,25)      \Text(165,10)[rc]{$\bar{f}_2'$}
    \ArrowLine(125,25)(150,40)      \Text(165,40)[rc]{$f_2$}
    \ArrowLine(150,60)(125,75)      \Text(165,60)[rc]{$\bar{f}_1'$}
    \ArrowLine(125,75)(150,90)      \Text(165,90)[rc]{$f_1$}
    \BBoxc(50,50)(5,5)
    \BBoxc(100,50)(5,5)
    \BBoxc(125,25)(5,5)
    \BBoxc(125,75)(5,5)
  \end{picture}}}
  }
%  #] DiagramMab : 
%  #[ DiagramMba :

\newcommand{\DiagramMba}{
  \vcenter{\hbox{
  %\SetScale{0.8}%
  \begin{picture}(150,100)(0,0)
    \ArrowLine(25,25)(50,50)        \Text(10,25)[lc]{$e^-$}
    \ArrowLine(50,50)(25,75)        \Text(10,75)[lc]{$e^+$}
    \Photon(50,50)(100,50){1}{9}    \Text(75,55)[bc]{$\gamma,Z$}
    \ArrowLine(120,30)(100,50)      \Text(100,40)[lt]{$f_1$}
    \ArrowLine(140,10)(120,30)      \Text(155,10)[rc]{$\bar{f}_1'$}
    \ArrowLine(100,50)(140,90)      \Text(155,90)[rc]{$f_1$}
    \Photon(120,30)(150,30){1}{7}   \Text(135,35)[bc]{$W$}
    \ArrowLine(175,15)(150,30)      \Text(190,15)[rc]{$\bar{f}_2'$}
    \ArrowLine(150,30)(175,45)      \Text(190,45)[rc]{$f_2$}
    \BBoxc(50,50)(5,5)
    \BBoxc(100,50)(5,5)
    \BBoxc(120,30)(5,5)
    \BBoxc(150,30)(5,5)
  \end{picture}}}
  }

%  #] DiagramMba : 
%  #[ DiagramMbb :

\newcommand{\DiagramMbb}{
  \vcenter{\hbox{
  %\SetScale{0.8}%
  \begin{picture}(150,100)(0,0)
    \ArrowLine(25,25)(50,50)        \Text(10,25)[lc]{$e^-$}
    \ArrowLine(50,50)(25,75)        \Text(10,75)[lc]{$e^+$}
    \Photon(50,50)(100,50){1}{9}    \Text(75,55)[bc]{$\gamma,Z$}
    \ArrowLine(120,70)(140,90)      \Text(155,90)[rc]{$f_1$}
    \ArrowLine(100,50)(120,70)      \Text(100,65)[lb]{$f_1'$}
    \ArrowLine(140,10)(100,50)      \Text(155,10)[rc]{$\bar{f}_1'$}
    \Photon(120,70)(150,70){1}{7}   \Text(135,65)[tc]{$W$}
    \ArrowLine(175,55)(150,70)      \Text(190,55)[rc]{$\bar{f}_2'$}
    \ArrowLine(150,70)(175,85)      \Text(190,85)[rc]{$f_2$}
    \BBoxc(50,50)(5,5)
    \BBoxc(100,50)(5,5)
    \BBoxc(120,70)(5,5)
    \BBoxc(150,70)(5,5)
  \end{picture}}}
  }

%  #] DiagramMbb : 
%  #[ DiagramMc :

\newcommand{\DiagramMc}{
  \vcenter{\hbox{
  %\SetScale{0.8}%
  \begin{picture}(150,100)(0,0)
    \ArrowLine(25,80)(50,75)        \Text(10,80)[lc]{$e^-$}
    \ArrowLine(50,75)(50,25)        \Text(35,50)[lc]{$\nu_e$}
    \ArrowLine(50,25)(25,20)        \Text(10,20)[lc]{$e^+$}
    \Photon(50,75)(80,80){1}{7}     \Text(65,72.5)[tc]{$W$}
    \Photon(50,25)(80,20){1}{7}     \Text(65,27.5)[bc]{$W$}
    \ArrowLine(105,65)(80,80)       \Text(120,65)[rc]{$\bar{f}_1'$}
    \ArrowLine(80,80)(105,95)       \Text(120,95)[rc]{$f_1$}
    \ArrowLine(105,5)(80,20)        \Text(120,5)[rc]{$\bar{f}_2'$}
    \ArrowLine(80,20)(105,35)       \Text(120,35)[rc]{$f_2$}
    \BBoxc(50,75)(5,5)
    \BBoxc(50,25)(5,5)
    \BBoxc(80,80)(5,5)
    \BBoxc(80,20)(5,5)
  \end{picture}}}
}

%  #] DiagramMc :
%  #[ DiagramMda :

\newcommand{\DiagramMda}{
  \vcenter{\hbox{
  %\SetScale{0.8}%
  \begin{picture}(150,100)(0,0)
    \ArrowLine(25,25)(50,50)        \Text(10,25)[lc]{$e^-$}
    \ArrowLine(50,50)(25,75)        \Text(10,75)[lc]{$e^+$}
    \GCirc(50,50){8}{0}
    \Photon(50,50)(100,50){1}{9}    \Text(75,55)[bc]{$B_1$}
    \Photon(100,50)(125,25){1}{7}   \Text(104,30)[lc]{$B_3$}
    \Photon(100,50)(125,75){1}{7}   \Text(104,73)[lc]{$B_2$}
    \GCirc(100,50){10}{0.5}
    \ArrowLine(150,10)(125,25)      \Text(165,10)[rc]{$\bar{f}_2$}
    \ArrowLine(125,25)(150,40)      \Text(165,40)[rc]{$f_2$}
    \GCirc(125,25){8}{0}
    \ArrowLine(150,60)(125,75)      \Text(165,60)[rc]{$\bar{f}_1$}
    \ArrowLine(125,75)(150,90)      \Text(165,90)[rc]{$f_1$}
    \GCirc(125,75){8}{0}
  \end{picture}}}
  }

%  #] DiagramMda : 
%  #[ DiagramMdb :

\newcommand{\DiagramMdb}{
  \vcenter{\hbox{
  %\SetScale{0.8}%
  \begin{picture}(150,100)(0,0)
    \ArrowLine(25,25)(50,50)        \Text(10,25)[lc]{$e^-$}
    \ArrowLine(50,50)(25,75)        \Text(10,75)[lc]{$e^+$}
    \Photon(50,50)(100,50){1}{9}    \Text(75,55)[bc]{$B_1$}
    \Photon(100,50)(125,25){1}{7}   \Text(104,30)[lc]{$B_3$}
    \Photon(100,50)(125,75){1}{7}   \Text(104,73)[lc]{$B_2$}
    \ArrowLine(150,10)(125,25)      \Text(165,10)[rc]{$\bar{f}_2$}
    \ArrowLine(125,25)(150,40)      \Text(165,40)[rc]{$f_2$}
    \ArrowLine(150,60)(125,75)      \Text(165,60)[rc]{$\bar{f}_1$}
    \ArrowLine(125,75)(150,90)      \Text(165,90)[rc]{$f_1$}
    \BBoxc(50,50)(5,5)
    \BBoxc(100,50)(5,5)
    \BBoxc(125,25)(5,5)
    \BBoxc(125,75)(5,5)
  \end{picture}}}
  }
%  #] DiagramMdb :  
%  #[ DiagramMea :

\newcommand{\DiagramMea}{
  \vcenter{\hbox{
  %\SetScale{0.8}
  \begin{picture}(150,100)(0,0)
    \ArrowLine(25,25)(50,50)        \Text(10,25)[lc]{$e^-$}
    \ArrowLine(50,50)(25,75)        \Text(10,75)[lc]{$e^+$}
    \Photon(50,50)(100,50){1}{9}    \Text(75,55)[bc]{$\gamma,Z$}
    \ArrowLine(120,30)(100,50)      \Text(100,40)[lt]{$f_1$}
    \ArrowLine(140,10)(120,30)      \Text(155,10)[rc]{$\bar{f}_1$}
    \ArrowLine(100,50)(140,90)      \Text(155,90)[rc]{$f_1$}
    \Photon(120,30)(150,30){1}{7}   \Text(135,35)[bc]{$\gamma,Z$}
    \ArrowLine(175,15)(150,30)      \Text(190,15)[rc]{$\bar{f}_2$}
    \ArrowLine(150,30)(175,45)      \Text(190,45)[rc]{$f_2$}
    \BBoxc(50,50)(5,5)
    \BBoxc(100,50)(5,5)
    \BBoxc(120,30)(5,5)
    \BBoxc(150,30)(5,5)
  \end{picture}}}
  }

%  #] DiagramMea : 
%  #[ DiagramMeb :

\newcommand{\DiagramMeb}{
  \vcenter{\hbox{
  %\SetScale{0.8}%
  \begin{picture}(150,100)(0,0)
    \ArrowLine(25,25)(50,50)        \Text(10,25)[lc]{$e^-$}
    \ArrowLine(50,50)(25,75)        \Text(10,75)[lc]{$e^+$}
    \Photon(50,50)(100,50){1}{9}    \Text(75,55)[bc]{$\gamma,Z$}
    \ArrowLine(120,70)(140,90)      \Text(155,90)[rc]{$f_1$}
    \ArrowLine(100,50)(120,70)      \Text(100,65)[lb]{$f_1$}
    \ArrowLine(140,10)(100,50)      \Text(155,10)[rc]{$\bar{f}_1$}
    \Photon(120,70)(150,70){1}{7}   \Text(135,65)[tc]{$\gamma,Z$}
    \ArrowLine(175,55)(150,70)      \Text(190,55)[rc]{$\bar{f}_2$}
    \ArrowLine(150,70)(175,85)      \Text(190,85)[rc]{$f_2$}
    \BBoxc(50,50)(5,5)
    \BBoxc(100,50)(5,5)
    \BBoxc(120,70)(5,5)
    \BBoxc(150,70)(5,5)
  \end{picture}}}
  }

%  #] Some figure definitions : 
%  #[ Title page :

\begin{flushright}
NIKHEF 96-031\\
PSI-PR-96-41  
\end{flushright}

\vspace{\fill}

\begin{center}

\def\demo{$\Delta\eta\mu \acute{o} \kappa \varrho \iota \tau o \varsigma$}
\vspace{\baselineskip}%
{\Large {\bf The fermion-loop scheme for finite-width effects \\[1ex]
in \boldmath{$e^+e^-$} annihilation into four fermions}}\\[2em]
\vspace{\baselineskip}%
Wim~Beenakker,
Geert Jan van Oldenborgh\\
{\it Instituut--Lorentz, Rijksuniversiteit Leiden, the Netherlands}\\[1mm]
Ansgar~Denner\\
{\it Paul-Scherrer-Institut, W\"urenlingen und Villigen,
Switzerland}\\[1mm]
Stefan~Dittmaier\\
{\it Institut f\"ur Theoretische Physik, Universit\"at Bielefeld, Germany}
\\[1mm]
Jiri~Hoogland\\
{\it NIKHEF-H, Amsterdam, the Netherlands}\\[1mm]
Ronald~Kleiss\\
{\it University of Nijmegen, Nijmegen, the Netherlands}\\[1mm]
Costas G.~Papadopoulos\\
{\it Institute of Nuclear Physics, NRCPS \demo, Athens, Greece}\\[1mm]
Giampiero~Passarino\\
{\it Dipartimento di Fisica Teorica, Universit\`a di Torino, Italy}\\
{\it INFN, Sezione di Torino, Italy}\\
\vspace{2\baselineskip}
\end{center}
{\bf Abstract}\\
We describe the gauge-invariant treatment of the finite-width 
effects of $W$ and $Z$ bosons in the fermion-loop scheme and
its application to the six-fermion (LEP2) processes $e^-e^+ \to\,$ four
fermions, with massless external fermions.
The fermion-loop scheme 
consists in including all fermio\-nic one-loop corrections in 
tree-level amplitudes and resumming the self-energies. 
We give explicit results for the unrenormalized fermionic one-loop 
contributions to the gauge-boson self-energies and the triple gauge-boson 
vertices, and perform the renormalization in a gauge-invariant way by 
introducing complex pole positions and running couplings. A simple effective 
Born prescription is presented, which allows for a relatively straightforward 
implementation of the fermion-loop scheme in LEP1 and LEP2 processes. We apply 
this prescription to typical LEP2 processes, i.e., \processccten, 
\processcceleven, and \processcctwenty\/, and give numerical comparisons 
with other gauge-invariance-preserving schemes in
the energy range of LEP2,
NLC and beyond.
\newpage
\pagestyle{plain}
\setcounter{page}{1}

%  #] Title page : 
%  #[ Introduction :

\section{Introduction}
\label{se:introduction}

The incorporation of finite-width effects in the theoretical predictions for
LEP2 processes and their implementation in the corresponding event
generators necessitate a careful treatment. Independently of how finite
widths of propagating particles are introduced, this requires (or at
least mimics) a resummation of the vacuum-polarization effects. However,
thereby the principle of gauge invariance must not be violated, i.e.,
the Ward identities have to be preserved; otherwise theoretical
uncertainties may get out of control.

In a previous article \cite{bhf1} we discussed several schemes that allow 
the incorporation of finite-width effects in tree-level amplitudes without 
spoiling gauge invariance. 
We argued that the preferable (fermion-loop) scheme consists in the 
resummation of the fermionic one-loop corrections to the vector-boson 
propagators and the inclusion of all remaining fermionic one-loop corrections,
in particular those to the Yang--Mills vertices. 
This resummation of one-particle-irreducible (1PI) fermionic $\ord{\al}$ 
corrections involves the closed set of all $\ord{[N^f_c \al/\pi]^i}$ 
(leading color-factor) corrections, and is as such
manifestly 
gauge-invariant. These corrections constitute the bulk of the
width effects for gauge bosons and an important
part of the complete set of weak corrections. 

In Ref.~\cite{bhf1} our main incentive was the discussion of the process
$e^-e^+ \to e^-{\bar\nu}_e u{\bar d}$ at small scattering angles and LEP2 
energies. 
Naive inclusion of the finite \PW-boson width breaks U(1) electromagnetic
gauge invariance and leads to a totally wrong cross-section 
in the collinear limit, as e.g.\ discussed in Ref.~\cite{Kurihara}. 
By taking into account in addition the imaginary
parts arising from cutting the massless fermion loops in the
triple-gauge-boson vertex, U(1) gauge invariance is restored and a sensible
cross-section is obtained.

After introducing the full fermion-loop (FL) scheme we restricted the
explicit discussion in Ref.~\cite{bhf1} to the minimal set of terms that
are necessary to solve the U(1) problem in $e^-e^+ \to e^-{\bar\nu}_e
u{\bar d}$, 
namely the imaginary parts of the contributions of massless fermions.
In this paper we give the details of the full-fledged fermion-loop scheme,
taking into account the complete fermionic one-loop
corrections including all real and imaginary parts, and all contributions of 
the massive top quark. 
We perform a proper
treatment of the neutral gauge-boson propagators by solving 
the Dyson equations for the photon, $Z$-boson, 
and mixed photon--$Z$ propagators.
This is necessary to guarantee the unitarity
cancellations at high energies. 
The top-quark contributions are particularly important for 
delayed-unitarity effects. In this respect also terms involving the
totally-antisymmetric $\varepsilon$-tensor (originating from vertex
corrections) are relevant.
While such terms are absent for complete generations of massless fermions
owing to the anomaly cancellations, they show up  for finite fermion
masses.  As the $\varepsilon$-dependent terms
satisfy the Ward identities by themselves, they can be left out in more
minimal treatments like the one used in Ref.~\cite{bhf1}.

We formulate
a renormalization of the fermion-loop corrections using the language of 
running couplings.
We show how to rewrite bare amplitudes in terms of these renormalized
couplings and demonstrate that the resulting renormalized amplitudes 
respect gauge invariance, i.e., that they fulfill the relevant Ward identities.
Moreover, we give the explicit analytical results for the fermionic
one-loop contributions to the gauge-boson self-energies and the triple 
gauge-boson vertices, which represent the necessary ingredients for
applying the FL scheme to LEP2 processes with massless
external fermions, i.e., $e^-e^+\to4f$.

The purpose of this paper is twofold. On the one hand it provides
the justification of the FL scheme: we show that it is fully consistent,
i.e., it is gauge-invariant, respects all relevant Ward identities, and
describes the finite-width effects correctly.
The full FL scheme includes the gauge-invariant subset of all
one-loop fermionic corrections.
However, our studies reveal that only relatively simple 
subsets of the fermionic corrections are required to arrive at a consistent 
description of finite-width effects for tree-level calculations.
On the other hand the full FL scheme is also a
starting point for the calculation of the ${\cal O}(\alpha)$ corrections
to six-fermion processes. It includes all fermionic corrections and
might therefore be used as a first approximation of the corrected 
cross-sections.
However, very often our experience has shown, 
especially at LEP1, that bosonic corrections may become 
sizeable and comparable to the fermionic ones \cite{LEPreport95}.
A large part of the bosonic corrections, as e.g.\ the leading-logarithmic 
corrections, factorize and can be treated by a convolution. Nevertheless
the remaining bosonic corrections can still be non-negligible, i.e.,
of the order of one percent at LEP2 \cite{LEP2WWreport} and even larger
at higher energies \cite{NLCreport}.
For the bosonic corrections a gauge-invariant treatment similar to
the FL scheme, i.e., a Dyson summation without violating Ward
identities, can 
be performed within the background-field method \cite{bfm}. 
However, the resulting matrix elements are gauge-parameter-dependent
at the loop level that is not completely taken into account.

The outline of the article is as follows: In Section~\ref{se:renormalization}
we describe the 
renormalization procedure and present the renormalized, resummed amplitudes 
for LEP1 processes, i.e., the four-fermion processes 
$e^+e^-\to f\bar{f}$ with massless external fermions.
In Section~\ref{se:VWW_vertex}
we discuss the fermionic one-loop corrections to the triple 
gauge-boson vertex and construct, from a set of basic matrix elements, the 
renormalized, resummed amplitudes for LEP2 processes, i.e., the
six-fermion processes $e^-e^+\to 4f$ with massless external fermions. 
By explicitly
checking the relevant Ward identities, we verify the gauge invariance of these
amplitudes.
Using the processes 
\processccten, \processcceleven, and \processcctwenty\/
as examples, 
we give a numerical comparison with other schemes and show
the numerical relevance of the resummed fermionic corrections 
at LEP2 energies and beyond in Section~\ref{se:process}.
Finally, a conclusion and outlook are given. 
In the appendices we give supplementary explicit formulae.

%  #] Introduction : 
%  #[ Renormalization :

\section{Renormalization}
\label{se:renormalization}

%  #[ Introduction :

In this section we address the issue of how to define the renormalization, 
while at the same time performing a resummation of 1PI fermionic $\ord{\al}$
corrections. In order to keep the expressions as compact as possible we 
renormalize the gauge-boson masses at their (gauge-invariant) complex pole
\cite{VeltmanUnstable}. This so-called
``complex'' scheme is also favored from a theoretical point of view 
\cite{bigwidth}.
It should be noted, however, that
the real part of the complex pole differs only by terms of order
$\Gamma^2/M^2$ from the gauge-boson mass in the
more familiar ``LEP1'' scheme, which is defined as the zero in the real
part of the inverse gauge-boson propagator.
As a result of the smallness of the decay widths of the 
gauge bosons this difference is marginal.
In fact, all expressions in this 
section can also be computed using the LEP1 language, for which the 
FL scheme works equivalently well 
(see App.~\ref{app:complex_vs_LEP1}). 
The computation of the renormalized parameters in the complex
renormalization scheme is discussed in 
App.~\ref{app:complex_scheme}.

Starting from the bare resummed propagators, the renormalization can be 
performed by first introducing renormalized masses and by subsequently 
rewriting all bare couplings in terms of running ones. The so-obtained 
renormalized resummed gauge-boson propagators automatically combine with the 
fermion--gauge-boson vertex functions to form dressed Born
matrix elements, involving running couplings and propagator functions.
This appealing feature is exemplified by the LEP1 processes, i.e.,
$e^+e^-\to f\bar{f}$ with massless external fermions.

%  #] Introduction : 
%  #[ Resumming the propagators :

\subsection{Resumming the propagators}

The natural way to incorporate the running-width effects of weak vector
bosons is the resummation of the corresponding self-energy corrections.
In the following we consider only fermionic one-loop corrections. 
As the  on-shell widths of the \PW\ and \PZ~bosons are entirely determined 
by fermionic decays, and the fermionic corrections are  separately
gauge-invariant and manifestly gauge-independent,
this is a sensible approach.
Solving the Dyson equations for the photon, \PZ, \PW, and mixed 
photon--\PZ~propagators, one finds the following expressions for the 
transverse parts of the dressed propagators
\begin{eqnarray}
\label{def:resummed}
\URPgg(p^2)   & = & { p^2 - \Umuz + \USzz(p^2) \over
                      \left[ p^2         + \USgg(p^2) \right]
                      \left[ p^2 - \Umuz + \USzz(p^2) \right]
                    - \left[ \USzg(p^2)               \right]^2 } \;,\nl
\URPzg(p^2)   & = & {   - \USzg(p^2) \over
                      \left[ p^2         + \USgg(p^2) \right]
                      \left[ p^2 - \Umuz + \USzz(p^2) \right]
                    - \left[ \USzg(p^2)               \right]^2 } \;,\nl
\URPzz(p^2)   & = & { p^2 + \USgg(p^2) \over
                      \left[ p^2         + \USgg(p^2) \right]
                      \left[ p^2 - \Umuz + \USzz(p^2) \right]
                    - \left[ \USzg(p^2)               \right]^2 } \;,\nl
\URPww(p^2)   & = & { 1 \over  p^2 - \Umuw + \USww(p^2) } \;,
\end{eqnarray}
where $X$ denotes the photon--\PZ~mixing, and $\Umuz$ and $\Umuw$ represent the
gauge-boson masses squared. Here and in the following, all parameters with a 
hat represent bare parameters. Functions with a hat depend on bare parameters.
As the top-quark mass appears in the considered processes only
at the one-loop level we need not explicitly distinguish between the bare
and the renormalized top-quark masses. All other fermion masses are
neglected whenever possible.

The self-energies $\hat{\Sigma}_{i}(p^2)$, with $i=Z,W$, are split into two 
pieces:
\begin{eqnarray} \label{eq:ZW_selfenergy_split}
\USzz(p^2) = \hat{\overline{\Sigma}}_{\ssZ}(p^2)
                        + { \Ugws \over \Uctws }\, T_{\ssZ}(p^2)\;,\nl
\USww(p^2) = \hat{\overline{\Sigma}}_{\ssW}(p^2)
                        + \Ugws\, T_{\ssW}(p^2)\;,
\end{eqnarray}
with $\Uctws \equiv 1-\Ustws \equiv \Umuw/\Umuz$, 
$\Ugws \equiv \Ue^2/(2\Ustws)$, 
and $\Ue$ is the bare electromagnetic coupling.
The first terms contain the universal contributions that have the
structure of the photon self-energy, 
the second terms contain extra contributions that depend on the fermion masses
and vanish with vanishing $m_f$. The dominant contributions to $T_{\ssZ}$ and 
$T_{\ssW}$ originate from the top quark.
The explicit form of
the self-energies can be found in App.~\ref{app:self-energies}.
We use the Feynman rules of Ref.~\cite{De93}.

As can be inferred from the explicit expressions, the universal parts of the 
self-energies are related as follows:
\begin{eqnarray}
\label{eq:relationsSigmas}
\hat{ \overline{\Sigma} }_\ssZ(p^2) & = & {\Uctws-\Ustws\over\Uctws}\,
                 \hat{\overline{\Sigma}}_\ssW(p^2) 
                 + {\Ustws\over\Uctws}\, \USgg(p^2) \;,\nl
\USzg(p^2) & = & -{\Ustw\over\Uctw}\lrbr 
                 \hat{\overline{\Sigma}}_\ssW(p^2) - \USgg(p^2) \rrbr \;.
\end{eqnarray}
These relations are in fact a consequence of gauge invariance.

%  #] Resumming the propagators : 
%  #[ Definition of renormalized parameters :

\subsection{Definition of renormalized parameters}

The \PW\ and \PZ\ bosons are renormalized at their {\it complex\/} pole% 
\footnote{Note that the real and imaginary parts of the complex vector-boson 
masses can be related by means of the optical theorem~\cite{bigwidth}.}.  
In this ``complex'' scheme the renormalized complex squared
masses $\mu_{\ssW,\ssZ}$ of the \PW\ and \PZ\ bosons are defined by
\begin{eqnarray}
\label{def:Wmassrenormalization}
\Umuw & = & \muw + \USww(\muw) \;, \\
\label{def:Zmassrenormalization}
\Umuz & = & \muz + \hat{Z}(\muz) \;,
\end{eqnarray}
with 
\begin{equation}
\label{def:Z(p^2)}
\hat{Z}(p^2) = \USzz(p^2) - { \USzg^2(p^2) \over p^2+\USgg(p^2) } \;.
\end{equation}
After mass renormalization, i.e., insertion of
\eqns{def:Wmassrenormalization}{def:Zmassrenormalization},
the resummed propagators in \eqn{def:resummed} can 
be rewritten as
\begin{eqnarray}\label{eq:resummed_propagators}
\URPgg(p^2) & = & \lrbr { p^2 \over p^2 + \hat{\Sigma}_\ssA(p^2) }\rrbr
                         { 1 \over p^2 }
                + \lrbr { \hat{\Sigma}_\ssX(p^2) \over p^2 
                          + \hat{\Sigma}_\ssA(p^2) }\rrbr^2 \URPzz(p^2) 
                \;,\nl
\URPzg(p^2) & = & \lrbr { -\hat{\Sigma}_\ssX(p^2) \over p^2 
                  + \hat{\Sigma}_\ssA(p^2) } \rrbr \URPzz(p^2) \;, \nl
\hat{P}_\ssZ(p^2) & = & \lrbr { p^2-\muz \over p^2-\muz + \hat{Z}(p^2) 
                        - \hat{Z}(\muz) }\rrbr { 1 \over p^2-\muz } \;,\nl
\hat{P}_\ssW(p^2) & = & \lrbr { p^2-\muw \over p^2-\muw 
                        + \hat{\Sigma}_\ssW(p^2) - \hat{\Sigma}_\ssW(\muw) } 
                        \rrbr { 1 \over p^2-\muw } \;.
\end{eqnarray}
Now we can renormalize the bare couplings by expressing them in terms of 
running couplings%
\footnote{It is not necessary to fix the input parameters in order to carry out
the renormalization. The running couplings can be adjusted to input
parameters later on (see App.~\ref{app:complex_scheme}).}:
\begin{eqnarray}
\label{def:runningalpha}
{ e^2(p^2)\over\hat{e}^2 } & = & { p^2 \over p^2 + \hat{\Sigma}_\ssA(p^2) }
                                 \;,\\
\label{def:runninggw}
{ {g}_w^2(p^2)\over\hat{g}_w^2 } & = & { p^2 \over p^2 
                                       + \hat{\overline{\Sigma}}_\ssW(p^2) }
                                       \;,\\
\label{def:runningstws}
{ \stws(p^2) \over \Ustws } & = & { p^2+\hat{\overline{\Sigma}}_\ssW(p^2) 
                                  \over p^2+\hat{\Sigma}_\ssA(p^2) } \;,
\end{eqnarray}
which fulfill, in analogy to the bare relation $\Ugws=\Ue^2 /(2 \Ustws)$, the 
relation
\begin{equation}
\label{eq:gws(p2)=es(p2)/2stws(p2)}
  \gws(p^2) = { e^2(p^2) \over 2\stws(p^2) } \;.
\end{equation}
This allows us to rewrite $\hat{Z}(p^2)$ using \eqn{eq:relationsSigmas}:
\begin{eqnarray}
\label{eq:Z(p^2)}
  { p^2+\hat{Z}(p^2) \over p^2 } & = & { \Ugws \over \Uctws }
                         \lrbr { \ctws(p^2) \over \gws(p^2) }
                         + { T_{\ssZ}(p^2) \over p^2 } \rrbr \;, 
\end{eqnarray}
with $\ctws(p^2) \equiv 1-\stws(p^2)$. Using \eqn{def:runninggw} and
\eqn{eq:Z(p^2)}, we can rewrite the mass-renormalization conditions
given in \eqns{def:Wmassrenormalization}{def:Zmassrenormalization} as
\begin{eqnarray}
\label{def:Wmassrenormalization2}
{ \Umuw \over \muw } & = & \Ugws \lrbr { 1 \over \gws(\muw) }
                           + { T_{\ssW}(\muw) \over \muw} \rrbr \;,\\
\label{def:Zmassrenormalization2}
{ \Umuz \over \muz } & = & { \Ugws \over \Uctws } 
                           \lrbr { \ctws(\muz) \over \gws(\muz) }       
                           + { T_{\ssZ}(\muz) \over \muz} \rrbr \;.
\end{eqnarray}

%  #] Definition of renormalized parameters : 
%  #[ Renormalized propagators and fermion--gauge-boson vertex functions :

\subsection{Renormalized propagators and fermion--gauge-boson vertex functions}
\label{se:LEP1_vertices}

Using the results of the previous subsection, the bare propagators 
\eqn{eq:resummed_propagators} can 
be expressed in terms of running couplings in the following way:
\begin{eqnarray}
\URPgg(p^2) & = & { e^2(p^2) \over \Ue^2 }\; { 1 \over p^2 } 
                  \;\;+\;\; { \Ustws \over \Uctws }
                  \lrbr { \stws(p^2) \over \Ustws } - 1 \rrbr^2 \URPzz(p^2)
                  \;,\nl
\URPzg(p^2) & = & {\Ustw \over \Uctw } \;
                  \lrbr {\stws(p^2) \over \Ustws } - 1 \rrbr \URPzz(p^2) \;,\nl
\URPzz(p^2) & = & { \gws(p^2) \over \Ugws } \; { \Uctws \over \ctws(p^2) } \;
                  { \chi_\ssZ(p^2) \over p^2 } \;,\nl
\URPww(p^2) & = & { \gws(p^2) \over \Ugws }\; { \chi_\ssW(p^2) \over p^2 } \;,
\label{eq:rewritten_bare_propagators}
\end{eqnarray}
where we define the propagator functions $\chi_{\ssZ,\ssW}(p^2)/p^2$ by
\begin{eqnarray}
\label{def:chiz}
\chi_\ssZ^{-1}(p^2) & = & 1 - { \gws(p^2) \over p^2\,\ctws(p^2) }\; 
                          \left[ { \muz\,\ctws(\muz) \over \gws(\muz) }
                                 - T_{\ssZ}(p^2) + T_{\ssZ}(\muz) 
                          \right] \;,\\
\label{def:chiw}
\chi_\ssW^{-1}(p^2) & = & 1 - { \gws(p^2) \over p^2} \; \left[              
                          { \muw \over \gws(\muw) } - T_{\ssW}(p^2) 
                          + T_{\ssW}(\muw) \right] \;.
\end{eqnarray}
The so-defined propagator functions are finite, as the ultraviolet (UV)
divergences in $T_{\ssW}$ and $T_{\ssZ}$ are independent of $p^2$ and hence 
cancel (see App.~\ref{app:self-energies}). 
The propagator functions are real%
\footnote{For time-like momenta ($p^2>0$) the original propagators contain a 
running width ($\propto p^2$). However, only a small and for large $p^2$ 
vanishing running-width component is retained in 
$p^2 \chi_{\ssZ,\ssW}^{-1}(p^2)$. This is a consequence of having pulled out 
the running couplings from the original propagators, merely leaving behind 
$T_{\ssZ,\ssW}(p^2)$.} 
for space-like $p$, i.e., 
$p^2<0$. This is easily seen, because the running couplings are real for 
$p^2<0$ owing to their definition, and the
terms in \eqns{def:chiz}{def:chiw} involving $\muz$ or $\muw$ combine to
a real quantity owing to 
\eqns{def:Wmassrenormalization2}{def:Zmassrenormalization2}.

The above expressions for the resummed propagators lead to very simple 
expressions for the 
photon--\PZ~system. 
{\samepage 
We write the fermionic corrections to the photon or \PZ\ boson 
coupled to massless fermions as
\begin{eqnarray}
\quad
\DiagramFermionToBosonFullWithMomenta{}{$f$}{$\bar{f}$}{$\ga$}
             {$p_1$}{$p_2$}{$p$}  & = &
             \DiagramFermionToBosonPropagator{}{$\gamma$}{$\gamma$}
           + \DiagramFermionToBosonPropagator[75]{}{$\gamma$}{$Z$} \;,\nl
\quad
\DiagramFermionToBosonFull{}{$Z$} & = &
             \DiagramFermionToBosonPropagator{}{$Z$}{$\gamma$}
           + \DiagramFermionToBosonPropagator[75]{}{$Z$}{$Z$} \nonumber \;.
\end{eqnarray}
}

\noindent
Here the open circles denote resummed propagators and the shaded circles
vertex functions. The dot on the right-hand side of the diagrams indicates
that the corresponding leg is not amputated, i.e., that the propagator is 
included. Since the fermion--gauge-boson vertex gets no fermionic 
corrections and thus enters only at Born level, the diagrams show 
the actual resummation of the propagators. 

In order to investigate the implications of our renormalization prescription
on these 
fermion--gauge-boson couplings, we introduce the unrenormalized 
fermion currents 
\begin{eqnarray}
\hat{\Gamma}^f_{_\gamma,\mu}(-p_2,p_1) & \equiv & 
    \hat{e}\;\bar{u}_f(-p_2)\; \gamma^\mu\,(-Q_f)\,u_f(p_1) \;,\nl
\hat{\Gamma}^f_{_Z,\mu}(-p_2,p_1) & \equiv & 
    \hat{e}\;\bar{u}_f(-p_2)\; \gamma^\mu\,( \hat{v}_f - \hat{a}_f \gamma^5 )
    \,u_f(p_1) \;,\nl
\hat{\Gamma}^f_{_W,\mu}(-p_2,p_1) & \equiv & 
    \Ugw\;\bar{u}_f(-p_2)\;\gamma^\mu\, 
    \omega_- u_{f'}(p_1) \;,
\end{eqnarray}
where $\omega_- = (1-\ga_5)/2$, 
a minus sign in the argument of a $u$ spinor indicates a $v$ spinor,
and $f'$ represents the iso-spin partner of $f$.
The electric charge fraction of the fermion $f$ is denoted by $Q_f$,
its iso-spin by $I^3_f$, and its bare couplings to the \PZ~boson read
\begin{equation}
\label{def:bareZffcoupling}
\hat{a}_f \equiv { I^3_f \over 2\Ustw\Uctw }\;\;,\qquad
\hat{v}_f \equiv \hat{a}_f - Q_f\,{\Ustw\over\Uctw}\;.
\end{equation}
The renormalized fermion currents are analogously given by
\begin{eqnarray}
\Gamma^f_{_\gamma,\mu}(-p_2,p_1) & \equiv & 
    e(p^2)\;\bar{u}_f(-p_2)\; \gamma^\mu\,(-Q_f)\,u_f(p_1) \;,\nl
\Gamma^f_{_Z,\mu}(-p_2,p_1) & \equiv & 
    e(p^2)\;\bar{u}_f(-p_2)\; \gamma^\mu \,[ v_f(p^2) - a_f(p^2) \gamma^5 ]\,
    u_f(p_1) \;,\nl
\Gamma^f_{_W,\mu}(-p_2,p_1) & \equiv & 
    \gw(p^2)\;\bar{u}_f(-p_2)\; \gamma^\mu\, 
    \omega_- u_{f'}(p_1) \;,
\end{eqnarray}
where $p=p_1+p_2$, 
and the running couplings $v_f(p^2)$ and $a_f(p^2)$ are defined as in
\eqn{def:bareZffcoupling} with $\Ustw$ and $\Uctw$ replaced by 
$\stw(p^2)$ and $\ctw(p^2)$, respectively.

Now we can rewrite the fermion--\PZ-boson coupling including the \PZ\ and 
mixing propagators:
\begin{eqnarray} \label{eq:Zvertexren}
\!\!\!\DiagramFermionToBosonFull{}{$Z$} { \Ue \over \Ustw\Uctw }\
  & = & \ \lrbr \hat{\Gamma}^{f,\mu}_\ssA(-p_2,p_1)\; \URPzg(p^2)
        + \hat{\Gamma}^{f,\mu}_\ssZ(-p_2,p_1)\; \URPzz(p^2) \rrbr
        { \Ue \over \Ustw\Uctw } \nl  
  & = & \ \Gamma^{f,\mu}_\ssZ(-p_2,p_1)\; { \chi_\ssZ(p^2) \over p^2 }\;
        { e(p^2) \over \ctw(p^2) \stw(p^2) } \nl
  & \equiv & \DiagramFermionToBosonEffective{}{$Z$}
             { e(p^2) \over \ctw(p^2) \stw(p^2) } \;,
\end{eqnarray}
where we denote the renormalized effective Born couplings of the 
fermions to the vector bosons by a box. The gauge bosons attached to such a
box are defined to have their propagators replaced by the corresponding 
renormalized propagator functions [\eqns{def:chiz}{def:chiw}].
The effect of the renormalization is apparently to change all bare
couplings to renormalized (running) ones and 
the \PZ~propagator to $\chi_\ssZ(p^2)/p^2$.

Similarly the fermion--photon interaction can be written as
\begin{eqnarray}
  \DiagramFermionToBosonFull{}{$\gamma$} \Ue\   
  & = & \ \lrbr \hat{\Gamma}^{f,\mu}_\ssA(-p_2,p_1)\; \URPgg(p^2)
        + \hat{\Gamma}^{f,\mu}_\ssZ(-p_2,p_1)\; \URPzg(p^2) \rrbr \Ue \nl  
  & = & \ \Gamma^{f,\mu}_\ssA(-p_2,p_1)\; { 1 \over p^2 }\; e(p^2) \nl
  &&{}+ \Gamma^{f,\mu}_\ssZ(-p_2,p_1)\; { \chi_\ssZ(p^2) \over p^2 }\; 
        {e(p^2) \over \ctw(p^2) \stw(p^2) }\;
        \parent{ \stws(p^2) - \Ustws } \nl
  & \equiv & \DiagramFermionToBosonEffective{}{$\gamma$} e(p^2) \nl
  &&{}+ \DiagramFermionToBosonEffective{}{$Z$}
        {e(p^2) \over \ctw(p^2) \stw(p^2) }\;
        \parent{ \stws(p^2) - \Ustws }\;. \nl
\end{eqnarray}
In this case some bare couplings are still left, but in specific 
processes involving massless fermionic final states (e.g.\ four-fermion 
processes at LEP1, six-fermion processes at LEP2) the fermion--photon
interactions usually appear in the following two combinations: 
\begin{eqnarray} \label{eq:gammaZvertexren1}
\lefteqn{ \DiagramFermionToBosonFull{}{$\gamma$} \hat{e}\ \ 
          - \DiagramFermionToBosonFull{}{$Z$} \hat{e}\;
          { \hat{c}_w\over\hat{s}_w }\ = } \nl
  & & \nl
  & = & \DiagramFermionToBosonEffective{}{$\gamma$} e(p^2)\ \ 
        - \DiagramFermionToBosonEffective{}{$Z$} e(p^2)\;
        { c_w(p^2)\over s_w(p^2) }
\end{eqnarray}
and
\begin{eqnarray} \label{eq:gammaZvertexren2}
\lefteqn{ \DiagramFermionToBosonFull{}{$\gamma$} \hat{e}\ \
          + \DiagramFermionToBosonFull{}{$Z$} \hat{e}\;
          { \hat{s}_w\over\hat{c}_w }\ = } \nl
  & & \nl
  & = & \DiagramFermionToBosonEffective{}{$\gamma$} e(p^2)\ \
        + \DiagramFermionToBosonEffective{}{$Z$} e(p^2)\;
        {s_w(p^2)\over c_w(p^2) } \;.
\end{eqnarray}
In these combinations all bare quantities have combined into renormalized 
(running) couplings.

The fermion--\PW-boson interaction yields in an analogous way
\begin{eqnarray} \label{eq:Wvertexren}
\DiagramFermionToBosonFullWithLabels{}{$W$}{$f'$}{$f$}{} \Ugw\ 
  & = & \ \hat{\Gamma}^{f,\mu}_\ssW(-p_2,p_1)\; \URPww(p^2)\,\Ugw \nl 
  & = & \ \Gamma^{f,\mu}_\ssW(-p_2,p_1)\; { \chi_\ssW(p^2) \over p^2 }\;
        \gw(p^2) \nl
  & & \nl
  & \equiv & \DiagramFermionToBosonEffective{}{$W$} \gw(p^2) \;.
\end{eqnarray}

The above effective interactions are essential ingredients for 
constructing renormalized, resummed amplitudes for LEP1 and LEP2 processes.

%  #] Renormalized propagators and fermion--gauge-boson vertex functions : 
%  #[ The process $e^+e^-\to f\bar{f}$:

\subsection{\boldmath{The process $e^+e^-\to f\bar{f}$}}
\label{se:eeff}

As an example of the renormalization procedure we discuss the 
four-fermion process {$e^+(p_2)e^-(p_1)\to f(q_1)\bar{f}(q_2)$} 
with massless external fermions. If the fermion $f$ is not in the same doublet
as the electron, the bare amplitude is 
given by the sum of the following sub-amplitudes ($p=p_1+p_2$):
\begin{eqnarray}
\label{def:amplitudeeetoff}
{\cal M}_1^{ee;\, ff}(-p_2,p_1;\,q_1,-q_2) & = & {\cal M}_{\ssA\ssA} 
  + {\cal M}_{\ssZ\ssA} + {\cal M}_{\ssZ\ssZ} + {\cal M}_{\ssA\ssZ} \;,\nl
{\cal M}_{\ssA\ssA} & = & \hat{\Gamma}^{e}_{\ssA,\mu}(-p_2,p_1)\,
  \URPgg(p^2)\,\hat{\Gamma}^{f,\mu}_\ssA(q_1,-q_2) \;,\nl  
{\cal M}_{\ssZ\ssA} & = & \hat{\Gamma}^{e}_{\ssZ,\mu}(-p_2,p_1)\,
  \URPzg(p^2)\,\hat{\Gamma}^{f,\mu}_\ssA(q_1,-q_2) \;,\nl  
{\cal M}_{\ssZ\ssZ} & = & \hat{\Gamma}^{e}_{\ssZ,\mu}(-p_2,p_1)\,
  \URPzz(p^2)\,\hat{\Gamma}^{f,\mu}_\ssZ(q_1,-q_2) \;,\nl  
{\cal M}_{\ssA\ssZ} & = & \hat{\Gamma}^{e}_{\ssA,\mu}(-p_2,p_1)\,
  \URPzg(p^2)\,\hat{\Gamma}^{f,\mu}_\ssZ(q_1,-q_2) \;.
\end{eqnarray}
Exploiting the results of the last subsection, 
these sub-amplitudes combine as follows 
\bqa
\lefteqn{ {\cal M}_1^{ee;\, ff}(-p_2,p_1;\,q_1,-q_2) \ = }\nonumber\\[3ex] 
     & \hspace*{7mm}= & 
             \DiagramFermionToBosonFullWithLabels{}{$\gamma$}{$e^-$}{$e^+$}{}  
             \hat\Gamma^{f,\mu}_\ssA(q_1,-q_2)\ \ 
           + \DiagramFermionToBosonFull{}{$Z$}
             \hat\Gamma^{f,\mu}_\ssZ(q_1,-q_2) \nl
     & \hspace*{7mm}= &   \DiagramFermionToBosonEffective{}{$\gamma$}
             \Gamma^{f,\mu}_\ssA(q_1,-q_2)\ \
           + \DiagramFermionToBosonEffective{}{$Z$}
             \Gamma^{f,\mu}_\ssZ(q_1,-q_2) \nl
     & \hspace*{7mm}= &   \DiagramMa \ +\  \DiagramMb \;. \\[3mm]\nonumber
\eqa
So, we arrive at a finite, effective Born amplitude
in terms of running couplings and propagator functions 
\bqa
\label{eq:renormalizedamplitudeeetoff}
{\cal M}_1^{ee;\, ff}(-p_2,p_1;\,q_1,-q_2) & = &
        \Gamma^{e,\mu}_\ssA(-p_2,p_1)\, { 1 \over p^2 }\,
        \Gamma^f_{_\gamma,\mu}(q_1,-q_2) \nl
&&{}  + \Gamma^{e,\mu}_\ssZ(-p_2,p_1)\,{ \chi_\ssZ(p^2) \over p^2 }\,
        \Gamma^f_{_Z,\mu}(q_1,-q_2) \;.
\eqa
The amplitude for Bhabha scattering, $e^+e^-\to e^+e^-$, is given in terms of
the above amplitudes as ${\cal M}_1^{ee;\,ee}(-p_2,p_1;\,q_1,-q_2)
- {\cal M}_1^{ee;\,ee}(-p_2,-q_2;\,q_1,p_1)$. 
In the second term the interchange of $-q_2$ and $p_1$ also applies to the 
definition of the gauge-boson momentum $p$. 

Processes that can also proceed via the charged current require an additional 
basic amplitude, involving the exchange of a $W$ boson $(q=p_2-q_2)$:
\begin{equation}
\label{eq:renormalizedamplitudeenutonue}
{\cal M}_2^{ff';\,\nu_e e}(-p_2,-q_2;\,q_1,p_1) =
    \Gamma^{f,\mu}_\ssW(-p_2,-q_2)\,{ \chi_\ssW(q^2) \over q^2 }\,
    \Gamma^{\nu_e}_{\ssW,\mu}(q_1,p_1) \;.
\end{equation}
Therewith the full renormalized amplitude for 
$e^+ e^- \to \nu_e \bar{\nu}_e$ is given by the combination
${\cal M}_1^{ee;\,\nu_e\nu_e}(-p_2,p_1;\,q_1,-q_2)
- {\cal M}_2^{e\nu_e;\,\nu_e e}(-p_2,-q_2;\,q_1,p_1)$.
Finally the renormalized amplitude for the muon decay 
$\mu(p)\to\nu_\mu(q_1) e(q_2)\bar\nu_e(q_3)$ 
reads ${\cal M}_2^{\nu_\mu\mu;\,e\nu_e}(q_1,p;\,q_2,-q_3)$.

%  #] The process $e^+e^-\to f\bar{f}$: 

%  #] Renormalization : 
%  #[ The triple gauge-boson vertex :

\section{The triple gauge-boson vertex}
\label{se:VWW_vertex}

%  #[ Introduction :

Besides the measurement of the $W$-boson mass, the main object of study 
in LEP2 processes will be the triple
gauge-boson vertex.
In this section we discuss the fermionic one-loop corrections to this vertex.
First the full gauge-boson vertex functions, including fermionic 
corrections, are split into bare couplings and finite coefficients. 
Then the Ward identities for the triple gauge-boson vertices are given,
from which the Ward identities for the finite coefficients and
subsequently those for 
the full renormalized gauge-boson vertex functions are
inferred. Using the results of Section~\ref{se:renormalization} for the
renormalized fermion--gauge-boson interactions, basic matrix elements for 
LEP2 processes are constructed. Finally, the gauge invariance of the 
renormalized LEP2 amplitudes is verified through these basic matrix elements.

%  #] Introduction : 
%  #[ The unrenormalized triple gauge-boson vertex :

\subsection{The unrenormalized triple gauge-boson vertex}
\label{se:VWW_unrenormalized}

Like the gauge-boson self-energies, the gauge-boson vertex functions can be 
split into a photonic piece, $G^{\ga}$, and an iso-spin part, $G^{I}$,  
that vanishes with vanishing fermion masses and that is dominated by the 
top-quark contribution:
\bqa \label{split_up_VWW}
  \UV^{\{\ga,Z\} W^+W^-}_{\mu\kappa\lambda}(q,p_+,p_-) & = &
       \left\{1,-{\Uctw\over\Ustw}\right\}\,\Ue\Ugw^2\,
       G^{\ga}_{\mu\kappa\lambda}(q,p_+,p_-) \nl
&&{} + \left\{0,1\right\}\,{ \Ue\Ugw^2 \over \Ustw\Uctw }\,
       G^I_{\mu\kappa\lambda}(q,p_+,p_-) \;.
\eqa
Here and in the following
all particles and momenta are defined to be incoming. 
The first term in the curly brackets refers to the 
photon, the second term to the \PZ~boson.
In LEP2 processes the vertex functions $\UV^{\ga WW}$ and $\UV^{ZWW}$
are contracted with external massless fermionic currents and appear in 
two distinct combinations. 
The combinations of couplings appearing in \eqn{split_up_VWW} are of the form 
described in \eqn{eq:gammaZvertexren1} and \eqn{eq:Zvertexren}.
Hence, if the coefficients $G^{\ga}$ and $G^I$ are 
finite, also the triple gauge-boson vertex functions yield finite contributions
to the LEP2 amplitudes, with all bare couplings replaced by renormalized 
(running) ones. 

The one-loop contributions to the coefficients $G^{\ga}$ and $G^I$ are
given in App.~\ref{app:vww}. The pure one-loop coefficient $G^{I,(1)}=G^I$ is 
finite. The coefficient $G^{\ga}$ can be decomposed into tree-level and 
one-loop contributions according to
\bqa 
  G^{\ga}_{\mu\kappa\lambda}(q,p_+,p_-) & = &
  { 1 \over \Ugw^2 }\,\Gamma_{\mu\kappa\lambda}(q,p_+,p_-)
  + G^{\ga,\,(1)}_{\mu\kappa\lambda}(q,p_+,p_-) \nl
                                        & = &
  \lrbr { 1 \over \gw^2(q^2) } - { \hat{\overline{\Sigma}}_{\ssW}(q^2) 
  \over q^2 \Ugw^2 } \rrbr\,\Gamma_{\mu\kappa\lambda}(q,p_+,p_-)
  + G^{\ga,\,(1)}_{\mu\kappa\lambda}(q,p_+,p_-) \;,\nl
\eqa
containing the lowest-order vertex tensor 
\bq \label{def:Gamma_tree}
  \Gamma_{\mu\kappa\lambda}(q,p_+,p_-) = (q-p_+)_\lambda \,g_{\mu\kappa}
      + (p_+-p_-)_\mu \,g_{\kappa\lambda} + (p_--q)_\kappa \,g_{\lambda\mu} \;.
\eq
The tree-level part of $G^\ga$ gives rise to an UV-divergent
term, indicated by $-\hat{\overline{\Sigma}}_{\ssW}(q^2)/(q^2 \Ugw^2)$. 
This term cancels the UV divergences contained in $G^{\ga,\,(1)}$ (see
App.~\ref{app:vww}).  So both $G^{\ga}$ and $G^{I}$ are finite.

%  #] The unrenormalized triple gauge-boson vertex : 
%  #[ Ward identities for the triple gauge-boson vertex :

\subsection{Ward identities for the triple gauge-boson vertex}
\label{se:ward_identities}

At the level of Green functions the underlying gauge symmetry manifests
itself in Ward identities, which in particular rule the gauge
cancellations. Since we only
deal with fermionic one-loop corrections we need those Ward identities that 
hold for these corrections and the tree-level expressions.
They can be obtained 
from the general Ward identities of gauge theories \cite{tHooftVeltman}
by omitting all other contributions. Alternatively, one can directly
take over the background-field Ward identities 
\cite{bfm} as the fermion-loop contributions are identical in the
conventional approach and the background-field formalism.

For the bare vertex functions $\UV^{\ga W^+W^-}$, $\UV^{ZW^+W^-}$, 
$\UV^{\ga W^+\phi^-}$, $\UV^{ZW^+\phi^-}$, 
and $\UV^{\chi W^+W^-}$ these Ward identities are given by:
\bqa
\label{eq:Ward-identity_VWW2}
\lefteqn{q^\mu\,\UV^{\{\ga,Z\} W^+W^-}_{\mu\kappa\lambda}(q,p_+,p_-)
         -i\,\sqrt{\Umuz}\,\left\{0,1\right\}
         \UV^{\chi W^+W^-}_{\kappa\lambda}(q,p_+,p_-) = } \nl[1mm]
&\hspace*{3mm}& \hat e \left\{1,-{\Uctw\over\Ustw}\right\}
                \biggl[ T_{\kappa\lambda}(p_-)\,\parent{p_-^2 + \USww(p_-^2)}
                + L_{\kappa\lambda}(p_-)\,\USwwL(p_-^2) \nl
&\hspace*{3mm}&{} \hphantom{\hat e \left\{1,-{\Uctw\over\Ustw}\right\}a}    
                  - T_{\kappa\lambda}(p_+)\,\parent{p_+^2 + \USww(p_+^2)}
                  - L_{\kappa\lambda}(p_+)\,\USwwL(p_+^2) \biggr] \;,
\\[2mm]
\label{eq:Ward-identity_VWW1}
\lefteqn{-p_-^\lambda\,\UV^{\{\ga,Z\}W^+W^-}_{\mu\kappa\lambda}(q,p_+,p_-)
         -\sqrt{\Umuw}\,\UV^{\{\ga,Z\}W^+\phi^-}_{\mu\kappa}(q,p_+,p_-) = } 
         \nl[1mm]
&\hspace*{3mm}& \hat e \left\{1,-{\Uctw\over\Ustw}\right\}
      \biggl[ T_{\mu\kappa}(q)\,\left(q^2 + {\hat{\Sigma}_{_{\{\ga,Z\}}}}(q^2)
      -\left\{{\Uctw\over\Ustw},{\Ustw\over\Uctw}\right\}\USzg(q^2)
      \right) \nl 
&\hspace*{3mm}&{} \hphantom{\hat e \left\{1,-{\Uctw\over\Ustw}\right\}a}
                  + \left\{0,1\right\}\,\left( L_{\mu\kappa}(q)\,\USzzL(q^2)
                  - g_{\mu\kappa}\,\Umuz \right) \nl
&\hspace*{3mm}&{} \hphantom{\hat e \left\{1,-{\Uctw\over\Ustw}\right\}a}
                   - T_{\mu\kappa}(p_+)\,\parent{p_+^2 + \USww(p_+^2)}
                   - L_{\mu\kappa}(p_+)\,\USwwL(p_+^2) + g_{\mu\kappa}\,\Umuw
                   \biggr] \;, \nl
\eqa
and an analogous Ward identity for the external $W^+$ leg.
While the longitudinal parts of the photon
self-energy and the photon--\PZ~mixing energy vanish, those of the  \PW\ and 
\PZ~self-energies, $\USwwL$ and $\USzzL$, appear explicitly in the Ward 
identities. Note that we have introduced the following transverse and 
longitudinal tensors:
\bq
  T_{\mu\nu}(p) \equiv g_{\mu\nu} - p_{\mu}p_{\nu}/p^2 ,
  \qquad 
  L_{\mu\nu}(p) \equiv p_{\mu}p_{\nu}/p^2 \;. 
\eq

The vertex functions involving the Higgs ghosts, $\chi$ and $\phi^\pm$, can be
split up in the same way as the gauge-boson vertex functions:
\bqa
  \sqrt{\Umuz}\,\UV^{\chi W^+W^-}_{\kappa\lambda}(q,p_+,p_-) & = & 
       { \Ue\Ugw^2 \over \Ustw\Uctw }\,
       X_{\kappa\lambda}(q,p_+,p_-) \;, \nl
  \sqrt{\Umuw}\,\UV^{\{\ga,Z\}W^+\phi^-}_{\mu\kappa}(q,p_+,p_-) & = &  
       \left\{1,-{\Uctw\over\Ustw}\right\}\,\Ue\Ugw^2\,
       F^\ga_{\mu\kappa}(q,p_+,p_-) \nl
&&{} + \left\{0,1\right\}\,{ \Ue\Ugw^2 \over \Ustw\Uctw }\,
       F^I_{\mu\kappa}(q,p_+,p_-) \;.
\eqa
As the fermion--Higgs-ghost couplings are proportional to the fermion
masses, these vertex functions involve apart from the tree-level contributions,
\bqa
  X_{\kappa\lambda}^{(0)}(q,p_+,p_-)   & = & 0 \;, \nl
  F^{\ga,\,(0)}_{\mu\kappa}(q,p_+,p_-) & = & F^{I,\,(0)}_{\mu\kappa}(q,p_+,p_-)
           = - { \Umuw \over \Ugw^2 }\,g_{\mu\kappa} \;,
\eqa
predominantly top-quark contributions.
Note that the ratio $\Umuw/\Ugw^2$ is not finite, its UV divergence equals the
one contained in $T_{\ssW}$ [see \eqn{def:Wmassrenormalization2}] and is
canceled by the top-mass dependent one-loop corrections 
$F^{\ga,\,(1)}$ and $F^{I,\,(1)}$. 

Using \eqn{eq:relationsSigmas}, the definitions of the running couplings
\eqns{def:runningalpha}{def:runningstws}, and the propagator functions
\eqns{def:chiz}{def:chiw}, we can rewrite the Ward identities in terms of the
coefficients:
\bqa
q^\mu\,G^{\ga}_{\mu\kappa\lambda}(q,p_+,p_-) & = & 
       { T_{\kappa\lambda}(p_-) \over \gw^2(p_-^2) }\, 
       { p_-^2 \over \chi_{\ssW}(p_-^2) }
     + L_{\kappa\lambda}(p_-)\,{ \USwwL(p_-^2)-\Umuw \over \Ugw^2 } \nl
&&{} - { T_{\kappa\lambda}(p_+) \over \gw^2(p_+^2) }\,  
       { p_+^2 \over \chi_{\ssW}(p_+^2) }  
     - L_{\kappa\lambda}(p_+)\,{ \USwwL(p_+^2)-\Umuw \over \Ugw^2 } \;,
       \nl[1mm]
q^\mu\,G^I_{\mu\kappa\lambda}(q,p_+,p_-) & = & 
       i\,X_{\kappa\lambda}(q,p_+,p_-) \;, \nl[1mm]
-p_-^{\lambda}\,G^{\ga}_{\mu\kappa\lambda}(q,p_+,p_-) & = & 
       F^\ga_{\mu\kappa}(q,p_+,p_-) + {T_{\mu\kappa}(q) \over \gw^2(q^2)}\,q^2
       \nl
&&{} - { T_{\mu\kappa}(p_+) \over \gw^2(p_+^2) }\,
       { p_+^2 \over \chi_{\ssW}(p_+^2) }  
     - L_{\mu\kappa}(p_+)\,{ \USwwL(p_+^2)-\Umuw \over \Ugw^2 } \;, \nl
-p_-^{\lambda}\,G^I_{\mu\kappa\lambda}(q,p_+,p_-) & = & 
       F^I_{\mu\kappa}(q,p_+,p_-) 
     - { T_{\mu\kappa}(q) \over \gw^2(q^2) }\,\ctws(q^2)\,\lrbr 
       { q^2 \over \chi_{\ssZ}(q^2) } - q^2 \rrbr \nl
&&{} - L_{\mu\kappa}(q)\,\Uctws\,{ \USzzL(q^2)-\Umuz \over \Ugw^2} \;.
\eqa
All terms in these Ward identities are finite, including the factors that 
multiply the longitudinal tensors. This is caused by the fact that the 
quantities $\Umuw/\Ugw^2$, 
$\USwwL/\Ugw^2$, and $\Uctws\,\USzzL/\Ugw^2$ contain the same UV divergence.

Defining the renormalized vertex functions as
\bqa
&&\!\!\!\!  V^{\{\ga,Z\} W^+W^-}_{\mu\kappa\lambda}(q,p_+,p_-) \equiv  
       \left\{1,-{\ctw(q^2) \over \stw(q^2) }\right\}\,
       e(q^2)\gw(p_+^2)\gw(p_-^2)\,G^{\ga}_{\mu\kappa\lambda}(q,p_+,p_-) \nl
&&{}   \hphantom{V^{\{\ga,Z\} W^+W^-}_{\mu\kappa\lambda}(q,p_+,p_-) 
                 \equiv } \!\!\!\!
     + \left\{0,1\right\}\,{ e(q^2)\gw(p_+^2)\gw(p_-^2) \over 
       \stw(q^2)\ctw(q^2) }\, G^I_{\mu\kappa\lambda}(q,p_+,p_-) \;, \nl
&&\!\!\!\! \sqrt{\muz}\,V^{\chi W^+W^-}_{\kappa\lambda}(q,p_+,p_-) \equiv  
       { e(q^2)\gw(p_+^2)\gw(p_-^2) \over \stw(q^2)\ctw(q^2) }\,
       X_{\kappa\lambda}(q,p_+,p_-) \;, \nl
&&\!\!\!\! \sqrt{\muw} V^{\{\ga,Z\}W^+\phi^-}_{\mu\kappa}(q,p_+,p_-) \!\equiv
       \!\left\{1,-{\ctw(q^2) \over \stw(q^2) }\right\}
       e(q^2)\gw(p_+^2)\gw(p_-^2) F^\ga_{\mu\kappa}(q,p_+,p_-) \nl
&&{}   \hphantom{\sqrt{\muw}\,V^{\{\ga,Z\}W^+\phi^-}_{\mu\kappa}(q,p_+,p_-) 
                 \!\equiv\! } \!\!\!\!
     + \left\{0,1\right\}\,{ e(q^2)\gw(p_+^2)\gw(p_-^2) \over 
       \stw(q^2)\ctw(q^2) }\,F^I_{\mu\kappa}(q,p_+,p_-) \;, \nl
\eqa
we can write down the Ward identities after renormalization. 
As we are going to investigate LEP2 processes, which involve only
massless fermionic currents, all uncontracted external legs
couple to conserved currents. As such all terms in the Ward identities
involving explicit momentum vectors drop out and only the terms proportional 
to the metric tensors survive. The final form of the Ward identities, to be 
used in the following, is then given by   
\bqa
\label{eq:renormalized_WI2}
\lefteqn{q^\mu\,V^{\{\ga,Z\} W^+W^-}_{\mu\kappa\lambda}(q,p_+,p_-)
         -i\,\sqrt{\muz}\,\left\{0,1\right\} 
         V^{\chi W^+W^-}_{\kappa\lambda}(q,p_+,p_-) = } \nl
&& e(q^2) \left\{1,-{\ctw(q^2)\over\stw(q^2)}\right\}\,\gw(p_+^2)\gw(p_-^2) 
   \biggl[ { g_{\kappa\lambda}\,p_-^2 \over \gw^2(p_-^2)\,\chi_{\ssW}(p_-^2) }
         - { g_{\kappa\lambda}\,p_+^2 \over \gw^2(p_+^2)\,\chi_{\ssW}(p_+^2) } 
   \biggr] \nl[1ex]
&& {} \quad + \mbox{terms vanishing for conserved external currents} \;,
\\[2.0ex]
\label{eq:renormalized_WI1}
\lefteqn{-p_-^\lambda\,V^{\{\ga,Z\}W^+W^-}_{\mu\kappa\lambda}(q,p_+,p_-)
         -\sqrt{\muw}\,V^{\{\ga,Z\}W^+\phi^-}_{\mu\kappa}(q,p_+,p_-) = } \nl
&& \! e(q^2) \left\{1,-{\ctw(q^2)\over\stw(q^2)}\right\}\gw(p_+^2)\gw(p_-^2) 
   \biggl[ { g_{\mu\kappa} \over \gw^2(q^2) }
           \left\{ q^2,{ q^2 \over \chi_{\ssZ}(q^2) } \right\}
         \!-\! { g_{\mu\kappa}\,p_+^2 \over \gw^2(p_+^2)\,\chi_{\ssW}(p_+^2) } 
   \biggr] \nl[1ex] 
&& {} \quad +\mbox{terms vanishing for conserved external currents} \;. 
\eqa
These Ward identities are linear in the vertex 
functions and the inverse propagator functions.

%  #] Ward identities for the triple gauge-boson vertex : 
%  #[ Basic matrix elements for LEP2 processes :

\subsection{Basic matrix elements for LEP2 processes}
\label{se:LEP2amplitudes}

Combining the above results with the renormalized fermion--gauge-boson 
vertices given in Section~\ref{se:renormalization}, we are now in the position 
to introduce basic matrix elements for LEP2 processes, i.e., 
the six-fermion processes
$e^-(p_1) e^+(p_2)\to f_1(q_1) f_2(q_2) \bar{f}_3(q_3) \bar{f}_4(q_4)$.
Here all external fermions are taken to be massless.
For the construction of the basic matrix elements it suffices to consider the 
situation that $e,f_1$, and $f_2$ are in different doublets. All other 
situations can be covered by interchanging particles in the basic matrix
elements that are given below. 

We first consider the basic matrix elements for \PW-pair-mediated LEP2 
processes,
i.e., $f_3=f'_1$ and $f_4=f'_2$. In order to identify the charge flow we assume
the fermion $f_1$ to have negative charge. Using a box to indicate the 
renormalized triple gauge-boson vertex functions, the corresponding three 
basic matrix elements are given by 
($p_i$ incoming, $q_i$ outgoing, $q=p_1+p_2$, $p_+=-q_1-q_3$, $p_-=-q_2-q_4$) 
\bqa
\label{def:building_block_WW}
\lefteqn{{\cal M}_1^{ee;\,f_1f'_1;\,f_2f'_2}(-p_2,p_1;\,q_1,-q_3;\,q_2,-q_4) 
         \ = } \nl[1mm]
&\hspace*{8mm}= & \DiagramMaa \nl
&\hspace*{8mm}= & \DiagramMab \nl[1mm]
&\hspace*{8mm}= & \!\!\sum_{\ssB=\ga,\ssZ} G^{e,\,\mu}_{\ssB}(-p_2,p_1)
     \,V^{B W^+W^-}_{\mu\kappa\lambda}(q,p_+,p_-)\,
     G^{f_1,\,\kappa}_{\ssW}(q_1,-q_3)\,G^{f_2,\,\lambda}_{\ssW}(q_2,-q_4) 
     \;, \nl[3mm]
\lefteqn{{\cal M}_2^{ee;\,f_1f'_1;\,f_2f'_2}(-p_2,p_1;\,q_1,-q_3;\,q_2,-q_4) 
         \ = } \nl[1mm]
&\hspace*{8mm}= & \hphantom{+} \DiagramMba \nl
&\hspace*{8mm}&{}  + \!\!\DiagramMbb \nl[1mm]
&\hspace*{8mm}= & - G^{f_2,\,\lambda}_{\ssW}(q_2,-q_4)\,
     G^{e,\,\mu}_{\ssB}(-p_2,p_1)\sum_{\ssB=\ga,\ssZ} \lrbr 
     \Gamma^{f_1 f'_1}_{\ssB\ssW,\,\mu\lambda}(q_1,q_1-q,-q_3) \right. \nl
&\hspace*{8mm}&{} \left. \hphantom{ - G^{f_2,\,\lambda}_{\ssW}(q_2,-q_4)\,
                         G^{e,\,\mu}_{\ssB}(-p_2,p_1)\sum_{\ssB=\ga,\ssZ} }
     + \Gamma^{f_1 f'_1}_{\ssW\ssB,\,\lambda\mu}(q_1,q_1-p_-,-q_3) \rrbr 
     \;, \nl[3mm]
\lefteqn{{\cal M}_3^{ee;\,f_1f'_1;\,f_2f'_2}(-p_2,p_1;\,q_1,-q_3;\,q_2,-q_4) 
         \ = } \nl[1mm]
&\hspace*{8mm}= & \hspace{1cm} \DiagramMc \nl[1mm]
&\hspace*{8mm}= & -\Gamma^{ee}_{\ssW\ssW,\,\lambda\kappa}(-p_2,-p_2-p_-,p_1)
     \,G^{f_1,\,\kappa}_{\ssW}(q_1,-q_3)\,G^{f_2,\,\lambda}_{\ssW}(q_2,-q_4)
     \;.
\eqa
Here we introduced a couple of shorthand notations in order to keep the
expressions as compact as possible ($B=\ga,Z$):
\bqa
G^{f,\,\mu}_{\ssB}(r,k)                    & = & 
   \Gamma^{f,\,\mu}_{\ssB}(r,k)\,{\chi_{\ssB}(K^2) \over K^2 } \;, \nl
G^{f,\,\mu}_{\ssW}(r,k)                    & = & 
   \Gamma^{f,\,\mu}_{\ssW}(r,k)\,{\chi_{\ssW}(K^2) \over K^2 } \;, \nl
\Gamma^{ff,\,\mu\nu}_{\ssW\ssW}(p,r,k)     & = &
   \gw(P^2)\,\gw(K^2)\,\bar{u}_f(p)\,\ga^\mu\,{\sla r \over r^2}\,\ga^\nu 
   \,\omega_- u_{f}(k) \;, \nl
\Gamma^{ff',\,\mu\nu}_{\ssB\ssW}(p,r,k)    & = &
   e(P^2)\,[v_f^{\ssB}(P^2)+a_f^{\ssB}(P^2)]\,\gw(K^2)\,\bar{u}_f(p)\,\ga^\mu 
   \,{\sla r \over r^2}\,\ga^\nu\,\omega_- u_{f'}(k) \;, \nl
\Gamma^{ff',\,\mu\nu}_{\ssW\ssB}(p,r,k)    & = &
   \gw(P^2)e(K^2)\,[v_{f'}^{\ssB}(K^2)+a_{f'}^{\ssB}(K^2)]\,\bar{u}_f(p) 
   \,\ga^\mu\,{\sla r \over r^2}\,\ga^\nu\,\omega_- u_{f'}(k) \;, \nl
\Gamma^{ff,\,\mu\nu}_{\ssB_1\ssB_2}(p,r,k) & = &
   e(P^2)e(K^2)\,\bar{u}_f(p)\,\ga^\mu\,
   [v_f^{\ssB_1}(P^2)-a_f^{\ssB_1}(P^2)\ga_5]\,
   {\sla r \over r^2}\, \nl
&& \hphantom{ e(P^2)e(K^2)\,A\,\,} \times\,\ga^\nu\,
   [v_f^{\ssB_2}(K^2)-a_f^{\ssB_2}(K^2)\ga_5]\,u_{f}(k) \;,
\eqa
with $\chi_\ga(K^2)=1$, $P=p-r$, $K=k-r$, and $v_f^{\ssB}$ and $a_f^{\ssB}$
denoting the vector and axial-vector couplings of the fermion $f$ to the 
neutral gauge boson $B$ ($v_f^{\ssZ}=v_f$, $a_f^{\ssZ}=a_f$,
$v_f^{\gamma}=-Q_f$, $a_f^{\gamma}=0$).

For the remaining, neutral-gauge-boson-mediated LEP2 processes, i.e.,
$f_3=f_1$ and $f_4=f_2$, 
two additional basic matrix elements are required. The first one originates 
from the triple gauge-boson vertices involving only neutral gauge bosons. 
These (C--odd and CP--even) vertices do not exist at lowest order and only 
receive contributions
from $\varepsilon$-tensor terms, entering through the fermionic one-loop
corrections. As all gauge bosons are neutral, no charge assignment is required.
Writing $k_1=p_1+p_2$, $k_2=-q_1-q_3$, and $k_3=-q_2-q_4$, the basic matrix 
element reads
\bqa
\label{def:building_block_ZZ1}
\lefteqn{{\cal M}_4^{ee;\,f_1f_1;\,f_2f_2}(-p_2,p_1;\,q_1,-q_3;\,q_2,-q_4) 
         \ = } \nl[1mm]
&\hspace*{8mm}= & \DiagramMda \nl
&\hspace*{8mm}= & \DiagramMdb \nl[1mm]
&\hspace*{8mm}= & \!\!\sum_{\ssB_i=\ga,\ssZ} G^{e,\,\mu}_{\ssB_1}(-p_2,p_1)
     \,V^{\ssB_1\ssB_2\ssB_3}_{\mu\kappa\lambda}(k_1,k_2,k_3)\,
     G^{f_1,\,\kappa}_{\ssB_2}(q_1,-q_3)\,G^{f_2,\,\lambda}_{\ssB_3}(q_2,-q_4) 
     \;.\nl 
\eqa
The renormalized vertex function 
$V^{\ssB_1\ssB_2\ssB_3}_{\mu\kappa\lambda}(k_1,k_2,k_3)$ is given by 
\eqn{eq:bbbmassive} in App.~\ref{app:bbb_massive} with $\Ue^3 \to e(k_1^2)\,
e(k_2^2)\,e(k_3^2)$, $\hat v_f^{\ssB_i} \to v_f^{\ssB_i}(k_i^2)$, and
$\hat a_f^{\ssB_i} \to a_f^{\ssB_i}(k_i^2)$. It obeys simple Ward identities of
the form:
\begin{equation}
  k_1^\mu\,V^{\ssB_1\ssB_2\ssB_3}_{\mu\kappa\lambda}(k_1,k_2,k_3) =
  i\,\sqrt{\muz}\,\delta_{\ssB_1 \ssZ}\,V^{\chi\ssB_2\ssB_3}_{\kappa\lambda}
  (k_1,k_2,k_3) \qquad \mathrm{for\ \ } B_i=\ga,Z \;.
\end{equation}

The second basic matrix element for neutral-gauge-boson-mediated LEP2 processes
is given by 
\begin{eqnarray}
\label{def:building_block_ZZ2}
\lefteqn{{\mathcal M}_5^{ee;\,f_1f_1;\,f_2f_2}(-p_2,p_1;\,q_1,-q_3;\,q_2,-q_4) 
         \ = } \nl[1mm]
&\hspace*{8mm}= &{} \hphantom{+} \DiagramMea \nl
&\hspace*{8mm}&{}  + \!\DiagramMeb \nl[1mm]
&\hspace*{8mm}=&{} - \!\!\sum_{\ssB_1,\ssB_2=\ga,\ssZ} 
       G^{e,\,\mu}_{\ssB_1}(-p_2,p_1)\,G^{f_2,\,\lambda}_{\ssB_2}(q_2,-q_4)
       \,\lrbr \Gamma^{f_1 f_1}_{\ssB_1\ssB_2,\,\mu\lambda}(q_1,q_1-k_1,-q_3)
       \right. \nl
&\hspace*{8mm}&{} \left. \hphantom{ - \sum_{\ssB_1\ssB_2=\ga,\ssZ} }
     + \Gamma^{f_1 f_1}_{\ssB_2\ssB_1,\,\lambda\mu}(q_1,k_1-q_3,-q_3) \rrbr \;.
\end{eqnarray}

{}From these basic matrix elements all amplitudes for massless six-fermion 
(LEP2) processes can be constructed (leaving out QCD diagrams).

%  #] Basic matrix elements for LEP2 processes : 
%  #[ Gauge invariance of renormalized LEP2 amplitudes :

\subsection{Gauge invariance of renormalized LEP2 amplitudes}
\label{se:LEP2_gauge_invariance}

In the previous subsection the basic building blocks are provided for the
construction of renormalized LEP2 amplitudes. As promised, we address now the
issue of gauge invariance. After all, we have performed a resummation and 
subsequent renormalization of 1PI fermionic $\ord{\al}$ corrections, which 
involves a different treatment of vertex corrections as compared to 
self-energy corrections. Such a procedure can be a possible
source of gauge-invariance-breaking effects.

The U(1) gauge cancellations become numerically very important
for electromagnetic interactions
in the collinear limit, where the electromagnetic current 
$\Gamma^\mu_\ga(r,r+k_\ga)$ becomes proportional to the momentum of the 
internal photon $k_\ga^\mu$ (with $k_\ga^2\ll E_{\scriptscriptstyle CM}^2$).
On the other hand,
SU(2) gauge cancellations become relevant in the high-energy limit
($k_{\ssW,\ssZ}^2\ll E_{\scriptscriptstyle CM}^2$) if an external current 
coupled to a massive internal gauge boson becomes approximately 
proportional to the gauge-boson momentum, in other words the gauge boson is
effectively longitudinal. From this it should be clear that in these regimes 
sensible theoretical 
predictions are only possible if the amplitudes with external currents 
replaced by the corresponding
gauge-boson momenta fulfill appropriate Ward identities.  When we talk 
about gauge invariance in this paper we always mean the validity of these Ward 
identities. 
The Ward identities explicitly read
\begin{equation}
  k^\mu{\cal M}^\gamma_\mu = 0, \qquad
  k^\mu{\cal M}^Z_\mu = i \sqrt{\muz} {\cal M}^\chi, \qquad
  k^\mu{\cal M}^{W^\pm}_\mu = \pm \sqrt{\muw} {\cal M}^{\phi^\pm}\;,
\end{equation}
where ${\cal M}^V_\mu$ ($V=\gamma,Z,W^\pm$) denote the amplitudes with
the corresponding currents amputated. The first identity expresses 
transversality with respect to external photon momenta,
the latter two imply the Goldstone-boson equivalence theorem.

The gauge invariance of the amplitudes that do not involve the triple 
gauge-boson vertex is evident. The same goes for the amplitudes of 
neutral-gauge-boson-mediated LEP2 processes. So, we restrict ourselves to
verifying the 
gauge invariance of the universal amplitudes for \PW-pair-mediated LEP2 
processes, i.e., the processes $e^- e^+ \to f_1 f_2 \bar{f}'_1 \bar{f}'_2\,$ 
with $e,f_1$, and $f_2$ assumed to be in different doublets. These universal 
amplitudes contribute to all \PW-pair-mediated processes. All other 
situations can be covered by adding extra universal 
amplitudes with interchanged particles. Hence, it suffices to check the gauge 
invariance of the universal amplitudes.

The universal amplitude, ${\cal M}_{\ssW\ssW}^{\scriptstyle \mathrm{univ}}$, 
can be 
written in terms of the basic matrix elements presented in the previous 
subsection:
\bqa \label{MWWuniv}
{\cal M}_{\ssW\ssW}^{\scriptstyle \mathrm{univ}} & = & \sum_{i=1}^3 
       {\cal M}_i^{ee;\,f_1f'_1;\,f_2f'_2}(-p_2,p_1;\,q_1,-q_3;\,q_2,-q_4) \nl
&&{} + {\cal M}_2^{ee;\,f_2f'_2;\,f_1f'_1}(-p_2,p_1;\,q_2,-q_4;\,q_1,-q_3) \;,
\eqa 
where as before the fermion $f_1$ has been assumed to have negative charge.
First we verify the gauge invariance with respect to the internal (incoming) 
$W^-$ boson by replacing $G^{f_2,\,\lambda}_{\ssW}(q_2,-q_4)$ by $p_-^\lambda$.
As the last term in \eqn{MWWuniv} only involves an (incoming) $W^+$ boson and,
hence, does not contain the $G^{f_2,\,\lambda}_{\ssW}$ current,
only the first set of three terms should be considered for this particular
gauge-invariance check.
Using \eqn{eq:renormalized_WI1} and the various definitions for the fermionic 
currents, one ends up with
\bqa
{\cal M}_{\ssW\ssW}^{\scriptstyle \mathrm{univ}} & \longrightarrow &
       p_-^\lambda\,\sum_{\ssB=\ga,\ssZ} G^{e,\,\mu}_{\ssB}(-p_2,p_1)
       \,V^{B W^+W^-}_{\mu\kappa\lambda}(q,p_+,p_-)\,
       G^{f_1,\,\kappa}_{\ssW}(q_1,-q_3)\nl
&&{}   \hspace*{-1.2cm} + G^{f_1,\,\kappa}_{\ssW}(q_1,-q_3)\,\gw(p_+^2)
       \gw(p_-^2)\,\bar{u}_e(-p_2)\,\ga_\kappa\,\omega_- u_e(p_1) \nl
&&{}   \hspace*{-1.2cm} - G^{e,\,\mu}_{\ga}(-p_2,p_1)\,
       G^{f_1}_{\ssW,\,\mu}(q_1,-q_3)\,e(q^2)\,
       { p_+^2 \over \chi_{\ssW}(p_+^2) }\,
       { \gw(p_-^2) \over \gw(p_+^2) } \nl
&&{}   \hspace*{-1.2cm}+ G^{e,\,\mu}_{\ssZ}(-p_2,p_1)\,
       G^{f_1}_{\ssW,\,\mu}(q_1,-q_3)\,e(q^2)\,{ \ctw(q^2) \over \stw(q^2) }
       { p_+^2 \over \chi_{\ssW}(p_+^2) }
       \,{ \gw(p_-^2) \over \gw(p_+^2) } \nl
&&{}   \hspace*{-1.2cm} =\; - \sum_{\ssB=\ga,\ssZ} G^{e,\,\mu}_{\ssB}(-p_2,p_1)
       \,\sqrt{\muw}\,V^{B W^+\phi^-}_{\mu\kappa}(q,p_+,p_-)\,
       G^{f_1,\,\kappa}_{\ssW}(q_1,-q_3) \;.
\eqa
The last expression exactly equals the contribution from the renormalized 
$\ga W\phi$ and $ZW\phi$ vertices. The gauge invariance with respect to the 
internal (incoming) $W^+$ boson can be verified in an analogous way.

The gauge invariance with respect to the internal neutral gauge 
bosons is verified by replacing $G^{e,\,\mu}_{\ssB}(-p_2,p_1)$ by $q^\mu$ for 
$B=\{\ga,Z\}$. In this case the matrix element ${\cal M}_3$ should be left out,
as it only involves $W$ bosons. Using \eqn{eq:renormalized_WI2} we find
\bqa
{\cal M}_{\ssW\ssW}^{\scriptstyle \mathrm{univ}} & \longrightarrow &
       q^\mu\,V^{\{\ga,Z\} W^+W^-}_{\mu\kappa\lambda}(q,p_+,p_-)\,
       G^{f_1,\,\kappa}_{\ssW}(q_1,-q_3)\,G^{f_2,\,\lambda}_{\ssW}(q_2,-q_4) 
       \qquad \nl
&&{}   \hspace*{-1.2cm} + G^{f_1,\,\kappa}_{\ssW}(q_1,-q_3)\,
       G^{f_2}_{\ssW,\,\kappa}(q_2,-q_4)\,
       e(q^2) \left\{1,-{\ctw(q^2)\over\stw(q^2)}\right\}\,
       { p_+^2 \over \chi_{\ssW}(p_+^2) }\,{ \gw(p_-^2) \over \gw(p_+^2) } \nl 
&&{}   \hspace*{-1.2cm} - G^{f_1,\,\kappa}_{\ssW}(q_1,-q_3)\,
       G^{f_2}_{\ssW,\,\kappa}(q_2,-q_4)\,
       e(q^2) \left\{1,-{\ctw(q^2)\over\stw(q^2)}\right\}\,
       { p_-^2 \over \chi_{\ssW}(p_-^2) }\,{ \gw(p_+^2) \over \gw(p_-^2) } \nl
&&{}   \hspace*{-1.2cm} =\; i\,\sqrt{\muz}\,\left\{0,1\right\} 
       V^{\chi W^+W^-}_{\kappa\lambda}(q,p_+,p_-)\,
       G^{f_1,\,\kappa}_{\ssW}(q_1,-q_3)\,G^{f_2,\,\lambda}_{\ssW}(q_2,-q_4)
       \;.
\eqa
For the \PZ~boson the last expression exactly equals the contribution from the 
renormalized $\chi WW$ vertex.

While for the amplitudes involving gauge bosons gauge cancellations are
essential, the amplitudes involving the Higgs ghosts behave properly
without cancellations.

%  #] Gauge invariance of renormalized LEP2 amplitudes : 

%  #] The triple gauge-boson vertex : 
\renewcommand{\floatpagefraction}{0.7}
%  #[ The process $e^-e^+\to e^-\bar{\nu}_eu\bar{d}$:

\section{\boldmath{Application of the fermion-loop scheme to typical
LEP2 processes}}
\label{se:process}

The fermion-loop (FL) scheme allows to introduce finite-width effects 
into all tree-level (non-QCD) six-fermion matrix elements.
In this section we illustrate the FL scheme using as examples the processes 
\processccten\/ (also called CC10),
\processcceleven\/ (also called CC11), and \processcctwenty\/ (also called CC20)
with massless external fermions.
The relevant terminology has been introduced in Ref.~\cite{LEP2MCreport}
and the CC class of processes comprises production of up (anti-up) and
anti-down (down) fermion pairs, $(U_i {\bar D}_i) + (D_j {\bar U}_j)$.
These reactions include some of the most interesting processes for
studies at LEP2 and beyond.

In all of them SU(2) gauge invariance is needed to guarantee the
unitarity cancellations at high energies. A violation of SU(2)
gauge invariance can be most easily seen in the simple CC10/11 processes,
where the dominant contribution comes 
from the \PW-pair-production diagrams
and the onset of a bad high-energy behavior is already appreciable around
$1$ TeV.

The CC20 process, on the other hand, is also sensitive to the breaking of U(1)
gauge invariance in the collinear limit, while the
SU(2)-gauge-invariance violation 
becomes sizeable only above 2 TeV.

%  #[ Amplitude :

\subsection{Amplitudes}

The amplitudes for the processes \processccten\/ and \processcceleven\/
are directly given by the universal
amplitude \eqn{MWWuniv} with $f_1=\mu, f_2=u$ and $f_1=s, f_2=u$,
respectively.

The amplitude for the process 
$e^-(p_1)e^+(k_1)\to e^-(p_2)\bar{\nu}_e(k_2)u(p_u)\bar{d}(p_d)$
is given by the sum of the universal amplitude \eqn{MWWuniv} and this universal
amplitude with the initial-state positron and final-state electron
interchanged,
\bqa \label{Meeenud}
{\cal M}^{ee\to\,e\nu_e ud} & = & 
  \sum_{i=1}^3 {\cal M}_i^{ee;\,e\nu_e;\,ud}(-k_1,p_1;\,p_2,-k_2;\,p_u,-p_d) 
  \nl
&&{} + {\cal M}_2^{ee;\,ud;\,e\nu_e}(-k_1,p_1;\,p_u,-p_d;\,p_2,-k_2) \nl
&&{} \hspace*{-3ex} -
     \sum_{i=1}^3 {\cal M}_i^{ee;\,e\nu_e;\,ud}(p_2,p_1;\,-k_1,-k_2;\,p_u,-p_d)
     \nl
&&{} - {\cal M}_2^{ee;\,ud;\,e\nu_e}(p_2,p_1;\,p_u,-p_d;\,-k_1,-k_2) \;.
\eqa

As all these matrix elements are just linear combinations of universal matrix 
elements, the gauge invariance follows directly from the discussion in 
Section~\ref{se:LEP2_gauge_invariance}.

For the 
CC20 process $e^-e^+\to e^-\bar{\nu}_eu\bar{d}$\/ 
U(1) gauge invariance becomes essential in the
region of phase space where the angle between the incoming and
outgoing electrons is small \cite{bhf1,Kurihara,ZeppenfeldBaur,Costas}.
In this region of phase space the superficial $1/q_\gamma^4$ divergence,
originating from the square of the photon propagator with momentum
$q_\gamma$,
is softened to $1/q_\gamma^2$ by U(1) gauge  
invariance.  In the presence of light fermion masses this gives rise  
to the familiar $\log(m_e^2/s)$ large logarithms.
In order to guarantee 
the softening of the $1/q_\gamma^4$ divergence,
which is necessary for a meaningful 
cross-section, U(1) gauge invariance is required.

From the numerical point of view the SU(2) gauge invariance is not of 
real relevance at LEP2 energies, but it becomes important for energies
reached at the next generation of linear colliders.
It is important for all processes that involve the \PW-pair-production
diagrams. %(also called CC03).
At high energies, the SU(2) gauge invariance
guarantees the gauge cancellations and thus ensures that the matrix
element respects unitarity.

%  #] Amplitude : 
%  #[ Input parameters :

\subsection{Input parameters}
\label{se:input}

First we fix the input parameters.
We work in a LEP2-like scheme \cite{LEP2WWreport}, which uses 
the input parameters%
\footnote{We have used the same masses as in the analysis of 
Ref.~\cite{LEP2MCreport}. The most recent values for $m_{_{Z,W}}$ are 
$91.1863(20)\,$GeV and $80.356(125)\,$GeV.}
$G_F, \Re\{\alpha_\ell(\mzs)^{-1}\}, m_{_Z}, m_{_W}$:
\begin{eqnarray} \label{parameters}
G_F    & = & 1.16639 \times 10^{-5} \GeV^{-2} \;, \nonumber\\
\Re\{\alpha_\ell(\mzs)^{-1}\} & = & 128.89 \;, \nonumber\\
m_{_W} & = & 80.26 \GeV \;, \nonumber\\ 
m_{_Z} & = & 91.1884 \GeV \;.
\end{eqnarray}
The electromagnetic coupling at the $Z$ mass $\alpha_\ell(\mzs)$ (see analysis
of Ref.~\cite{alpz}) differs slightly from our $\alpha(\mzs) = e^2(\mzs)/4\pi$,
as top-quark effects are not included in the former.\footnote{These effects
are well approximated by
        $\alpha(m_{_Z}^2)^{-1}-\alpha_\ell(m_{_Z}^2)^{-1} =
        4m_{Z}^2/(45\pi{m_t^2})\approx
0.0282942\,({m_{_Z}^2}/{m_t^2})$.}
Furthermore, $m_{_W}$ and $m_{_Z}$ are defined in the usual (LEP1) way, 
in which the 
renormalized mass is fixed by the zero of the real part of the 
inverse 
propagator.
The masses of the light fermions (except top) are neglected.

With these input parameters we compute all bare parameters
for some arbitrary value of the infinity $\Delta$,
taking into account that some input parameters are given
in a different renormalization scheme.
The knowledge of the bare parameters allows us to compute all 
running (renormalized) couplings of \eqns{def:runningalpha}{def:runninggw} 
and the renormalized complex pole positions $\muw$ and $\muz$.
An explicit computational scheme to accomplish this is given in 
App.~\ref{app:complex_scheme}, where the LEP2 scheme is described within our 
fermion-loop approach, i.e., without any bosonic and QCD corrections.
%  #[ table renormalization :
\begin{table}%[htb]
\begin{center}
$$
\setlength{\arraycolsep}{\tabcolsep}
\renewcommand\arraystretch{1.3}
\begin{array}{|l|rrr|}
\hline\hline
m_{_W}                         & 80.10     & 80.26     & 80.42   \\
\hline\hline
m_t
                               & 104.768   & 132.185   & 157.195 \\
\sqrt{\ \Re\muw}               
                               & 80.074    & 80.234    & 80.393  \\
-\Im\muw/\sqrt{\ \Re\muw}\quad 
                               & 2.0377    & 2.0509    & 2.0636  \\
\sqrt{\ \Re\muz}               
                               & 91.1552   & 91.1550   & 91.1548 \\
-\Im\muz/\sqrt{\ \Re\muz}      
                               & 2.4538    & 2.4610    & 2.4688  \\
\hline
\Re e(\mws)                 
                               &  0.311967 &  0.311979 &  0.311986 \\
\Im e(\mws)                 
                               & -0.002685 & -0.002685 & -0.002685 \\
\Re g_w(\mws)               
                               &  0.459802 &  0.460576 &  0.461400 \\
\Im g_w(\mws)               
                               & -0.006450 & -0.006482 & -0.006516 \\
\hline
\hline
\end{array}
$$
\caption{Values of the {\it effective} top-quark mass, pole positions, and 
         effective couplings at $\mws$ for three different $W$ masses.
         All masses and pole-position-related quantities are given in GeV.}
\label{tab:renormalization}
\end{center}
\end{table}
%  #] table renormalization : 
Note that the mass of the top quark is not an independent quantity in this 
approach. In Table~\ref{tab:renormalization} we give the {\it effective} 
top-quark mass, the renormalized complex pole positions, and the complex 
couplings at $\mws$ for three input $W$-boson masses $m_{_W}$. 

The top-quark mass cannot be seen as more than an effective value, which partly 
compensates for the missing bosonic and QCD corrections, and hence it cannot 
be compared with the direct measurement~\cite{CDFD0top}.
Indeed the $m_t$--$m_{_H}$ connection plays a fundamental role and when we 
perform a more complete $m_t$-determination by including all the available 
radiative corrections in the on-shell scheme~\cite{LEP2WWreport}, we find a 
quite remarkable agreement with the experimental data.

The real and imaginary parts of the $W$ and $Z$ poles are described well by 
the approximative formulae $\Re\mu \approx m^2 - \Gamma^2$ and 
$\Im\mu \approx -m\Gamma+\Gamma^3/m$, respectively 
(see App.~\ref{app:complex_vs_LEP1}). Here $\Gamma$ is the gauge-boson width
including fermionic corrections, calculated in the LEP1 scheme. For the $W$
boson this corrected width is about 0.8\% larger than the lowest-order width
in the $G_F$ parameterization [$=3 G_F m_{_W}^3/(2\sqrt{2}\pi)$]. This is in
agreement with the arctangent term in Eq.(9) of Ref.~\cite{bigwidth}.

%  #] Input parameters :
%  #[ Schemes :

\subsection{Schemes}
\label{se:schemes}

In the following we present numerical results obtained with the FL
scheme and compare these with the results obtained with other schemes.
The following schemes are considered in our analysis:
\begin{description}
\item[Running width:] The cross-section is computed using the tree-level 
   amplitude. The massive gauge-boson propagators acquire a running 
   width for $p^2>0$: 
   $1/(p^2-m^2+ip^2\Gamma/m) = 1/([1+i\Gamma/m][p^2-m^2/(1+i\Gamma/m)])$. 
   Thus, for $p^2>0$ this scheme is nothing more than a fixed-width scheme 
   with modified pole and residue. This scheme violates U(1) and SU(2)
   gauge invariance.
\item[Fixed width:] The cross-section is computed using the 
   tree-level amplitude. The massive gauge-boson propagators are given by 
   $1/(p^2-m^2+im\Gamma)$.
   This gives an unphysical width for $p^2<0$, but retains U(1) gauge 
   invariance in the CC20 process~\cite{bhf1}.
   For the considered six-fermion processes
   the SU(2) gauge violation is suppressed by
   $m \Gamma/s$ (at the matrix-element level) at high energies 
   and therefore  the high-energy behavior is consistent with unitarity.
   This scheme will exhibit a bad high-energy behavior when applied to
   processes involving more intermediate
   gauge bosons, e.g.~processes with six fermions in the final state.
\item[Running width + U(1)-invariance-restoring \boldmath{$\gamma WW$} 
      vertex factor:]  \hfil\qquad 
    \linebreak
   This scheme was proposed in Ref.~\cite{bhf1} as a simple and sufficiently 
   accurate approach for LEP2 generators. It 
   involves naive running widths for the massive gauge bosons,
   supplemented by a simple multiplicative factor for the 
   $\gamma WW$ vertex 
   (as derived in the limit $q_\ga^2 \to 0$). 
   By construction, U(1) gauge invariance is retained, but the 
   SU(2) Ward identities are violated, rendering this scheme not appropriate
   for the high-energy regime.
\item[Minimal FL scheme:] 
    A simplified  
    minimal approach to incorporate the finite width,
    while ensuring both U(1) and  SU(2) gauge invariance,
     consists in taking
    into account only  the imaginary parts of the
    fermionic corrections in the massless 
    limit.
    The top-quark mass is kept only in step functions that switch off the
    (massless) top contributions below the top thresholds
    \cite{Costas}.
\item[Imaginary-part FL scheme:] 
    Now the full imaginary part of the fermion-loop corrections is used, 
    leaving out all $\varepsilon$-tensor terms. The fermion masses are neglected
    except for the top-quark mass. This is the version of the 
    FL scheme that was originally used in the analysis%
\footnote{For the sake of clarity, the actual discussion and numbers presented 
in Ref.~\cite{bhf1} were restricted to a U(1)-invariant subset of four 
diagrams involving near-collinear space-like photons. Moreover, the top
contributions were not required in the case considered there.}
    of Ref.~\cite{bhf1}.
    Once the massive top contributions are included this scheme is in
    fact not much easier than the complete FL scheme.
\item[Full FL scheme:] We take all fermion-loop contributions into account,
   but neglect all fermion masses except for the top-quark mass.
The implementation of the fermionic one-loop corrections is rather 
straightforward (see Section~\ref{se:LEP2amplitudes}). 
The tree-level couplings are replaced by running couplings at the 
appropriate momenta and the massive gauge-boson propagators by
the functions $\chi_{\{\ssW,\ssZ\}}(p^2)/p^2$. 
The vertex coefficients $G^\ga$ and $G^I$, entering through the Yang--Mills 
vertex, contain the lowest-order couplings as well as the one-loop
fermionic vertex corrections given in App.~\ref{app:vww}. 
\end{description}

In all schemes apart from the full FL scheme no non-imaginary
higher-order corrections are included. In those schemes the values of
the couplings have been
 determined from $G_F$, $m_{_Z}$ and $m_{_W}$, as given in 
\eqn{parameters}, in order to account for the most important universal 
corrections, i.e.,
\bq \label{eq:improvedcouplings}
\gws = 2\sqrt{2}G_F m_{_W}^2 = 0.212514, \qquad
\alpha = \frac{\gws}{2\pi} \left(1-\frac{m_{_W}^2}{m_{_Z}^2}\right) = 1/131.2145.
\eq
In the non-FL schemes the $W$-boson width has been
determined in Born approximation
without any QCD correction, giving $\Gamma_W = 2.03595\,$GeV{}.
In the FL schemes the top-quark mass has been set to
$m_t=132\GeV$, in accordance with the discussion in Section \ref{se:input}.

%  #] Schemes :
%  #[ Numerical results :

\subsection{Numerical results}

The theoretical results of the previous sections have been put into 
practice in computer programs.
All the numbers presented in this section have been
generated with three 
independently written Fortran programs: ERATO \cite{ERATO}, WTO \cite{WTO}
and WWF \cite{WWF}. 
While the minimal FL scheme is only implemented in ERATO and the
imaginary-part fermion-loop scheme only in WTO, the other schemes are
implemented in all three generators.

A detailed and tuned comparison was well beyond the purpose of the present 
work so that our numerical analysis, although reliable, is still far from the
high-precision level reached in Ref.~\cite{LEP2MCreport}. 
Whenever referring to the
same setup (kinematical cuts etc.) we have found numerical results with a
satisfactory agreement, 
within the quoted integration errors.

In this respect it should be noted that owing to the unitarity
cancellations 
at high energies several digits, e.g., 6 digits at 10 TeV, 
cancel between
the contribution of the neutrino-exchange and the other contributions of
the \PW-pair-production diagrams.
As compared with the standard formulation of the processes $e^-e^+\to 4f$,
also the average amount of needed CPU time has increased considerably due
to the complexity of the new calculation.
This fact alone is a justification
of the relatively lower technical precision of our results when compared
to the results presented in Ref.~\cite{LEP2MCreport}.
During the present analysis we have been mainly concerned in showing the
feasibility of the project and paid 
less attention to the phenomenological
side, thus most of our numerical results will be given without initial-state
QED radiation (ISR).
If not stated otherwise the canonical LEP2 cuts~\cite{LEP2MCreport} are applied
and no ISR is included.

In the discussion of the numerical results we focus on the two regions
of phase-space where we expect gauge-invariance issues to 
play a role: small space-like $q_\ga^2$ (collinear electron) and large 
$s=E_{\scriptscriptstyle CM}^2$.  
Apart from the experimentally relevant cross-sections at 
LEP2 energies with the canonical LEP2 cuts applied \cite{LEP2MCreport},
we also give results for \processcctwenty\ with only a cut on the angle 
between the final-state electron and either beam.
This we do because most of the experimental analysis is 
currently done without imposing an energy threshold on the outgoing electron
\cite{mgpc}. For very small scattering angles, however, it is mandatory to use 
a fully massive phase space (and possibly matrix element), since the electron 
mass becomes essential% 
\footnote{WWF includes masses to the necessary approximation \cite{WWFmasses}.}
in the limit of very small initial-final invariant mass $(p_2-p_1)^2$.
In this respect it should be mentioned that the CC20 process is also needed
to study the background to Higgs-boson production. 
This involves the so-called single-$W$ events, with the outgoing electron
lost in the beam pipe. In this case both a gauge-invariance-preserving scheme
and the finite electron mass are needed.

%Moreover, for 
For large $s$ the $W$ boson becomes effectively 
massless and $W$-bremsstrahlung becomes increasingly important. 
The matching negative virtual term, involving the exchange of $W$ bosons,
is not accounted for in our program
and therefore we are obliged to make a cut on the outgoing $W$, or the 
highly correlated outgoing electron angle.
We did not look into this any further since we are mainly interested 
in the SU(2) gauge cancellations that occur in $W$-{\it pair\/} production.

In the schemes described in Section~\ref{se:schemes} 
and for the CC20 process \processcctwenty, we
have computed all quantities compared in Ref.~\cite{LEP2MCreport};
the relevant ones are the first few moments of the angle of the $W^-$  
and outgoing lepton to the electron beam, and the shift in the peak position
of the $W^\pm$ invariant-mass distributions. 
Disregarding the naive running-width scheme, it appears that only
in the total cross-section one sees any effect of the different schemes.
For most other quantities the deviations are less than the integration 
accuracies.

We now discuss the various schemes in a situation where U(1)
electromagnetic gauge invariance is crucial. To this end we consider the
cross-section for \processcctwenty\ allowing for almost collinear
incoming and outgoing electrons.%
\footnote{If the angle between the incoming and outgoing electron
          is allowed to become arbitrarily
          small, the fermion masses in the loops and $\Gamma_\ga^{f,\,\mu}$
          can no longer be neglected.}
In Table~\ref{tab:totalxsec175} we give 
the total cross-section without canonical
LEP2 cuts except for a beam-angle cut of the final-state electron to either
beam ($\theta^{\mathrm{min}}_{e^-,\mathrm{beam}}$) 
for the typical LEP2 energy of $\sqrt{s}=175$ GeV{}.
Satisfactory agreement is found in general between ERATO and WWF
although some discrepancy is present for small
$\theta^{\mathrm{min}}_{e^-,\mathrm{beam}}$
in the running-width scheme
owing to the unphysical $1/q^4_\gamma$ peak.
%  #[ table clean xsection :
\begin{table}%[t]
\begin{center}
$$
\setlength{\arraycolsep}{\tabcolsep}
\renewcommand\arraystretch{1.3}
\def\tstrut{\vrule width0pt height2.5ex}
\def\bstrut{\vrule width0pt depth1.0ex}
\begin{array}{|l|
r@{}lr@{}lr@{}l|}
\hline
\hline
\theta^{\mathrm{min}}_{e^-,\mathrm{beam}}
          & 0.1^\circ&  & 1^\circ&   & 10^\circ&  \\
\hline
\hline
\mbox{Running width} 
          & 1.380&(6) & 0.6284&(8) & 0.5904&(7) \\
\mbox{} 
          & 1.426&(3) & 0.6332&(7) & 0.5918&(6) \\
\hline
\mbox{Fixed width}   
          & 0.6444&(9)  & 0.6214&(8) & 0.5904&(7) \\
\mbox{}   
          & 0.6443&(3)  & 0.6210&(3) & 0.5909&(2) \\
\hline
\mbox{Running width}
          & 0.6448&(9)  & 0.6219&(8) & 0.5912&(7) \\
\mbox{$+$ $\gamma WW$ vertex factor \bstrut}
          & 0.6456&(7)  & 0.6214&(7) & 0.5916&(6) \\
\hline
\mbox{Minimal FL scheme}
          & 0.6463&(9) & 0.6218&(8) & 0.5910&(7) \\
\mbox{} &&&&&&\\
\hline
\mbox{Full FL scheme} 
          & 0.6507&(11) & 0.6280&(9) & 0.6002&(8) \\
\mbox{}
          & 0.6514&(9)  & 0.6298&(7) & 0.5992&(7) \\
\hline\hline
\end{array}
$$
\caption[]{Total cross-sections (in pb) for the CC20 process at 
     $\sqrt{s}=175\GeV$ without ISR and without cuts
     except on the angle between the outgoing 
     electron and either beam, % ($\theta^{\mathrm{min}}_{e^-,\mathrm{beam}}$),
     as predicted by ERATO (first entry) and by WWF (second entry).}
\label{tab:totalxsec175}
\end{center}
\end{table}
%  #] table clean xsection : 
%  #[ tchannel :
Breaking U(1) gauge invariance clearly 
gives wrong results in the
collinear
region, as was already argued in Ref.~\cite{bhf1}.
%  #[ table clean xsection WTO:
\begin{table}%[t]
\small
\begin{center}
$$
\setlength{\arraycolsep}{\tabcolsep}
\renewcommand\arraystretch{1.3}
\def\tstrut{\vrule width0pt height2.5ex}
\def\bstrut{\vrule width0pt depth1.0ex}
\begin{array}{|l|
r@{}l
r@{}l
r@{}lr@{}lr@{}l|}
\hline
\hline
\theta^{\mathrm{min}}_{e^-,\mathrm{beam}}
          & 0.1^\circ&  & 0.5^\circ&   & 1^\circ&  & 5^\circ& & 10^\circ&  \\
\hline \hline
\mbox{Running width} 
  & 1.36&(1)  & 0.654&(3)  & 0.6249&(4) & 0.60167&(6) & 0.58907&(5) \\
\mbox{Fixed width}   
  & 0.641&(1) & 0.6256&(8) & 0.6188&(5) & 0.6020&(2)  & 0.58915&(5) \\
\parbox{6.8em}{\tstrut Running width \\ $+$ $\gamma WW$ vertex factor\bstrut}  
  & 0.641&(1) & 0.6245&(5) & 0.6187&(5) & 0.60190&(2) & 0.589242&(8) \\
\mbox{Full FL scheme }   
  & 0.642&(3) & 0.630&(1)  & 0.6247&(4) & 0.6100&(2)  & 0.59771&(7) \\
\hline\hline
\end{array}
$$
\caption[]{Total cross-sections (in pb) for the CC20 process
           at $\sqrt{s}=175\GeV$ with the default setup
           except for a variable cut on the angle between the
           outgoing electron and either beam, as predicted by WTO.}
\label{tab:totalxsec175WTO}
\end{center}
\end{table}
%  #] table clean xsection WTO : 

In Table \ref{tab:totalxsec175WTO} we list the same cross-section as in
Table  \ref{tab:totalxsec175}, but now with canonical LEP2 cuts applied 
(the cut on the angle between the outgoing electron and either beam is
still varied).
This demonstrates that the U(1) violation is only related to collinear 
electrons 
and that, down to approximately $0.1^\circ$, there 
are no sizeable effects from including a cut on the
energy of the outgoing electron (at least, compared with 
                     the expected experimental precision).
%  #[ figure tchannel :
\begin{figure}
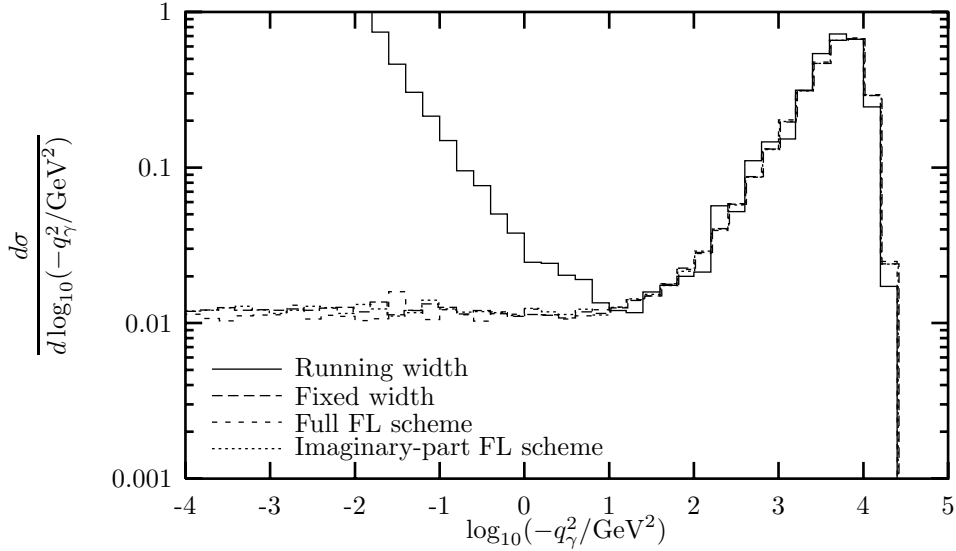

\begin{center}
 \input newtchan.tex
\end{center}
  \caption[]{The effect of gauge-breaking terms in the CC20 process 
             at $\sqrt{s}=175\GeV$ as a function of the virtuality $q_\ga^2$ of 
             the photon without ISR and without cuts, as predicted by WWF.}
\label{fig:tchannel}
\end{figure}
%  #] figure tchannel :
This collinear behavior is made more pronounced in \fig{fig:tchannel},
where the cross-section without any cuts is shown as a function of the 
virtuality of the photon [$q_\gamma^2=(p_2-p_1)^2$].

The deviations between the schemes other than the U(1)-violating running-%
width scheme are small. In adding the full fermionic contribution one observes 
nevertheless an increase in cross-section compared with the other 
U(1)-invariance-preserving schemes.
   At $10^\circ$ ERATO/WWF give a result in the full FL scheme which 
   is 1.7\%/1.4\% higher than the one obtained in the fixed-width 
   scheme but the effect is decreasing at small scattering angles.
This is not related to the collinear-electron limit,
but is rather a direct consequence of the incorporation of the
complete fermionic ${\mathcal O}(\alpha)$ effects, which
are only partly taken into account in the other schemes by using
improved Born couplings.
In the full FL scheme the fermionic corrections enter via the
universal running couplings, via the propagator functions
$\chi_{_W,_Z}$, and via the vertex corrections.

%  #] tchannel :  

%  #[ figure couplings :
%  #] figure couplings : 

For instance, the
running of the couplings is such that between 100 and $1000\GeV$ they 
increase by several percent.
This provides a qualitative estimate of the size of the fermionic
corrections that are not included in the effective couplings 
\eqn{eq:improvedcouplings}.
Because these couplings are fixed at the scale of the \PW- and \PZ-boson
masses, this effect is small at LEP2 energies.
However, the increase of the couplings as a function of $p^2$ will tend to
increase the cross-section in the full FL scheme at high
energies with respect to the other schemes. 
Of course, the neglected  
bosonic contributions will temper or even reverse the running of the couplings
above the boson thresholds [in view of the asymptotic freedom of $g_w^2(p^2)$],
so the computed values will in general be too high.

%  #[ table real xsection :
\begin{table}%[b]
\begin{center}
$$
\setlength{\arraycolsep}{\tabcolsep}
\renewcommand\arraystretch{1.3}
\def\tstrut{\vrule width0pt height2.5ex}
\def\bstrut{\vrule width0pt depth1.0ex}
\begin{array}{|l|r@{}lr@{}lr@{}l|}
\hline \hline
\sqrt{s}                            & \multicolumn{2}{c}{161\GeV} 
& \multicolumn{2}{c}{175\GeV} & \multicolumn{2}{c|}{190 \GeV} \\
\hline \hline
\mbox{Running width}                & 0.1191&(2) & 0.4865&(6) & 0.6127&(7) \\
\mbox{Fixed width}                  & 0.1174&(2) & 0.4862&(6) & 0.6124&(6) \\
\parbox{3.6cm} {\tstrut Running width \\ $+$ $\gamma WW$ vertex factor \bstrut} 
                                    & 0.1192&(2) & 0.4864&(6) & 0.6125&(7) \\
\mbox{Minimal FL scheme}  & 0.1191&(2) & 0.4864&(6) & 0.6128&(7) \\
\mbox{Full FL scheme}     & 0.1208&(3) & 0.4930&(7) & 0.6220&(6) \\
\hline
\mbox{Running width}                & 0.11942&(3) & 0.48681&(4) & 0.6135&(7) \\
\mbox{Fixed width}                  & 0.11764&(3) & 0.48674&(4) & 0.61314&(9) \\
\parbox{3.6cm} {\tstrut Running width \\ $+$ $\gamma WW$ vertex factor \bstrut} 
                                    & 0.11941&(3) & 0.48690&(4) & 0.6136&(7) \\
\mbox{Imaginary-part FL scheme}& 0.1195&(2) & 0.4861&(3) & 0.6125&(9) \\
\mbox{Full FL scheme}     & 0.1215&(2) & 0.4934&(3) & 0.6223&(9) \\
\hline
\mbox{Running width}                & 0.1195&(2) & 0.4878&(6) & 0.6119&(7) \\
\mbox{Fixed width}                 & 0.11757&(5) & 0.48636&(16) & 0.61271&(22)\\
\parbox{3.6cm} {\tstrut Running width \\ $+$ $\gamma WW$ vertex factor
\bstrut}
                                    & 0.1196&(2) & 0.4875&(6) & 0.6138&(8) \\
\mbox{Full FL scheme}     & 0.1213&(2) & 0.4946&(6) & 0.6229&(8) \\
\hline \hline
\end{array}
$$
\caption[]{Total cross-sections (in pb) for the CC20 process 
           at LEP2 energies with ISR and canonical cuts, 
         % which include a $10^\circ$ beam-angle cut for the electron.}
    as predicted by ERATO (rows 1--5), WTO (rows 6--10) and WWF (rows 11--14).}
\label{tab:physicalxsec175}
\end{center}
\end{table}
%  #] table real xsection : 
In Table~\ref{tab:physicalxsec175} we present a more realistic  
calculation at three typical LEP2 energies: 
the total cross-section with ISR and canonical LEP2 cuts.
The ISR has been implemented in the leading-logarithmic approximation,
in the same way
as it has been used for the tuned comparisons in  Ref.~\cite{LEP2MCreport}.
The numbers from the various generators agree rather well.

The difference between the minimal FL scheme (implemented in
ERATO) and the imaginary-part FL scheme (implemented in WTO) is
approximately 0.3\% at $\sqrt{s} = 161\,$GeV and below 0.1\% above threshold,
thus at the level of (or below) the integration errors.
Technically the imaginary-part FL scheme is obtained from the full FL scheme
when all non-imaginary corrections are set to zero, however, the most relevant 
universal (real) corrections are accounted for 
by fixing $\alpha$ as in 
\eqn{eq:improvedcouplings}, giving a satisfactory agreement with the 
minimal approach.
As for the full FL predictions, the differences between the three programs
are always below 0.6\%.
On threshold, the fixed-width scheme underestimates the cross-section 
by 1.4--1.6\% compared with the other schemes.

In adding the full fermionic contribution one observes a 1.6\%
increase in cross-section at the $W$-pair threshold, compared with the other 
schemes. The deviation remains practically constant up to $190\,$GeV{}.
Comparing the values at $\sqrt{s}=175\GeV$ for the case with and without ISR,
the cross-section is seen to decrease by approximately 17\%, 
which is reasonable since we are close to the $W$-pair 
threshold and ISR lowers the effective~$s$. 

In order to study the effects of SU(2) violation we have to consider
high energies. These effects are present in all processes
that involve the \PW-pair-production diagrams.
In Tables \ref{tab:highenergyxsecCC10} and \ref{tab:highenergyxsecCC11}
we give the cross-sections for the processes CC10 and CC11,
respectively. The quality of agreement between ERATO and WTO is satisfactory,
with perhaps some marginal discrepancy for the running-width scheme at
the highest energy.
%  #[ table real high-energy xsection CC10:
\begin{table}%[b]
\small
\begin{center}
$$
\setlength{\arraycolsep}{\tabcolsep}
\def\tstrut{\vrule width0pt height2.5ex}
\def\bstrut{\vrule width0pt depth1.0ex}
\renewcommand\arraystretch{1.3}
\begin{array}{|l|r@{}lr@{}lr@{}lr@{}lr@{}lr@{}l|}
\hline \hline
\sqrt{s}                            & \multicolumn{2}{c}{200\GeV}
& \multicolumn{2}{c}{500\GeV} & \multicolumn{2}{c}{1 \TeV}
& \multicolumn{2}{c}{2\TeV} 
& \multicolumn{2}{c}{5\TeV} & \multicolumn{2}{c|}{10 \TeV} \\
\hline \hline
\mbox{Running width}                &
672.96&(3) & 225.45&(3) & 62.17&(1) & 33.06&(1) & 123.759&(8) & 481.18&(5) \\
\mbox{}                             &
673.1&(6)  & 225.5&(3)  & 62.17&(10)& 33.05&(4) & 123.75&(8)  & 485.7&(3)  \\ 
\hline
\mbox{Fixed width}                  &
673.08&(4) & 224.05&(3) & 56.90&(1) & 13.19&(1) &   2.212&(6) & 0.591&(4) \\
\mbox{}                             &
673.3&(6)  & 224.2&(3)  & 56.93&(10)& 13.17&(3) &   2.209&(10) & 0.584&(5) \\
\hline
\mbox{Imaginary-part}  &
673.1&(1) & 224.5&(7) & 56.8&(1)  & 13.18&(4) & 2.24&(3) & 0.597&(6) \\
\mbox{FL scheme}  &
&&&&&&&&&&&\\
\hline
\mbox{Minimal}  &
&&&&&&&&&&&\\
\mbox{FL scheme}  &
672.7&(6) & 223.9&(3) & 56.85&(10)& 13.09&(4) & 2.215&(11) &  0.584&(5) \\
\hline
\mbox{Full FL scheme}  &
683.7&(1) & 227.9&(2) & 58.0&(1) & 13.57&(4) & 2.34&(3) & 0.632&(6) \\
\mbox{}                &
685.0&(6) & 228.1&(3) & 57.89&(9)& 13.59&(4) & 2.290&(11) &  0.621&(5) \\
\hline \hline
\end{array}
$$
\caption[]{Total cross-sections (in fb) for the CC10 process at high energies 
           (default setup),
           as predicted by WTO (first entry) and ERATO (second entry).}
\label{tab:highenergyxsecCC10}
\end{center}
\end{table}%
%  #] table real high-energy xsection CC10:
%  #[ table real high-energy xsection CC11:
\begin{table}%[b]
\small
\begin{center}
$$
\setlength{\arraycolsep}{\tabcolsep}
\def\tstrut{\vrule width0pt height2.5ex}
\def\bstrut{\vrule width0pt depth1.0ex}
\renewcommand\arraystretch{1.3}
\begin{array}{|l|r@{}lr@{}lr@{}lr@{}l|}
\hline
\hline
\sqrt{s}                            & \multicolumn{2}{c}{200\GeV}
& \multicolumn{2}{c}{500\GeV} & \multicolumn{2}{c}{1 \TeV}
& \multicolumn{2}{c|}{2\TeV}  \\
\hline
\hline
\mbox{Running width}                &
2.0395&(1) & 0.8113&(2) & 0.3254&(2) & 0.1671&(9) \\
\mbox{Fixed width}                  &
2.0399&(1) & 0.8072&(2) & 0.3096&(2) & 0.1075&(9) \\
\parbox{2.2cm}{\tstrut Imaginary-part \\ FL scheme\bstrut }  &
2.0397&(5) & 0.8067&(9)  & 0.3087&(8)  & 0.1073&(8) \\ 
\parbox{2.2cm}{\tstrut Full FL scheme\bstrut }  &
2.0720&(3) & 0.8212&(9)  & 0.3145&(8)  & 0.1097&(9)  \\
\hline
\hline
\end{array}
$$
\caption[]{Total cross-sections (in pb) for the CC11 process at high energies
           (default setup),
           as predicted by WTO.}
\label{tab:highenergyxsecCC11}
\end{center}
\end{table}%
%  #] table real high-energy xsection CC11:
The cross-section for CC11 is larger by about
a factor three owing to the color factor.
%  #[ figure CC10 :
\begin{figure}%[p]
\vspace{0.1cm}
\begin{center}
\unitlength 1cm
\begin{picture}(12,9.0)%(-.5,0) %gunplot size 1,0.85 16pt
\put(-0.7,-2.8){\includegraphics{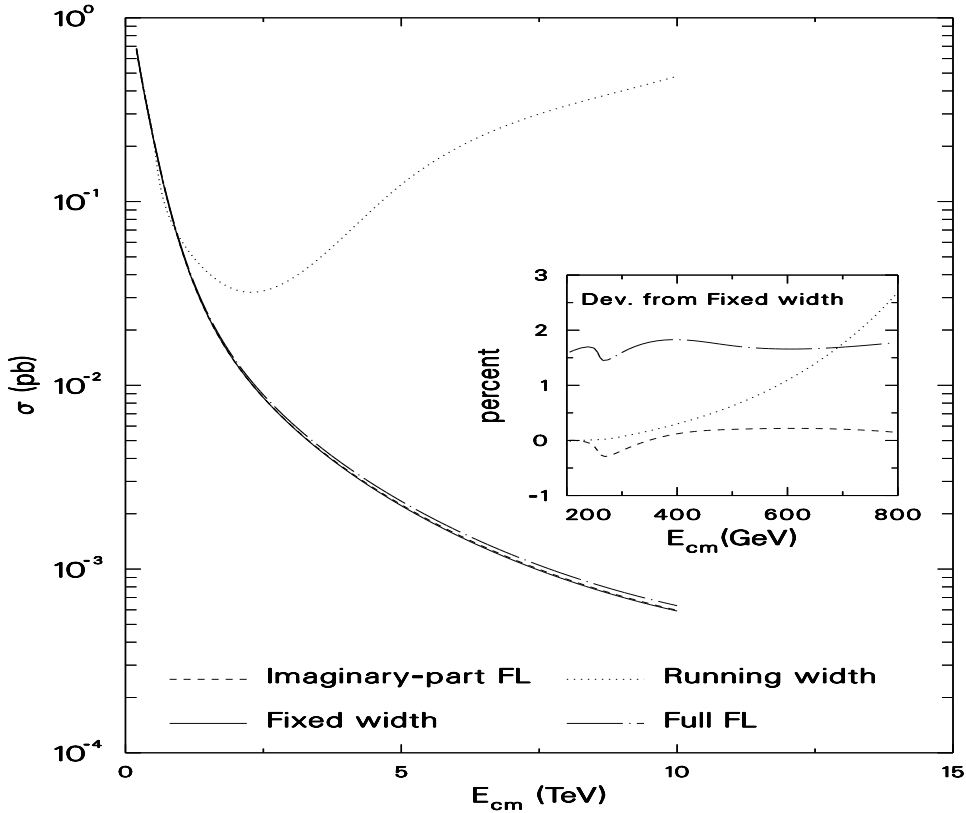}}
\end{picture}
\end{center}
\vspace{0.1cm}
\caption[]{Total cross-sections for the CC10 process (default setup),
           as predicted by WTO, and deviations of the various schemes with 
           respect to the fixed-width scheme in percent.}
\label{fig:bhfa}
\end{figure}
%  #] figure CC10 :
From these results it is clear that for large $s$
the fixed-width scheme and the FL schemes yield reasonable 
results, whereas the running-width scheme diverges.
The latter deviates already at $1\TeV$ significantly and yields wrong results 
at higher energies.
The fixed-width scheme and the imaginary-part FL scheme agree
within $2\,\sigma$ (thus at the percent level) up to
high energies.
Although both the fixed-width scheme and FL schemes respect unitarity 
for large $s$, only the FL schemes satisfy SU(2) gauge invariance.
As mentioned above, 
the gauge-invariance-violating terms that are introduced by the fixed-width 
scheme
in  the cross-sections  for $e^-e^+\to 4f$ are only of order 
${\mathcal O}(m \Gamma/s)$. Therefore, 
the high-energy behavior in six-fermion processes is consistent with unitarity,
although gauge invariance is broken.

At energies above $1\TeV$ the full FL scheme deviates from the fixed-width 
scheme and the minimal (imaginary-part) FL scheme by 1.7--6.3\% (1.9--6.9\%), 
according
to ERATO (WTO). Note that this is just the order of magnitude that is to be 
expected from the running of the couplings.
This rather large effect should be interpreted with care owing
to the omission of QCD and bosonic corrections.

The high-energy behavior of the different schemes as predicted by WTO
is also shown in 
Fig.~\ref{fig:bhfa} for the CC10 cross-section. In the same figure we have
illustrated the behavior of the corrections in a region of energy accessible
to the NLC. 
Deviations from the fixed-width results are 
given, showing the
${\bar t} t$ threshold at approximately $264\GeV$ and the relatively minor
impact of the imaginary-part FL scheme. In the whole range of energy between
$200\,$GeV and $1\TeV$ the imaginary-part FL scheme deviates by less than
0.3\% (only reached around the $\bar t t$ threshold) 
from the fixed-width scheme.

Since the running-width scheme reaches a 9.3\% deviation in the same region, we 
may conclude that the imaginary-part FL scheme offers here the same quantitative
behavior as the fixed-width calculation, but within a fully self-consistent
approach. So does the minimal FL scheme, at least away from the ${\bar t} t$
threshold (where the top-mass effects play a role).

We register a difference between running-width and fixed-width results of
9.3\% (CC10) and 5.1\% (CC11) at $1\TeV$. So, a gauge-invariance-preserving
scheme is of actual relevance already for the NLC, and not only for
theoretical speculations at extremely high energies.
Moreover, a realistic estimate of the radiative corrections%
---within 1\% accuracy---in the phenomenologically interesting region
cannot avoid the proper inclusion of the bosonic corrections. 

%  #[ total xs :
%  #] total xs : 
%  #[ table real high-energy xsection CC20 :
\begin{table}%[b]
\small
\begin{center}
$$
\setlength{\arraycolsep}{\tabcolsep}
\def\tstrut{\vrule width0pt height2.5ex}
\def\bstrut{\vrule width0pt depth1.0ex}
\renewcommand\arraystretch{1.3}
\begin{array}{|l|r@{}lr@{}lr@{}lr@{}lr@{}l|}
\hline
\hline
\sqrt{s}                            & \multicolumn{2}{c}{200\GeV} 
& \multicolumn{2}{c}{500\GeV} & \multicolumn{2}{c}{1 \TeV} 
& \multicolumn{2}{c}{2\TeV} & \multicolumn{2}{c|}{10 \TeV} \\
\hline
\hline
\mbox{Running width}                & 
0.7058&(10) & 0.3728&(11) & 0.2751&(17) & 0.2163&(19) & 0.5135&(10) \\
\mbox{}                  & 
0.7070&(4)  & 0.374&(1)   & 0.281&(4)   &       &     &       &     \\
\hline
\mbox{Fixed width}                  & 
0.7062&(8)  & 0.3734&(11) & 0.2709&(17) & 0.1965&(19) & 0.0305&(6) \\
\mbox{}                  & 
0.70719&(6) & 0.372&(1)   & 0.275&(4)   &       &     &       &    \\
\hline
\parbox[t]{2.2cm}
{\tstrut Running width \\ $+$ $\gamma WW$ vertex factor\bstrut} & 
0.7045&(10) & 0.3736&(11) & 0.2763&(17) & 0.2176&(19) & 0.5261&(10) \\
\hline
\mbox{Minimal FL}  &   
0.7060&(6)  & 0.3713&(11) & 0.2705&(17) & 0.2000&(19) & 0.0306&(6) \\
\mbox{scheme} &&&&&&&&&& \\
\hline
\parbox{2.2cm}{\tstrut Full FL scheme\bstrut }  & 
0.7177&(7)  & 0.3776&(11) & 0.2797&(17) & 0.2076&(22) & 0.0326&(6) \\
\mbox{}                  & 
0.7186&(3)  & 0.377&(5)   & 0.280&(9)   &       &     &       &    \\
\hline
\hline
\end{array}
$$
\caption[]{Total cross-sections (in pb) for the CC20 process at high energies 
           (default setup), as predicted by ERATO (first entry) and,
           in the NLC energy range, by WTO (second entry).
           }
\label{tab:highenergyxsec}
\end{center}
\end{table}
%  #] table real high-energy xsection CC20 : 
In Table~\ref{tab:highenergyxsec} we list the total cross-section 
for the process \processcctwenty\
at NLC and higher energies.
The predictions by ERATO 
have been confirmed in the
NLC energy region by WTO within the integration errors.
This table shows the same features that were already discussed for
CC10/11 before.
The cross-section of this process is more sizeable than that of the
CC10 process owing to the presence of photonic $t$-channel diagrams.
The relative importance of the \PW-pair-production diagrams is suppressed
and therefore the bad high-energy behavior of the running-width schemes, which
originates from these diagrams, occurs at higher energies.
Note that also the running-width scheme with
the U(1)-invariance-preserving $\gamma WW$-vertex factor diverges.

The net effect of the full FL scheme is quite large,
reaching $5.6\%$ at $2\TeV$ and $6.9\%$ at $10\TeV$.
This effect can be understood by noticing that in the fixed-width scheme we
use $G_F$ everywhere (except for ISR). In the TeV range the
$t$-channel photon-exchange contributions become dominant and in Born 
approximation these diagrams are evaluated for $\alpha^{-1} = 131.2$.
In the full FL scheme $\alpha$ is effectively replaced by
$\alpha(|t_{min}|) = \alpha(s\,[1-\cos(10^\circ)]/2)$, 
which is larger
($\alpha^{-1} = 131.2$ roughly corresponds to $q^2=[18\,$GeV$]^2$).
A rough estimate of the effect gives $[\alpha(|t_{min}|)/131.2]^2 \approx
1.050, 1.090$ for $2,10\TeV$.

%  #] Numerical results : 

%  #] The process $e^-e^+\to e^-\bar{\nu}_eu\bar{d}$: 
%  #[ Conclusions :

\section{Conclusions}

In Ref.~\cite{bhf1} we had introduced the fermion-loop (FL) scheme for the
gauge-invariant treatment of the finite-width effects of $W$ and $Z$
bosons. This scheme consists in including all fermionic one-loop
corrections in tree-level amplitudes and resumming the self-energies.
In this article we have presented the justification and an extension of the 
FL scheme. We have performed 
the full resummation and renormalization of the fermionic one-loop corrections 
to six-fermion processes, including virtual massive top-quark effects and 
$\varepsilon$-tensor contributions.
A simple {\it effective Born\/} prescription has been presented, which
allows for a relatively straightforward implementation of the full fermionic 
one-loop corrections to LEP2 six-fermion processes.
We have given explicit formulae that are sufficient for all such non-QCD
processes and discussed the gauge 
invariance of the amplitudes. We have implemented the full FL scheme
in three different Monte-Carlo generators, which were used to compute the 
cross-section for the processes \processccten, \processcceleven, and
\processcctwenty, and we have compared the results with other schemes,
including simplified variants of the full FL scheme.

The fixed-width and FL schemes behave properly in the 
collinear and high-energy regions of phase space, as required. 
The other schemes are seen to diverge for large $s$.
It should be noted, however, that only the FL schemes 
satisfy SU(2) gauge invariance.
The gauge-invariance-violating terms in the fixed-width scheme are suppressed 
for large $s$ in six-fermion processes. 
For processes with more intermediate gauge bosons the fixed-width scheme 
gives rise to a bad high-energy behavior.

In order to include the finite width into tree-level matrix elements it
is sufficient to use a minimal subset of the fermionic corrections given by
the imaginary parts of the fermion loops for massless
fermions. This significantly simplifies the expressions and 
increases the speed of the computation.
Moreover, 
in practical LEP2 calculations it is even sufficient to use running widths
and to supplement these by incorporating the imaginary parts of the fermionic 
one-loop corrections to the $\gamma WW$ vertex in the limit
$q_\ga^2\to0$.

In contrast to the schemes that are only based on the imaginary parts of the 
fermion loops the full FL scheme includes the complete
fermionic ${\mathcal O}(\alpha)$ corrections.
Indeed, the full FL scheme can be viewed as a first attempt to include
all ${\mathcal O}(\alpha)$
radiative corrections.
As a consequence, the corresponding results differ 
from the ones obtained with the other schemes by some percent at LEP2, 
and much more for higher energies.  
To give an idea of this effect we consider our most complete
set of predictions, i.e., 
the CC20 process with ISR at $\sqrt{s} = 161\,$GeV, $175\,$
GeV and $190\,$GeV{}. The difference 
amounts 
to approximately 1.3--1.6\%. At higher (NLC) energies
(up to $2\TeV$) the differences range from  
1.6--2.0\% for CC11 to  1.5--3\% for CC10 and 1--5\% for CC20.
This is a non-negligible effect from the experimental point of view.
Once more we stress that the 
relatively large effects induced by the 
full FL scheme should be interpreted with 
due caution. 

The full FL scheme,
including the fermionic corrections with resummation, already has some of the 
features that the ultimate event generator should have. Therefore it represents
a justification of
the scheme of Ref.~\cite{bhf1} and a starting point for 
the evaluation of all ${\mathcal O}(\alpha)$
radiative corrections in a situation where we know that
the impact of the 
bosonic corrections cannot be neglected.
It is therefore important to know the bosonic one-loop
corrections and to see whether the increase in the cross-section 
owing to the fermionic corrections is significant for
experiments. Inclusion of the bosonic one-loop
corrections will possibly temper the increase, in 
particular at high energies.
The large impact of the bosonic corrections
is mainly due to corrections of the form $\alpha/\pi\log^2(s/m_{_W}^2)$
as explicitly calculated for on-shell \PW~bosons. 
Another large part of the
bosonic corrections, as e.g.~the leading-logarithmic corrections 
(initial-state radiation), factorizes and can be treated by a convolution.
A proper treatment of the complete 
bosonic one-loop corrections is still 
lacking.

Simultaneous inclusion of bosonic corrections and finite-width effects
leads in general to problems with gauge invariance.
Manifestly gauge-independent amplitudes 
can be obtained upon an expansion about the complex pole.
This so-called pole 
scheme is based on a systematic expansion 
according to the degree of resonance and thus allows an efficient calculation 
of the most important higher-order corrections. 
However, one version of the pole scheme is only applicable sufficiently far
above the \PW-pair production threshold, whereas other versions do
not preserve the SU(2) Ward identities (for the same reason as the 
fixed-width scheme).

A direct generalization
of the FL scheme to the complete (fermionic and bosonic) corrections
is provided by the background-field formalism, where Ward identities
hold also after Dyson summation.  However, also in this formalism the Dyson
summation introduces a gauge dependence, starting at the loop level which
is not completely taken into account. This gauge dependence is nothing but a 
reflection of the fact that any resummation is arbitrary to some extent.

Another possibility would be to use a hybrid scheme, i.e., using the
FL scheme and adding the bosonic corrections by means of the pole
scheme. 
In any case, the proper inclusion of the bosonic corrections needs
further investigation.

\section*{Note added}
Shortly before finishing this paper a preprint related to our work
appeared \cite{Beuthe}. In this paper the absorptive part of the
$\gamma WW$ vertex is evaluated for massive fermions
and the photon on shell. It is stressed that the
axial contribution of this vertex is not zero. 
We would like to emphasize that this is not
in contradiction to our previous work \cite{bhf1}. In Ref.~\cite{bhf1}
we restricted ourselves to massless fermions and even issued a warning
that (massive) top-quark effects were not included. Moreover, as the
axial
contribution satisfies the Ward identities on its own it is
not required for a gauge-invariant description of finite-width effects.
Once the axial contribution is taken into account the non-absorptive
parts should be included as well: then one arrives at 
the full fermion-loop scheme that
we are presenting in this paper.

\section*{Acknowledgments}    
This research has been partly supported by the EU under contract
numbers CHRX-CT-92-0004, CHRX-CT-93-0319, and CHRX-CT-94-0579.
W. Beenak\-ker is supported  by a fellowship of the Royal Dutch Academy 
of Arts and Sciences.
G.J. van Oldenborgh, J. Hoogland and R.~Kleiss are supported by FOM.
C.G. Papadopoulos would like to thank the Department of Physics of the
University of Durham, UK, where part of this work was done,
for its kind hospitality.
G.~Passarino gratefully acknowledges the help of A.~Ballestrero and
R.~Pittau in several comparisons performed during this work.
We would also like to thank the CERN Liquid Support Division.

%  #] Conclusions : 
\appendix
%  #[ Appendix: Gauge-boson self-energies :

\section{Gauge-boson self-energies} 
\label{app:self-energies}

In this appendix we give explicit formulae for the fermionic one-loop 
contributions to the gauge-boson self-energies. First we give the results for
arbitrary fermion masses, subsequently we take the limit $m_f\to 0$ for 
$f\ne t$. The electric charge fraction of the fermion $f$ is denoted by $Q_f$, 
its vector and axial-vector couplings to the \PZ~boson by $v_f$ and $a_f$, its
iso-spin by $I_f^3$, its left-handed hypercharge by $Y_f^L = 2(Q_f - I^3_f)$ 
[$-1$ for leptons and $1/3$ for quarks], and its color factor by $N_c^f$.

We do not include tadpole contributions, i.e., we implicitly assume that the
tadpoles are removed by appropriate counter terms.

The fermionic contributions to the photon self-energy and photon--\PZ~mixing 
energy read
\begin{eqnarray} \label{eq:gg_gz_ses}
\USgg(p^2) & = & {\Ual \over 3\pi }\, \sum\limits_f\,
    N_c^f \, Q_f^2\, \biggl\{ p^2 \left[B_0(p^2,0,0) - \frac{1}{3}\right]
  + F(p^2,m_f) \biggr\} \;, \nl
\USzg(p^2) & = & - {\Ual \over 3\pi }\, \sum\limits_f\,
    N_c^f \, \Uv_f Q_f\, \biggl\{ p^2 \left[B_0(p^2,0,0) - \frac{1}{3}\right]
  + F(p^2,m_f) \biggr\} \;, 
\eqa
where $\Ual = \Ue^2/(4\pi)$, $B_0$ is the one-loop scalar 2-point function 
\cite{tHooftVeltman}, and
\bq
 F(p^2,m_f) = (p^2+2m_f^2)\,B_0(p^2,m_f,m_f) - 2 m_f^2\,B_0(0,m_f,m_f)
              - p^2 \,B_0(p^2,0,0) \;.
\eq
The function $F(p^2,m_f)$ is finite and vanishes for $m_f=0$ as well as for 
$p^2=0$. For $|p^2| \ll m_f^2$ we obtain for 
$\UPgg(p^2) \equiv \USgg(p^2)/p^2$ 
\begin{eqnarray}
\UPgg(p^2) & \approx & { \Ual \over 3\pi } \sum\limits_f N_c^f 
    \,Q_f^2\,\Bigl[ B_0(0,m_f,m_f) + \frac{p^2}{5m_f^2} \Bigr]
\;.
\end{eqnarray}
The running of the coupling is given by
\bq
    p^2\frac{\partial}{\partial p^2} \frac{\hat{\alpha}}{\alpha(p^2)} = 
    p^2\frac{\partial}{\partial p^2} \UPgg(p^2) 
                {\stackrel{p^2\to0}{\ \longrightarrow\ }} 0
\;.
\eq
As a result of the decoupling, the fermion $f$ does not contribute to the  
$p^2$ evolution of the running electromagnetic coupling $\alpha(p^2)$ if 
$|p^2| \ll m_f^2$. Hence, the masses of the light fermions cannot be 
neglected for very low momentum transfers, $|p^2|\lesssim m_f^2$, as is the 
case for 
near-collinear photon emission from light particles.

The fermionic contributions to the \PZ\ and \PW~self-energies are given by
\eqn{eq:ZW_selfenergy_split}
with 
\bqa 
\hat{ \overline{\Sigma} }_\ssZ(p^2) &=& {\Ual \over 3\pi }\,
        \sum\limits_f\, N_c^f \, \biggl\{ (\Uv_f^2+\Ua_f^2)\, p^2 
        \left[B_0(p^2,0,0) - \frac{1}{3}\right] \nl
&&{} \hphantom{ {\Ual \over 3\pi }\sum\limits_f\, N_c^f \,A} 
      + (\Uv_f^2+\Ua_f^2+4 I_f^3 Y_f^L \Ua_f^2)\,F(p^2,m_f)\biggr\} \;,\nl[1mm]
\hat{\overline{\Sigma}}_\ssW(p^2) &=& {\Ugws \over 48\pi^2 }\,
        \sum\limits_f\, N_c^f \, \biggl\{ p^2 \left[B_0(p^2,0,0) 
      - \frac{1}{3}\right] + (1+2 I_f^3 Y_f^L)\,F(p^2,m_f) \biggr\} \:, \nl
\eqa
and
\bqa
T_{\ssZ}(p^2) &=& -\sum\limits_f\, {N_c^f \over 48\pi^2}\,
        \biggl[ 3m_f^2 \,B_0(p^2,m_f,m_f) + 2 I_f^3 Y_f^L\,F(p^2,m_f) \biggr] 
        \;, \nl
T_{\ssW}(p^2) &=&  T_{\ssZ}(p^2) + \sum\limits_{\mathrm{doublets}}
        \, {N_c^f \over 48\pi^2}\,\biggl\{ \left(2p^2-m_f^2-m_{f'}^2\right)
        B_0(p^2,m_f,m_{f'}) \nl[1mm]
&&{}    - (p^2-m_f^2)\,B_0(p^2,m_f,m_f)
        - (p^2-m_{f'}^2)\,B_0(p^2,m_{f'},m_{f'}) \nl[1mm] 
&&{}    + {m_f^2-m_{f'}^2 \over p^2}\,\biggl[
        m_f^2\,B_0(0,m_f,0) - m_{f'}^2\,B_0(0,m_{f'},0) \nl[1mm]
&&{}    \hphantom{+ {m_f^2-m_{f'}^2 \over p^2}A} 
        + (m_{f'}^2-m_f^2)\,B_0(p^2,m_f,m_{f'}) \biggr] \biggr\}\;.  
\eqa
Here $(f,f')$ represent the iso-spin partners within an SU(2) doublet.
The terms involving $F(p^2,m_f)$ in $\hat{ \overline{\Sigma} }_\ssZ$
and  $\hat{ \overline{\Sigma} }_\ssW$ are determined such that the
relations \eqn{eq:relationsSigmas}, which hold for massless fermions,
hold also in the massive case.
At zero momentum transfer the \PZ\ and \PW~self-energies are given by the 
non-universal terms only:
\begin{eqnarray}\label{eq:sigma0f}
{ \hat{\Sigma}_\ssZ(0) \over \Umuz } = 
             { \Ugws \,T_{\ssZ}(0) \over \Uctws\,\Umuz }
     & = & - \sum_f { \Ugws \over  16\pi^2\Uctws }\,N_c^f \,{ m_f^2 \over \Umuz }\, 
             B_0(0,m_f,m_f) \;, \nl 
{ \hat{\Sigma}_\ssW(0) \over \Umuw } = 
             { \Ugws \,T_{\ssW}(0) \over \Umuw } 
     & = & - \sum_f {\Ugws \over 16\pi^2 }\,N_c^f \,{ m_f^2 \over \Umuw }\,
             \biggl[ B_0(0,m_f,m_f) + { 1 \over 2 } \nl
     &&{} \hspace{3cm} 
           + \frac{m_{f'}^2}{ m_{f'}^2- m_f^2} \log\frac{m_f}{ m_{f'}} \biggr] \;.
\end{eqnarray}

At momentum transfers of the order of the LEP1/2 energies the
fermion masses can be neglected, with the exception of the top-quark mass.
In this limit all contributions of the form $F(p^2,m_f)$ drop out for $f\neq t$
and the universal self-energies are proportional to the photon self-energy:   
\bqa 
\USzg(p^2) & = & {\Ustw \over \Uctw }\,\parent{ 1 - {3 \over 8\Ustws} }\,
                 \USgg(p^2) \;, \nl[1mm] 
\hat{ \overline{\Sigma} }_\ssZ(p^2) & = & {\Ustws \over \Uctws}\,
                 \parent{ 1 - {3 \over 4\Ustws} 
                 + {3 \over 8\Ustw^4} }\,\USgg(p^2) \;, \nl[1mm]
\hat{\overline{\Sigma}}_\ssW(p^2) & = & {3 \over 8\Ustws}\,\USgg(p^2)\;.
\eqa
The non-universal contributions $T_{\ssZ}$ and $T_{\ssW}$ take on the form
\bqa
T_{\ssZ}(p^2) &=& -{N_c^t \over 144\pi^2}\,
        \biggl[ 9m_t^2 \,B_0(p^2,m_t,m_t) + F(p^2,m_t) \biggr] \;, \nl
T_{\ssW}(p^2) &=&  {N_c^t \over 48\pi^2 }\,
        \biggl[ \left(2p^2 - m_t^2 - \frac{m_t^4}{p^2}\right)B_0(p^2,m_t,0)\nl 
&&{}    \hphantom{{N_c^t \over 48\pi^2 }a}
      - \frac{4}{3}\left(p^2 + 2m_t^2\right) B_0(p^2,m_t,m_t)
      - \frac{2}{3} p^2 \,B_0(p^2,0,0) \nl 
&&{}    \hphantom{{N_c^t \over 48\pi^2 }a}
      + \frac{2}{3} m_t^2 \,B_0(0,m_t,m_t) 
      + \frac{m_t^4}{p^2} \,B_0(0,m_t,0) \biggr] \;.
\eqa
At zero momentum transfer \eqn{eq:sigma0f} simplifies to
\begin{eqnarray}\label{eq:sigma0t}
{ \hat{\Sigma}_\ssZ(0) \over \Umuz } = 
             { \Ugws \,T_{\ssZ}(0) \over \Uctws\,\Umuz }
     & = & - { \Ugws \over  16\pi^2\Uctws }\,N_c^t \,{ m_t^2 \over \Umuz }\, 
             B_0(0,m_t,m_t) \;, \nl 
{ \hat{\Sigma}_\ssW(0) \over \Umuw } = 
             { \Ugws \,T_{\ssW}(0) \over \Umuw } 
     & = & - {\Ugws \over  16\pi^2 }\,N_c^t \,{ m_t^2 \over \Umuw }\,
             \lrbr B_0(0,m_t,m_t) + { 1 \over 2 } \rrbr \;.
\end{eqnarray}

We note that when evaluating $B_0(\mu_{\ssZ,\ssW},m_1,m_2)$ 
care has to be taken,
because the complex pole does not lie on the usual physical (first) Riemann 
sheet.
Its location is determined by the fact that it should smoothly approach
the value for a stable gauge boson when the coupling tends to zero.

%  #] Appendix: Gauge-boson self-energies : 
%  #[ Appendix: Fermion-loop corrections to the triple gauge-boson vertex :

\section{Fermion-loop corrections to the triple gauge-boson vertices}
\label{app:vww}

At one loop, the fermion-loop corrections to the triple gauge-boson vertices 
consist of two sets of contributions, which are generically depicted in the
two diagrams in Fig.~\ref{fig:vww} for the vertices involving W~bosons. 
In the diagram on the right-hand side $f$ is a down-type fermion, in the one 
on the left-hand side $f$ is up-type.
%  #[ figure BWW vertex :
\begin{figure}
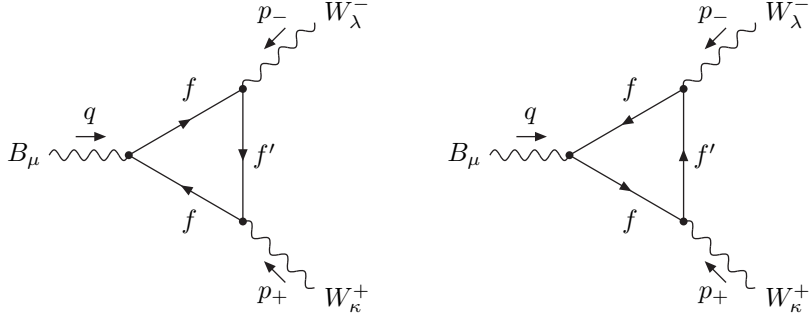

\begin{center}
\DiagramVWWa\hspace*{2em}\DiagramVWWb
\end{center}
\caption{Feynman diagrams for the fermion-loop corrections to the
         $BW^+W^-$ vertex ($B=\gamma,Z$). All particles are assumed to be
incoming.}
\label{fig:vww}
\end{figure}
%  #] figure BWW vertex :

\subsection{Vertices involving charged bosons}
\label{app:vww_massive}

In this appendix we give the fermion-loop corrections to the $ZWW$ and 
$\gamma WW$ vertex functions for arbitrary fermion masses. All results are
presented in terms of one-loop tensor-integral coefficient functions: 
%  #[ expression stefan:
\begin{eqnarray}
\lefteqn{ \hat V^{(1),\,\{\ga,Z\} W^+W^-,\,\mu\kappa\lambda}(q,p_+,p_-) \;=\;}
&& \nonumber\\[.3em]
&&   \biggl\{     \frac{\Ue\Ugws}{32\pi^2} \, \sum_f N^f_c \, 
          \lrbr -\,|Q_f|\,\left\{1,-{\Uctw \over \Ustw} \right\}
          + { 1-2|Q_f| \over 2\Ustw\Uctw }\,\{0,1\} \rrbr\,
\nonumber\\[.3em] 
& & \times\Bigl\{ q^\mu g^{\kappa\lambda} \, \Bigl[ B_0(p_+^2)
        + p_+^2\,(C_0+C_1-C_2) - (q^2+2m^2_{f}+2m^2_{f'})\,C_1 
        +8\,C_{001} \Bigr]
\nonumber\\[.3em] && {}
        + (p_+-p_-)^\mu g^{\kappa\lambda} \, \Bigl[ -B_0(p_+^2)-q^2\,C_1
\nonumber\\[.3em] && \hspace{3em} {}
        + (p_+^2-m^2_{f}-m^2_{f'})\,(C_0+C_1+C_2)+4\,C_{00}+8\,C_{001} \Bigr]
\nonumber\\[.3em] && {}
        + p_-^\lambda g^{\mu\kappa} \, \Bigl[ 2\,B_0(p_-^2)+4\,B_1(p_-^2)
                                              -B_0(p_+^2)+B_0(q^2) 
\nonumber\\[.3em] && \hspace{3em} {}
        + (p_-^2-2p_+^2+m^2_{f}+m^2_{f'})\,C_0 + (p_-^2-p_+^2-q^2)\,(2C_1+C_2) 
\nonumber\\[.3em] && \hspace{3em} {}
        -8\,C_{00}-16\,C_{001}-8\,C_{002} \Bigr]
\nonumber\\[.3em] && {}
        + (q-p_+)^{\lambda} g^{\mu\kappa} \, \Bigl[ - B_0(q^2)-B_0(p_+^2)
\nonumber\\[.3em] && \hspace{3em} {}
        + (p_-^2-m^2_{f}-m^2_{f'})\,C_0 + (p_-^2-p_+^2-q^2)\,C_2
        -8\,C_{002} \Bigr]
\nonumber\\[.3em] && {}
        - q^\mu p_+^\kappa p_-^\lambda \, 2\,[C_1+3\,C_{11}+2\,C_{111}
        +3\,C_{112}]
\nonumber\\[.3em] && {}
        - (p_+-p_-)^\mu p_+^\kappa p_-^\lambda \, 
        [6\,C_1+10\,C_{11}+11\,C_{12}+4\,C_{111}+14\,C_{112}]
\nonumber\\[.3em] && {}
        - q^\mu p_+^\kappa (q-p_+)^{\lambda} \, 
        2\,[C_2+C_{12}-C_{22}+C_{112}+C_{122}-2\,C_{222}]
\nonumber\\[.3em] && {}
        - q^\mu (q-p_+)^{\lambda} (q-p_-)^{\kappa} \, 2\,C_{112} 
\nonumber\\[.3em] && {}
        - (p_+-p_-)^\mu p_-^\lambda (q-p_-)^{\kappa} \, 
        2\,[C_1+3\,C_{11}+2\,C_{12}+2\,C_{111}+3\,C_{112}+C_{122}]
\nonumber\\[.3em] && {}
        - (p_+-p_-)^\mu (q-p_+)^{\lambda} (q-p_-)^{\kappa} \, 
        [C_{12}+C_{112}+C_{122}]
\Bigr\}
\nonumber\\[.3em] 
&&
{} + \quad (\,p_+\leftrightarrow-p_-,\,q\rightarrow-q,
            \,\kappa\leftrightarrow\lambda\,) \quad \biggr\} %} 
\nl[.5em]
\lefteqn{ {}+ \biggl\{ \frac{\Ue\Ugws}{32\pi^2} \, \sum_f N^f_c \, 
          { 1 \over \Ustw\Uctw }\,\{0,1\}\,m^2_{f} \, \Bigl[ 
          q^\mu g^{\kappa\lambda} \, C_1 
          +(p_+-p_-)^\mu g^{\kappa\lambda} \, C_1
        } &&
\nonumber\\[.0em] && {}
          +p_-^\lambda g^{\mu\kappa} \, (2\,C_1+C_2)
          +(q-p_+)^{\lambda} g^{\mu\kappa} \, C_2 \; \Bigr]
\nonumber\\[.3em] 
&&
{} + \quad (\,p_+\leftrightarrow-p_-,\,q\rightarrow-q,
            \,\kappa\leftrightarrow\lambda\,) \quad \biggr\} %} 
\nl[.5em]
\lefteqn{ {} + \frac{\Ue\Ugws}{16\pi^2} \, \sum_f N^f_c \,
          \lrbr -\,Q_f\,\left\{1,-{\Uctw \over \Ustw} \right\}
          + { I_f^3-Q_f \over \Ustw\Uctw }\,\{0,1\} \rrbr\,
          }
&& \nonumber\\[.3em] 
&& \times \Bigl\{
 i \varepsilon^{\mu\lambda\alpha\beta}p_-^\kappa p_{+,\alpha} p_{-,\beta}
\, 4\,C_{12}
-i \varepsilon^{\mu\lambda\alpha\beta}p_+^\kappa p_{+,\alpha} p_{-,\beta}
\, 4\left[C_2+C_{22}\right]
\nonumber\\[.3em] && {}
-i \varepsilon^{\mu\kappa\alpha\beta}p_+^\lambda p_{+,\alpha} p_{-,\beta}
\, 4\,C_{12}
+i \varepsilon^{\mu\kappa\alpha\beta}p_-^\lambda p_{+,\alpha} p_{-,\beta}
\, 4\left[C_1+C_{11}\right]
\nonumber\\[.3em] && {}
+i \varepsilon^{\mu\kappa\lambda\alpha} q_\alpha \,
\left[ (p_+^2+m_f^2-m_{f'}^2)\,C_2 - (p_-^2+m_f^2-m_{f'}^2)\,C_1 \right] 
\; \Bigr\}
\nonumber\\[.5em] 
\lefteqn{ {}+\frac{\Ue\Ugws}{8\pi^2} \, \sum_f N^f_c \, 
          { I_f^3 \over \Ustw\Uctw }\,\{0,1\}\,m^2_{f} \, \Bigl[ 
-i \varepsilon^{\mu\kappa\lambda\alpha}p_{-,\alpha} \,C_1
+i \varepsilon^{\mu\kappa\lambda\alpha}p_{+,\alpha} \,C_2
\; \Bigr] \;, } && 
\label{eq:vwwmassive}
\end{eqnarray}
%  #] expression stefan :
where $m_{f}$ denotes the mass of the fermion that couples to the
neutral gauge boson and $m_{f'}$ the mass of its iso-spin partner.
The totally-antisymmetric $\varepsilon$-tensor is fixed according to 
$\varepsilon^{0123} = 1$.
The conventions for the arguments of the tensor-integral coefficient functions 
$C_{ijk}$ follow Ref.~\cite{De93},
\begin{eqnarray} \label{def:tensor_coefficients}
B_i(q^2)  &=& B_i(q,m_f,m_f) \;,   \qquad
B_i(p_\pm^2) =  B_i(p_\pm,m_{f'},m_{f}) \;,
\nonumber\\
C_{ijk} &=& C_{ijk}(p_-,-p_+,m_{f'},m_{f},m_{f}) \;.
\end{eqnarray}
In \eqn{eq:vwwmassive}
the substitution $(\,p_+\leftrightarrow-p_-,\,q\rightarrow-q\,)$ also
applies to the arguments of the tensor coefficients. For the 
$C_{ijk},\,C_{ij}$, and $C_i$ this is equivalent to the interchange of
``1'' and ``2'' in the indices, leaving the arguments unaffected. 

{}From the above expression for $\hat V^{(1),\,\{\ga,Z\} W^+W^-}$
one can easily read off the one-loop coefficients $G^{\ga,\,(1)}$ and 
$G^{I,\,(1)}$. The UV divergence contained in $\hat V^{(1)}$ can be
written as
\begin{eqnarray}
  \hat V^{(1),\,\{\ga,Z\} W^+W^-}_{\mu\kappa\lambda}(q,p_+,p_-) &=&
  { \Ue\Ugws \over 48\pi^2} \,\left\{1,-{\Uctw \over \Ustw}\right\}\, 
  \Gamma_{\mu\kappa\lambda}(q,p_+,p_-)\,
  \sum_f N^f_c \, B_0(q^2) 
\nl \lefteqn{ \phantom{=}
\hspace{3.2cm} {}+\mbox{UV-finite terms}  
            } \hspace{3cm}
\nl \lefteqn{ = 
\Ue \left\{1,-{\Uctw \over \Ustw}\right\}\, 
{ \hat{\overline{\Sigma}}_{\ssW}(q^2) \over q^2 } \,
\Gamma_{\mu\kappa\lambda}(q,p_+,p_-)
+\mbox{UV-finite terms} \;,
            } \hspace{3cm}
\end{eqnarray}
with $ \Gamma_{\mu\kappa\lambda}$ defined in \eqn{def:Gamma_tree}.

The vertex correction  $\hat V^{(1)}$ contains $\varepsilon$-terms, 
which contribute to the triangle anomaly. 
The cancellation of the anomaly requires the
contribution of the massive top quark.
Writing
\begin{equation}
\hat V^{(1)} = \sum_{f} \hat V^{(1)}_f(m_f=0) +
\left[\hat V^{(1)}_{\mathrm{top}}(m_t\ne0) -\hat V^{(1)}_{\mathrm{top}}(m_t=0)
\right],
\end{equation}
it follows from the anomaly-cancellation conditions that all
$\varepsilon$-terms disappear in the (massless) sum on the right-hand side.  
The remainder contains $\varepsilon$-terms and is $m_t$-dependent. 
This $m_t$ dependence is known
to produce effects of delayed unitarity cancellation,
which may become relevant at high energies. 

Introducing the auxiliary projective momenta and tensors 
\begin{eqnarray} \label{def:covariants}
\bar q^\mu &=& \parbox{4.5cm}{$ \displaystyle
\frac{1}{q^2}\Bigl[ (qp_-)p_+^\mu-(qp_+)p_-^\mu \Bigr]\;, $}
\bar g^{\kappa\lambda}_0 =
g^{\kappa\lambda} - \frac{p_+^\lambda p_-^\kappa}{(p_+p_-)}\;, 
\nonumber\\
\bar p_\pm^\nu &=& \parbox{4.5cm}{$ \displaystyle
p_\mp^\nu-\frac{(p_+p_-)}{p_\pm^2}p_\pm^\nu\;, $}
\bar g^{\mu\nu}_\pm = g^{\mu\nu} - \frac{q^\nu p_\mp^\mu}{(qp_\mp)}\;,
\end{eqnarray}
the terms in \eqn{eq:vwwmassive} involving no $\varepsilon$-tensors
can be rewritten in the following way
\begin{eqnarray}
\lefteqn{
\hat V^{(1),\,\{\ga,Z\} W^+W^-,\,\mu\kappa\lambda}(q,p_+,p_-) =
q^\mu p_+^\kappa p_-^\lambda X_{0+-}
+\bar q^\mu \bar p_+^\kappa \bar p_-^\lambda X^{0+-}
}
\nonumber\\[.3em] && \hspace{2em} {} 
+ \bar q^\mu p_+^\kappa p_-^\lambda X^0_{+-}
+ q^\mu p_+^\kappa \bar p_-^\lambda X^-_{0+}
+ q^\mu \bar p_+^\kappa p_-^\lambda X^+_{0-}
\hspace{2cm}
\nonumber\\[.3em] && \hspace{2em} {} 
+ q^\mu \bar p_+^\kappa \bar p_-^\lambda X^{+-}_0
+ \bar q^\mu p_+^\kappa \bar p_-^\lambda X^{0-}_+
+ \bar q^\mu \bar p_+^\kappa p_-^\lambda X^{0+}_-
\nonumber\\[.3em] && \hspace{2em} {} 
+ q^\mu \bar g^{\kappa\lambda}_0 X_0
+ p_+^\kappa \bar g^{\mu\lambda}_+ X_+
+ p_-^\lambda \bar g^{\mu\kappa}_- X_-
\nonumber\\[.3em] && \hspace{2em} {} 
+ \bar q^\mu \bar g^{\kappa\lambda}_0 X^0
+ \bar p_+^\kappa \bar g^{\mu\lambda}_+ X^+ 
+ \bar p_-^\lambda \bar g^{\mu\kappa}_- X^- 
\nonumber\\[.3em] && \hspace{2em} {}
+ \mbox{terms with $\varepsilon$-tensors} \;  %\Bigr]\; 
\label{eq:vwwdecomp}
\end{eqnarray}
with appropriate coefficient functions $X$.
The covariants introduced in \eqn{def:covariants} are constructed such that
\begin{eqnarray}
(q\bar q) &=& 0, \qquad
(p_\pm\bar p_\pm)=0, \nonumber\\
p_+^\kappa\,\bar g_{0,\kappa\lambda} =
p_-^\lambda \,\bar g_{0,\kappa\lambda} &=& 0, \qquad
q^\mu \,\bar g_{\pm,\mu\nu} = 
p_\mp^\nu \,\bar g_{\pm,\mu\nu} =0. 
\end{eqnarray}
This implies that for a given external leg the ``barred'' quantities in 
$\hat V^{(1),\mu\kappa\lambda}$ drop out upon contraction 
with the corresponding momentum, e.g.\ 
$\bar q^\mu$ and $\bar g^{\mu\nu}_\pm$ vanish after contraction with
$q_\mu$. Hence the barred terms are not required for the restoration of
the Ward identities; this must be done by the ``unbarred'' terms.
On the other hand, each unbarred term contains at least one factor 
that vanishes upon contraction with a conserved
current. At first sight, this situation looks paradoxical: 
the terms that restore gauge invariance do not explicitly contribute
to the amplitude.
However, the decomposition becomes singular for 
$q^2/E_{\scriptscriptstyle CM}^2 \to 0$, 
$p_\pm^2/E_{\scriptscriptstyle CM}^2 \to 0$, 
and as a result of the analyticity of the amplitudes
barred and unbarred terms are related in these limits.
But these limits correspond exactly to the kinematic situations 
where the Ward identities are required to guarantee well-behaved
amplitudes.

\subsection{Vertices involving only neutral bosons}
\label{app:bbb_massive}

In this appendix we give the fermion-loop contributions to the (C--odd and 
CP--even) triple gauge-boson vertices involving only neutral gauge bosons, 
which do not exist at lowest order.
For each fermion the two diagrams of Fig.~\ref{fig:vww} with $f'=f$
contribute, and all terms cancel apart from those involving
$\varepsilon$-tensors. We find for the generic, finite $B_1B_2B_3$ vertex
($B_i=\gamma,Z$):
%  #[ expression stefan :
\begin{eqnarray}
\lefteqn{ \hat V^{(1),\,\mu\kappa\lambda}_{B_1 B_2 B_3}(q_1,q_2,q_3)
\;=\; \frac{\Ue^3}{8\pi^2} \, \sum_f N^f_c \, \hat{a}_{123}^{+++}
          }
&& \nonumber\\[.3em] 
&& \times \Bigl\{
 i \varepsilon^{\mu\lambda\alpha\beta}q_3^\kappa q_{2,\alpha} q_{3,\beta}
\, 4\,C_{12}
-i \varepsilon^{\mu\lambda\alpha\beta}q_2^\kappa q_{2,\alpha} q_{3,\beta}
\, 4\left[C_2+C_{22}\right]
\nonumber\\[.3em] && {}
-i \varepsilon^{\mu\kappa\alpha\beta}q_2^\lambda q_{2,\alpha} q_{3,\beta}
\, 4\,C_{12}
+i \varepsilon^{\mu\kappa\alpha\beta}q_3^\lambda q_{2,\alpha} q_{3,\beta}
\, 4\left[C_1+C_{11}\right]
\nonumber\\[.3em] && {}
-i \varepsilon^{\mu\kappa\lambda\alpha}q_{3,\alpha} 
\left[ (2m_f^2-q_3^2)\,C_1 + q_2^2\,C_2 \right] 
\nonumber\\[.3em] && {}
+i \varepsilon^{\mu\kappa\lambda\alpha}q_{2,\alpha} 
\left[ (2m_f^2-q_2^2)\,C_2 + q_3^2\,C_1 \right] 
\; \Bigr\}
\nonumber\\[.5em] 
\lefteqn{ {}+\frac{\Ue^3}{4\pi^2} \, \sum_f N^f_c \, m_f^2\,\hat{a}_{123}^{-++}
          \, \Bigl[ 
i \varepsilon^{\mu\kappa\lambda\alpha}q_{3,\alpha} \,C_1
-i \varepsilon^{\mu\kappa\lambda\alpha}q_{2,\alpha} \,C_2
\; \Bigr]  } \nonumber\\[.3em] 
\lefteqn{ {}+\frac{\Ue^3}{4\pi^2} \, \sum_f N^f_c \, m_f^2\,\hat{a}_{123}^{+-+}
          \, \Bigl[ 
i \varepsilon^{\mu\kappa\lambda\alpha}q_{3,\alpha} \,(C_0+C_1)
-i \varepsilon^{\mu\kappa\lambda\alpha}q_{2,\alpha} \,C_2
\; \Bigr]  } \nonumber\\[.3em] 
\lefteqn{ {}+\frac{\Ue^3}{4\pi^2} \, \sum_f N^f_c \, m_f^2\,\hat{a}_{123}^{++-}
          \, \Bigl[ 
i \varepsilon^{\mu\kappa\lambda\alpha}q_{3,\alpha} \,C_1
-i \varepsilon^{\mu\kappa\lambda\alpha}q_{2,\alpha} \,(C_0+C_2)
\; \Bigr] \;, } \nonumber\\[.3em] && 
\label{eq:bbbmassive}
\end{eqnarray}
with
\begin{eqnarray} \label{def:tensor_coefficients2}
C_{ij} &=& C_{ij}(q_3,-q_2,m_{f},m_{f},m_{f}) \;,
\end{eqnarray}
and
\begin{eqnarray}
\hat{a}_{123}^{\sigma_1\sigma_2\sigma_3} &=& 
(\hat v_f^{\ssB_1} + \sigma_1\hat a_f^{\ssB_1})
(\hat v_f^{\ssB_2} + \sigma_2\hat a_f^{\ssB_2})
(\hat v_f^{\ssB_3} + \sigma_3\hat a_f^{\ssB_3})
\nl && {}
- (\hat v_f^{\ssB_1} - \sigma_1\hat a_f^{\ssB_1})
(\hat v_f^{\ssB_2} - \sigma_2\hat a_f^{\ssB_2})
(\hat v_f^{\ssB_3} - \sigma_3\hat a_f^{\ssB_3}) 
\;.
\end{eqnarray}
The vector and axial-vector couplings $\hat v_f^{\ssB_i}$ and 
$\hat a_f^{\ssB_i}$ of the Z~boson ($B_i=Z$)
to fermions have been defined in \eqn{def:bareZffcoupling}, those of the 
photon ($B_i=\gamma$) to fermions are given by $\hat a_f^{\gamma} = 0$,
$\hat v_f^\gamma=-Q_f$. As expected from the C-invariance of electromagnetic
interactions, the $\gamma\gamma\gamma$ vertex function vanishes.
The result for the \PZ-boson--gluon--gluon vertex can be obtained from the
previous formulae by substituting $\hat v_f^{\ssB_i}\to 1$, 
$\hat a_f^{\ssB_i}\to0$ for the
gluon--fermion couplings and $\Ue^3 \to \Ue g_s^2 \delta^{ab}/2$, where
$g_s$ is the strong coupling constant, $a$ and $b$ the colors of the gluons, 
and $\delta^{ab}$ the unit matrix in color space.

%  #] Appendix: Fermion-loop corrections to the triple gauge-boson vertex : 
%  #[ Appendix: The complex scheme :

\section{The complex scheme}
\label{app:complex_scheme}

%  #[ Renormalization conditions :

In this appendix we give an algorithm to fix the renormalized parameters 
in the complex renormalization scheme in terms of the input parameters.
We discuss two options for the input parameters, the LEP1 
input-parameter scheme and the LEP2 input-parameter scheme.

As a first option we 
take the input parameters of the LEP1 scheme, i.e.,
the real LEP1 \PZ-boson mass $\mz$, the Fermi coupling $G_F$, 
the effective electromagnetic coupling $\Re\{\alpha_\ell(\mzs)^{-1}\}$ at 
the LEP1 \PZ~peak, and the top-quark mass $m_t$. 

First we compute the bare parameters for given $\Delta$ and $\mu_0$,
where $\Delta$ is the dimensional regularized infinity in the loop corrections 
and $\mu_0$ the squared mass parameter of dimensional regularization.

The bare electromagnetic coupling is calculated through the relation
\begin{equation}
\label{alpharenormalization}
\frac{1}{\Ual} = \frac{1}{\alpha(p^2)} - \frac{\SA(p^2,\mut)}{p^2} \;,
\end{equation}
at $p^2=\mzs$,
where we have split the photon self-energy according to
\bq\label{S_gamma_definition}
\USgg(p^2) \equiv \Ual\,\SA(p^2,\mut).
\eq
Our $\alpha(\mzs)$ differs from the commonly used effective
electromagnetic
coupling $\alpha_\ell(\mzs)$ that is extracted from the ratio
$R(s)=\sigma(e^+e^-\to\mbox{hadrons}) / \sigma(e^+e^-\to\mu^+\mu^-)$
via a dispersion relation \cite{alpz}. 
The difference is due to the top-quark contribution to the photon
self-energy, which is included in $\alpha(\mzs)$ but not in
$\alpha_\ell(\mzs)$, where the index ``$\ell$'' refers to ``light fermions''.
Since both running couplings coincide at zero-momentum transfer,
$\alpha(0)=\alpha_\ell(0)$, we have
\begin{equation}
\label{alpha_alphal_relation}
\frac{1}{\alpha(\mzs)} = \frac{1}{\alpha_\ell(\mzs)}
+ \frac{\SAt(\mzs,\mut)}{\mzs} 
- \left. \frac{\partial \SAt(p^2,\mut)}{\partial p^2} \right|_{p^2=0} 
\;,
\end{equation}
where $\SAt$ includes only the contribution of the top-quark loop of
$\SA$, which can be easily read off from \eqn{eq:gg_gz_ses}.

The knowledge of $\Re\{\alpha_\ell(\mzs)^{-1}\}=128.89\pm0.09$ 
\cite{alpz} 
is sufficient to fix $\Re\{\alpha(\mzs)^{-1}\}$ by
\eqn{alpha_alphal_relation}.
Then the real part of \eqn{alpharenormalization} determines the (real)
bare coupling $\Ual$, and the imaginary part of
\eqn{alpharenormalization} yields
\bq \label{eq:Im_alpha}
\Im\{\alpha(\mzs)^{-1}\} = \Im\frac{\SA(\mzs,\mut)}{\mzs}\;,
\eq
so that also $\alpha(\mzs)$ is completely known.
As already mentioned, a precise determination of $\alpha(p^2)$ requires
the knowledge of the non-perturbative hadronic part of $\SA$. Given 
$\Re\{\alpha_\ell(\mzs)^{-1}\}$ as experimental data point, we obtain
$\alpha(p^2)$ through a perturbative evolution with massless fermions
(apart from the top quark) in order
to match the vertex corrections which also
have massless fermions and no QCD corrections. In order to test the
numerical relevance of our procedure we have verified that the resulting
$\alpha(p^2)$ agrees, over a wide range of $p^2$, to better than $0.1\%$
with the one that has been evolved from
 $\alpha(0)$ through a non-perturbative
parameterization of the hadronic vacuum polarization \cite{alpz}.

The Fermi condition is used to fix the ratio of the bare \PW-boson
mass $\sqrt{\Umuw}$ and the weak coupling $\Ugw$, 
where massive fermions, i.e., in practice only the top quark, give 
an extra contribution:
\bq \label{eq:Fermi_condition}
  2\sqrt{2}\,G_F = \frac{\Ugws}{\Umuw-\Ugws \,\TW(0)} 
\ \longrightarrow\ \gsovermu \equiv \frac{\Ugws}{\Umuw} 
                 = \frac{2\sqrt{2}\,G_F}{1+2\sqrt{2}\,G_F\,\TW(0)} \;.
\eq
The bare \PW-boson mass is reconstructed by inverting the LEP1
mass-renormali\-zation condition $\Umuz =  \mz^2 + \Re \hat{Z}(\mz^2)$
for the
\PZ-boson mass.
Using \eqn{eq:relationsSigmas},
\eqns{alpharenormalization}{S_gamma_definition},
the decomposition
 \bq\label{S_W_definition}
 \hat{\overline{\Sigma}}_\ssW(p^2) \equiv \Ugws\,\SW(p^2,\mut),
 \eq
and
\bq \label{bare_parameter-relations}
\Ugws=\gsovermu\,\Umuw, \quad \Ustws = 2\pi\Ual/\Ugws, \quad
\Uctws = 1-\Ustws, \quad \Umuz = \Umuw/\Uctws,
\eq
this leads to
\begin{eqnarray} \label{Umuwdetermination1}
  A     & = & 1 - \gsovermu\,\Re\TZ(\mzs)
              - \gsovermu \Re\lrbr \SW(\mzs,\mut)
              - { 2\pi\,\alpha(\mzs)\,\SW^2(\mzs,\mut) \over \mz^2 } \rrbr\;,\nl
  B     & = & - \mzs  + 4 \pi \,\Re\biggl[\alpha(\mzs) \, \SW(\mzs,\mut)
             \biggr]\;, \nl[1mm]
  C     & = & { 2\pi \over \gsovermu }\,\mzs \Re \alpha(\mzs) \;,\nl[1mm]
  \Umuw & = & \lrbr \,-B+\sqrt{B^2-4AC}\, \rrbr/(2A) \;.
\end{eqnarray}
For $m_f=0$, $f\ne t$, $\SW(m_{_Z}^2,\mut)$ can be replaced by
$(3/16\pi)\SA(\mzs,\mut)$ in \eqn{Umuwdetermination1}.
Now from \eqn{bare_parameter-relations} all bare parameters can be
computed.

At this point we choose to work in the complex renormalization scheme. 
The running couplings are computed using 
\eqns{def:runningalpha}{def:runningstws}.
Using \eqn{alpharenormalization} for $1/\alpha(p^2)$ and a similar expression 
derived from \eqn{def:runninggw} for $1/\gws(p^2)$, it can be proven that 
the running couplings, deduced from the input in the above way, are finite. 
From this the finiteness of the propagator functions and vertex functions 
follows. This also holds for the masses of the weak vector bosons, which are 
computed by iteration.

We start off the iteration by initializing the complex masses of the
\PZ\ and \PW~boson by
\bq
  \muz^{\scriptstyle \mathrm{ini}} = \mz^2-i\mz\gmz \;, \qquad
  \muw^{\scriptstyle \mathrm{ini}} = \mw^2-i\mw\gmw \;,
\eq
with $\mz,\gmz$ the real LEP1 \PZ-boson mass and width, and $\mw,\gmw$ 
reasonable estimates for the real \PW-boson mass and width.
The iteration is based on the equations
\begin{eqnarray}
\muw  & = & \gws(\muw)
  \lrbr \frac{1}{2\sqrt{2}\,G_F}-\TW(\muw)+\TW(0) \rrbr \;,\nl
\muz  & = & \frac{\gws(\muz)}{\ctws(\muz)}
  \lrbr \frac{1}{2\sqrt{2}\,G_F}-\TZ(\muz)+\TW(0) \rrbr \;,
\label{eq:iterationmasses}
\end{eqnarray}
which follow from the complex mass-renormalization conditions
\eqns{def:Wmassrenormalization2}{def:Zmassrenormalization2} and 
\eqn{eq:Fermi_condition}.
We have verified that the iteration yields finite values for the 
complex masses, independent of the regularization
parameters $\Delta$ and $\mu_0$.

As a check, we applied the optical theorem to the processes 
$e^+\nu_e \to e^+\nu_e$ and $e^+e^- \to e^+e^-$, and verified that the computed
complex gauge-boson masses satisfy the resulting relations.

For experiments at LEP2 it is more natural to use $\mw$ as an input
parameter. Following Ref.~\cite{LEP2WWreport} this can be done by
determining $m_t$ from 
$\Re\{\alpha_\ell(\mzs)^{-1}\}$, $\mz$, $G_F$, and $\mw$.
The real \PW-boson mass $\mw$ obeys the condition $\Umuw =  \mw^2 + \Re
\USww(\mw^2)$. Using the relations mentioned before 
\eqn{Umuwdetermination1} this gives
\bq\label{Umuwdetermination2}
\Umuw = \mw^2\left[1-\gsovermu\Re\TW(\mws) 
 - \gsovermu \Re \SW(\mws,\mut)\right]^{-1}.
\eq
Equating this to $\Umuw$ from \eqn{Umuwdetermination1} and 
using $\alpha(\mzs)$ as derived from \eqns{alpha_alphal_relation}{eq:Im_alpha} 
yields a relation between $\mz$, $G_F$, $\Re\{\alpha_\ell(\mzs)^{-1}\}$, $\mw$,
and $m_t$. This can be solved iteratively for $m_t$.
The rest is done as above.

%  #] Renormalization conditions : 

%  #] Appendix: The complex scheme : 
%  #[ Appendix: The complex scheme vs the LEP1 scheme :

\section{The complex scheme versus the LEP1 scheme}
\label{app:complex_vs_LEP1}

\newcommand{\mv}{m_{_V}}
\newcommand{\bmv}{\bar m_{_V}}
\newcommand{\Gv}{\Gamma_{_V}}
\newcommand{\bGv}{\bar\Gamma_{_V}}
\newcommand{\USvvrem}{\hat{\Sigma}_\ssV^{\mathrm{rem}}}

In this appendix we give a short description of the perturbative
relation between the complex-mass scheme and the more familiar on-shell (LEP1)
scheme.
We consider the unrenormalized $V$ self-energy, $\USvv$, which contains 
possible tadpole contributions. If $\Umuv$ is the bare $V$-boson mass squared 
then the complex pole $\muv$ is defined by
\begin{equation}
\muv - \Umuv + \USvv(\muv;\Umuv) = 0 \; .
\label{eq:muvdef}
\end{equation}
The complex pole is a basic property of the $S$-matrix, and therefore 
gauge-invariant. 
Here we stress that in the full theory one should actually write 
$\USvv(s;\mu)$, where $\mu$ denotes the squared 
mass to be used in the propagators 
for the internal $V$ lines. In the following the second argument of 
$\USvv(s;\mu)$ is omitted, since we always understand that 
bare masses and couplings are inserted in the loop calculations. 
Alternatively, one could of course insert renormalized quantities inside 
the loops, but then additional counter-term contributions would have to be taken
into account, which become relevant at the two-loop level.

For clarity, we first develop the formalism up to and including
two-loop order accuracy in the relations for the $V$-boson mass and width, 
and turn to the special case of fermionic one-loop corrections afterwards.
The complex pole is rewritten in terms of real quantities $\bmv$ and
$\bGv$ as
\bq \label{def:mvbar}
\muv = \bmv^2-i\,\bGv\bmv\;,
\eq
so that \eqn{eq:muvdef} yields
\begin{eqnarray}
\bmv^2-i\,\bGv\bmv 
&=& \Umuv-\USvv(\bmv^2-i\,\bGv\bmv)  \nl
&=& \Umuv-\USvv(\bmv^2) + i\,\bGv\bmv \USvv'(\bmv^2)
+\frac{1}{2}\bGv^2\bmv^2 \USvv''(\bmv^2) + \dots \; .
\nl
\label{eq:muv}
\end{eqnarray}
Taking the real and imaginary parts of \eqn{eq:muv}, we get 
relations for the mass,
\bq
\bmv^2 = \Umuv-\Re\USvv(\bmv^2)-\bGv\bmv\Im\USvv'(\bmv^2) + \dots \; ,
\label{eq:bmv}
\eq
and the width,
\bq
\bGv\bmv = \Im\USvv(\bmv^2)-\bGv\bmv\Re\USvv'(\bmv^2)
-\frac{1}{2}\bGv^2\bmv^2\Im\USvv''(\bmv^2) + \dots \; .
\label{eq:bGv}
\eq
The width $\bGv$ can be eliminated from the right-hand sides of 
\eqn{eq:bmv} and \eqn{eq:bGv} by iteration,
\begin{eqnarray}
\bmv^2 &=& \Umuv-\Re\USvv(\bmv^2)
-[\Im\USvv(\bmv^2)][\Im\USvv'(\bmv^2)] + \dots \; ,
\label{eq:bmv2}
\\[.2em]
\bGv\bmv &=& \Im\USvv(\bmv^2)
\biggl\{1-\Re\USvv'(\bmv^2)+[\Re\USvv'(\bmv^2)]^2 \nl
&& \phantom{\Im\USvv(\bmv^2)\biggl\{} {}  %}
-\frac{1}{2}[\Im\USvv(\bmv^2)][\Im\USvv''(\bmv^2)] 
\biggr\} + \dots \; .
\end{eqnarray}
We now introduce the usual on-shell mass $\mv$,
\bq
\mv^2 = \Umuv-\Re\USvv(\mv^2) \; .
\label{eq:mv}
\eq
Eliminating $\Umuv$ from \eqn{eq:bmv2} and \eqn{eq:mv}, $\bmv^2$ can be
calculated from $\mv^2$ by iteration
\bq
\bmv^2 = \mv^2-[\Im\USvv(\mv^2)][\Im\USvv'(\mv^2)] + \dots \; ,
\label{eq:bmv3}
\eq
i.e., $\bmv^2$ and $\mv^2$ coincide at the one-loop level, but differ by
two-loop corrections. 
The on-shell $V$-boson width is defined as
\begin{eqnarray}
\Gv\mv &=& {\Im\USvv(\mv^2)  \over 1+\Re\USvv'(\mv^2) }
\quad \nl
&=& \Im\USvv(\mv^2)
\biggl\{1-\Re\USvv'(\mv^2)+[\Re\USvv'(\mv^2)]^2+ \dots \biggr\} \; , \quad
\label{eq:Gv3}
\end{eqnarray}
giving
\begin{eqnarray}
\bGv\bmv &=& \Gv\mv-[\Im\USvv(\mv^2)][\Im\USvv'(\mv^2)]^2  \nl
&&{}-\frac{1}{2}[\Im\USvv(\mv^2)]^2[\Im\USvv''(\mv^2)] + \dots \; .
\label{eq:Gv2}
\end{eqnarray}

As a next step, we specialize the above formulae to fermionic one-loop
corrections and to the case where the $V$ particle decays exclusively into
massless fermions. Then, we have the relations
\bq
\Im\USvv(s) = \frac{\Gv}{\mv}s \; , \qquad
\Im\USvv'(s) = \frac{\Gv}{\mv} \; ,
\label{eq:imsev}
\eq
in the vicinity of $s=\mv^2$, so that \eqn{eq:bmv3} and \eqn{eq:Gv2} yield
\begin{eqnarray}
\bmv^2 &=& \mv^2-\Gv^2 + \dots \; , 
\label{eq:bmv4}
\\
\qquad \frac{\bGv}{\bmv}&=&\frac{\Gv}{\mv} + \dots \qquad \mbox{or} \qquad
\bGv = \Gv\left(1-\frac{\Gv^2}{2\mv^2}\right)  + \dots \; .
\label{eq:bGv2}
\end{eqnarray}
Equations (\ref{eq:bmv4}-\ref{eq:bGv2}) represent the perturbative
solutions of the basic equations given in App.~\ref{app:complex_scheme}
and are the basis of the so-called fixed-width scheme that we have used
in Section~\ref{se:process}.

Finally we consider the $V$ propagator, which can be rewritten as
\bq
\URPvv(s) = [s-\Umuv+\USvv(s)]^{-1} 
= [s-\muv+\USvv(s)-\USvv(\muv)]^{-1} \; .
\label{eq:propv}
\eq
In the vicinity of the resonance it is possible to extract the
``running-width contributions'' from $\USvv(s)$ according to
\eqn{eq:imsev} as follows,
\bq
\USvv(s) = \USvvrem(s) + i\frac{\Gv}{\mv}s.
\label{eq:usvrem}
\eq
Inserting \eqn{eq:usvrem} into 
\eqn{eq:propv} and using the relations \eqn{def:mvbar}, \eqn{eq:bmv4},
and \eqn{eq:bGv2}, we arrive at
\bq
\URPvv(s) = \left[s-\mv^2+i\frac{\Gv}{\mv}s
+\USvvrem(s)-\USvvrem(\muv)\right]^{-1} \; ,
\label{eq:propv2}
\eq
showing the appearance of the familiar line-shape parameters for the
process $e^+e^-\to V\to\bar ff$. The
``remainder'' $\USvvrem(s)$ consists of the complete real part of the 
self-energy $\USvv(s)$ and the contribution of heavy fermions to the
imaginary part, such as the top-quark contributions for $V=W,Z$, which
become relevant at $s>(m_t+m_b)^2$ and $s>4m_t^2$, respectively. 
Equation (\ref{eq:propv2}) explicitly illustrates that all fermionic self-energy
corrections are resummed.

The same translation dictionary between complex-mass scheme and on-shell
scheme will work in the more realistic case where we consider the full neutral
sector of the Standard Model, therefore including also $\gamma$--$\gamma$ and
$Z$--$\gamma$ transitions. 
Owing to the photon--\PZ-boson mixing, the \PZ-boson self-energy,
$\hat{\Sigma}_Z(p^2)$, is effectively replaced by $\hat{Z}(p^2)$ 
[see \eqn{def:Z(p^2)}].

%  #] Appendix: The complex scheme vs the LEP1 scheme : 
%%  #[ Appendix: BHF1 treatment of e e-> e nu u d:
%%  #] Appendix: BHF1 treatment of e e-> e nu u d: 
%  #[ Bibliography :

%  #] Bibliography : 

\newpage

\begin{thebibliography}{99}

\bibitem{bhf1} 
E.N. Argyres et al., Phys.\ Lett.\ {\bf B358} (1995) 339.

\bibitem{Kurihara}
Y. Kurihara, D. Perret-Gallix and Y. Shimizu, Phys.\ Lett. {\bf B349}
(1995) 367.

\bibitem{LEPreport95} 
D. Bardin, W. Hollik and G. Passarino (eds.), {\sl Reports of the working
group on precision calculations for the \PZ~resonance}
(CERN 95-03, Gen\`eve, 1995).

\bibitem{LEP2WWreport}
W.~Beenakker et al., in {\sl Physics at LEP2},
eds.\ G.~Altarelli, T.~Sj\"ostrand and F.~Zwirner,
(CERN 96-01, Gen\`eve, 1996) Vol.~1, p.~79,
hep-ph/9602351.

\bibitem{NLCreport} 
P.M. Zerwas (ed.), {\sl $e^+e^-$ collisions at 500 GeV: The physics
potential} (DESY 93-123C, Hamburg, 1993);\\
W.~Beenakker and A.~Denner, Int. J. Mod. Phys. A9 (1994) 4837.

\bibitem{bfm}
A. Denner, S. Dittmaier and G. Weiglein, Nucl. Phys. {\bf B440} (1995) 95;\\
A. Denner and S. Dittmaier, Phys. Rev. {\bf D54} (1996) 4499.

\bibitem{VeltmanUnstable}
M. Veltman, Physica {\bf 29} (1963) 186;\\
R.~G. Stuart, Phys.~Lett. {\bf B262} (1991) 113.

\bibitem{bigwidth} W.~Beenakker, G.~J.~van Oldenborgh, J.~Hoogland and
R.~Kleiss, Phys.Lett. {\bf B376} (1996) 136. 

\bibitem{De93}
A. Denner, Fortschr. Phys. {\bf 41} (1993) 307.

\bibitem{tHooftVeltman}
G. 't Hooft and M. Veltman, Nucl. Phys. {\bf B50} (1972) 318;\\
for our conventions see e.g.\ Ref.~\cite{De93}.

\bibitem{LEP2MCreport}
D.~Bardin et al., in {\sl Physics at LEP2},
eds.\ G.~Altarelli, T.~Sj\"ostrand and F.~Zwirner,
(CERN 96-01, Gen\`eve, 1996) Vol.~2, p.~3.

\bibitem{ZeppenfeldBaur}
U. Baur and D. Zeppenfeld, Phys.\ Rev.\ Lett. {\bf 75} (1995) 1002.

\bibitem{Costas}
C.~G. Papadopoulos, Phys.\ Lett. {\bf B352} (1995) 144.

\bibitem{alpz} 
S.~Eidelman and F.~Jegerlehner,
Z.\ Phys.\ {\bf C67} (1995) 585;\\
H.~Burkhardt and B.~Pietrzyk,
Phys.\ Lett.\ {\bf B356} (1995) 398.

\bibitem{CDFD0top}
F.~Abe et al., Phys.\ Rev.\ Lett.\ {\bf 74} (1996) 2626; 
and contributed paper to ICHEP96, PA-08-018;\\
D0 Collaboration, S.~Abachi et al., contributed papers to ICHEP96,
PA-05-027 and PA-05-028;\\
P. Grannis, talk presented at ICHEP96, to appear in the proceedings.

\bibitem{ERATO}
C.~Papadopoulos, ERATO: event generator for four-fermion production at
LEP2 energies and beyond, hep-ph/9609320, to appear in Comp. Phys.
Commun.

\bibitem{WTO}
G.~Passarino, 
Comp.\ Phys.\ Comm. {\bf 97} (1996) 261;\\
G.~Passarino, Acta\ Phys.\ Pol. {\bf B27} (1996) 1605.

\bibitem{WWF}
G.~J.~van Oldenborgh, P.~Franzini, A.~Borrelli,
Comp.\ Phys.\ Comm. {\bf 83} (1994) 14.

\bibitem{mgpc} M.~Gr\"unewald, private communication.

\bibitem{WWFmasses}
J. Hoogland and G.J. van Oldenborgh, in preparation.

\bibitem{Beuthe}
M. Beuthe, R. Gonzalez Felipe, G. L\'opez Castro and J. Pestieau,
preprint UCL-IPT-96-20, hep-ph/9611434.

\end{thebibliography}
\end{document}